\documentclass[10pt,letterpaper,twocolumn,english,showpacs,prl,amsfonts,amssymb,balancelastpage,notitlepage,preprint]{revtex4}
\pdfoutput=1
\usepackage[]{fontenc}
\usepackage[latin1]{inputenc}
\usepackage{amsmath}
\usepackage{amssymb}
\usepackage{graphicx}
\usepackage{esint}

\makeatletter

\@ifundefined{textcolor}{}
{%
 \definecolor{BLACK}{gray}{0}
 \definecolor{WHITE}{gray}{1}
 \definecolor{RED}{rgb}{1,0,0}
 \definecolor{GREEN}{rgb}{0,1,0}
 \definecolor{BLUE}{rgb}{0,0,1}
 \definecolor{CYAN}{cmyk}{1,0,0,0}
 \definecolor{MAGENTA}{cmyk}{0,1,0,0}
 \definecolor{YELLOW}{cmyk}{0,0,1,0}
}

\usepackage{amsfonts}\usepackage{babel}
\usepackage{babel}
\providecommand{\U}[1]{\protect\rule{.1in}{.1in}}

\@ifundefined{textcolor}{}{
\definecolor{BLACK}{gray}{0}
 \definecolor{WHITE}{gray}{1}
 \definecolor{RED}{rgb}{1,0,0}
 \definecolor{GREEN}{rgb}{0,1,0}
 \definecolor{BLUE}{rgb}{0,0,1}
 \definecolor{CYAN}{cmyk}{1,0,0,0}
 \definecolor{MAGENTA}{cmyk}{0,1,0,0}
 \definecolor{YELLOW}{cmyk}{0,0,1,0}
}

\makeatother

\usepackage{babel}
\begin{document}

\title{Understanding and controlling $N$-dimensional quantum walks via
dispersion relations. Application to the 2D and 3D Grover walks: Diabolical
points and more}

\author{Germán J. de Valcárcel$^{(1)}$, Margarida Hinarejos$^{(2)}$, Eugenio
Roldán$^{(1)}$, Armando Pérez$^{(2)}$, and Alejandro Romanelli$^{(3)}$}

\affiliation{$^{(1)}$Departament d'Òptica, Universitat de València, Dr. Moliner
50, 46100-Burjassot, Spain, EU.}

\affiliation{$^{(2)}$Departament de Física Teòrica and IFIC, Universitat de València-CSIC,
Dr. Moliner 50, 46100-Burjassot, Spain, EU.}

\affiliation{$^{(3)}$Instituto de Física, Facultad de Ingeniería, Universidad
de la República, C.C. 30, C.P. 11000, Montevideo, Uruguay.}
\begin{abstract}
The discrete quantum walk in $N$ dimensions is analyzed from the
perspective of its dispersion relations. This allows understanding
known properties, as well as designing new ones when spatially extended
initial conditions are considered. This is done by deriving wave equations
in the continuum, which are generically of the Schrödinger type, and
allow devising interesting behaviors, such as ballistic propagation
without deformation, or the generation of almost flat probability
distributions, what is corroborated numerically. There are however
special points where the energy surfaces display intersections and,
near them, the dynamics is entirely different. Applications to the
two- and three-dimensional Grover walks are presented.
\end{abstract}


\maketitle

\section{Introduction}

Quantum walks (QWs) are becoming widespread in physics. Originally
introduced as quantum versions of random classical processes \cite{Aharonov93}
and quantum cellular automata \cite{Meyer96}, they have been deeply
investigated, specially in connection with quantum information science
\cite{Watrous01,Nayak00,Ambainis03,Kempe03,Kendon06,Kendon06b,Konno08},
and have been recently shown to constitute a universal model for quantum
computation \cite{prl102,pra81}. Remarkably, along the last decade
this simple quantum diffusion model has found connections with very
diverse phenomena such as optical diffraction and other wave phenomena
\cite{Knightc,German}, quantum games \cite{Flitney12}, Anderson
localization and quantum chaos \cite{Buerschaper,Wojcik,Romanelli,Romanelli2012},
and has even been identified as a generalized quantum measurement
device \cite{Kurzynski12}. Moreover, through proposals for and actual
implementations, QWs have found connections with the dynamics of quite
different physical systems such as linear optical systems and optical
cavities \cite{Bouwmester,Knight,Jeong03,Knightb,Roldan,Do,Banuls,Perets08,Schreiber},
optical lattices \cite{Dur02,Eckert05,Chandrashekar08,Science}, trapped
ions \cite{Travaglione02,Schmitz,Zaehringer}, and Bose-Einstein condensates
\cite{Chandrashekar06} among others. Along these recent years the
experimental capabilities in performing QWs have increased steadily
and nowadays there are devices that, e.g., implement two dimensional
QWs with a large number of propagation steps \cite{Schreiber12}.
For recent reviews on the field see \cite{Buzek11,Venegas12}.

Recently, some of us have been insisting in the interest of analyzing
the dynamics of coined QWs from the perspective of their dispersion
relation \cite{German,Roldan12}, which has been also used for purposes
different to ours (see, e.g., \cite{Kempf09,Stefanak12}). We think
this is an interesting approach that is particularly suitable for
non localized initial conditions, i.e., when the initial state is
described by a probability distribution extending on a finite region
in the lattice. When this region is not too small, the evolution of
the QW can be derived, to a good degree of approximation, from familiar
linear differential wave equations. Certainly the use of extended
initial conditions has been rare up to now \cite{Machida}, but we
think it is worth insisting in that they allow to reach final probability
distributions which, interestingly, can be tailored to some extent.
It is also worth mentioning that extended initial conditions also
make a connection with multiparticle quantum walks for noninteracting
particles \cite{Rohde11}.

In \cite{Roldan12} we have recently applied the above viewpoint in
studying \textit{alternate} QWs \cite{DiFranco11} in $N$ dimensions.
Alternate QWs are a simpler version of multidimensional QWs than the
standard one \cite{Mackay} as they only require a single qubit as
a coin (used $N$ times per time step) \cite{DiFranco11} instead
of using a 2$N$-qudit as the coin (used once per time step), as originally
defined in \cite{Mackay}. Here we shall carry out, from the dispersion
relation viewpoint, the study of multidimensional QWs in their standard
form. Below we make a general treatment, valid for $N$ dimensions,
and illustrate our results with the special cases of the Grover two-
and three-dimensional QWs. We notice that the 2D Grover-QW has received
most of the theoretical attention paid to multidimensional QWs \cite{Mackay,Tregenna03,Shenvi,Inui04,Carneiro,Adamczak,Marquezino,Venegas2,Watabe,Oliveira,Omar,Stefanak10,Stefanak08}.

We shall derive continuous wave equations, whose form depend on the
probability distribution at time $t=0$. To the leading order, these
equations are partial differential equations that can be written in
the form
\begin{equation}
\mathrm{i}\frac{\partial A^{\left(s\right)}\left(\mathbf{X},t\right)}{\partial t}=-\frac{1}{2}\sum_{i,j=1}^{N}\varpi_{ij}^{\left(s\right)}\frac{\partial^{2}A^{\left(s\right)}}{\partial X_{i}\partial X_{j}}+\cdots,\label{Sch}
\end{equation}
with $A^{\left(s\right)}$ a continuous amplitude probability, coefficients
$\varpi_{ij}^{\left(s\right)}$ determined by the QW dispersion relation
properties, and $X_{i}$ are the spatial coordinates in a reference
frame moving with the group velocity (see below for full details).
There is a lot of accumulated knowledge regarding the solutions of
the above continuous equations and this knowledge allows, to some
extent, for a qualitative knowledge of the long term probability distribution
of the QW for particular initial conditions. This qualitative knowledge
is particularly appealing to us, as we are interested, not in obtaining
approximate continuous solutions to the discrete QW, but in getting
a quick intuition of the QW evolution for different initial probability
distributions. This allows, in particular, to reach a desired asymptotic
distribution by suitably tailoring the initial (extended)\ state
\cite{German}. Of course, this qualitative knowledge will be only
approximate (as the continuous solutions are strictly valid only for
infinitely extended initial distributions) but, as we show below,
even for relatively narrow initial distributions the qualitative knowledge
turns out to be quite accurate. In each case, we will present an exact
numerical simulation, obtained from the discrete map of the QW, that
illustrates the agreement with the qualitative analysis described
above. As far as we know, this is the first time that continuous approximations
to the discrete QW are derived for a multidimensional QW.

The continuous equations we derive, however, cannot apply near degeneracies
in the dispersion relations, as in their vicinity the eigensolutions
of the QW vary wildly from point to point. Hence, close to degeneracies
a specific treatment is necessary and we develop such a treatment
for a particular type of degeneracies, namely conical intersections.
We apply this treatment to the two--dimensional Grover walk that exhibits
such conical intersections in the dispersion relation, a fact that
seems to have been noticed only recently \cite{Roldan12} and that
we study here in mathematical detail. These conical intersections,
called \textit{diabolical points} \cite{Berry1,Berry2,Berry3}, determine
a type of dynamics similar to that of massless Dirac fermions or of
electrons in graphene \cite{graphene1,graphene2,Ohta,Pleti,CastroNeto}
and appear in very different systems: quantum triangular billiards
\cite{Berry1}, conical refraction in crystal optics \cite{Berry2,Berry3},
the already mentioned graphene electrons \cite{graphene1,graphene2,Ohta,Pleti,CastroNeto},
the spectra of polyatomic molecules \cite{Mead79,Cederbaum03}, optical
lattices \cite{Zhang} or acoustic surface waves \cite{Torrent12},
just to mention a few. In all cases the remarkable physical properties
exhibited by these systems are directly associated with the existence
of the diabolical points. In the three--dimensional Grover walk we
also find degeneracies, but these are different to the diabolical
points. Diabolical points on the dispersion relations play also a
role on the analysis of topological properties of the QW. Configurations
with different topologies or Chern numbers can be manufactured, for
example, by introducing a spatial variation of one of the coin angles
in alternate QW's, such that these regions possess different topologies.
At the boundaries delimiting those regions the gap in the dispersion
relation closes, and one can observe the formation of bound states
or unidirectionally propagating modes, which are protected by topological
invariants and by the existence of the gap away from the boundary
\cite{Kitagawa}. 

The rest of the paper is organized as follows. In the next section
we introduce the $N$-dimensional QW and solve it formally. In Section
III we formally derive continuous equations in the general $N$--dimensional
case, and we discuss how to treat conical intersections. In Sections
IV and V we apply our results to the two- and three--dimensional Grover
QWs, respectively. Finally, in Sect. VI we give our main conclusions.

\section{N-dimensional discrete quantum walks. Generalities}

In the $N$-dimensional QW the walker moves at discrete time steps
$t\in\mathbb{N}$ across an $N$-dimensional lattice of sites $\mathbf{x}\equiv\left(x_{1},\ldots,x_{N}\right)\in\mathbb{Z}^{N}$.
The walker is endowed with a $2N$-dimensional coin which, after a
convenient unitary transformation, determines the direction of displacement.
The Hilbert space of the whole system (walker+coin) has then the form
\begin{equation}
\mathcal{H}=\mathcal{H}_{\mathrm{P}}\otimes\mathcal{H}_{\mathrm{C}},\label{espacio}
\end{equation}
where the position space, $\mathcal{H}_{\mathrm{P}}$, is spanned
by $\left\{ \left\vert \mathbf{x}\right\rangle \equiv\left\vert x_{1},\ldots,x_{N}\right\rangle :x_{\alpha}\in\mathbb{Z};\,\,\alpha=1,\ldots,N\right\} $
($\left\langle \mathbf{x}\right.\left\vert \mathbf{x}^{\prime}\right\rangle =\delta_{\mathbf{x,x}^{\prime}}$),
and the coin space, $\mathcal{H}_{\mathrm{C}}$, is spanned by $2N$
orthonormal quantum states $\left\{ \left\vert \alpha_{\eta}\right\rangle :\alpha=1,\ldots,N;\eta=\pm\right\} $.
Note that $\alpha$ is associated with the axis and $\eta$ with the
direction. For example, in the popular one dimensional QW ($N=1$)
we would have just $\left\vert 1_{+}\right\rangle $ and $\left\vert 1_{-}\right\rangle $,
which are the equivalent to the $\left\vert \mathrm{R}\right\rangle $
and $\left\vert \mathrm{L}\right\rangle $ (for right and left) states
commonly used in the literature. This notation is introduced in order
to deal easily with an arbitrary number of dimensions.

The state of the total system at time $t$ is represented by the ket
$\left\vert \psi_{t}\right\rangle $, which can be expressed in the
form 
\begin{equation}
\left\vert \psi_{t}\right\rangle =\sum_{\mathbf{x}}\sum_{\alpha=1}^{N}\sum_{\eta=\pm}\psi_{\mathbf{x},t}^{\alpha,\eta}\ \left\vert \mathbf{x}\right\rangle \otimes\left\vert \alpha_{\eta}\right\rangle ,\label{psi}
\end{equation}
where the projections 
\begin{equation}
\psi_{\mathbf{x},t}^{\alpha,\eta}=\left(\left\langle \alpha_{\eta}\right\vert \otimes\left\langle \mathbf{x}\right\vert \right)\left\vert \Psi_{t}\right\rangle ,\label{proj}
\end{equation}
are wave functions on the lattice. We find it convenient to define,
at each point $\mathbf{x}$, the following ket 
\begin{equation}
\left\vert \psi_{\mathbf{x},t}\right\rangle =\left\langle \mathbf{x}\right.\left\vert \psi_{t}\right\rangle =\sum_{\alpha=1}^{N}\sum_{\eta=\pm}\psi_{\mathbf{x},t}^{\alpha,\eta}\left\vert \alpha_{\eta}\right\rangle ,\label{psi_x}
\end{equation}
which is an (unnormalized) coin state, so that $\psi_{\mathbf{x},t}^{\alpha,\eta}=\langle\alpha_{\eta}\mid\psi_{\mathbf{x},t}\rangle$.
As $\left\vert \psi_{\mathbf{x},t}^{\alpha,\eta}\right\vert ^{2}=\left\vert \left(\left\langle \alpha_{\eta}\right\vert \otimes\left\langle \mathbf{x}\right\vert \right)\left\vert \psi_{t}\right\rangle \right\vert ^{2}$
is the probability of finding the walker at $\left(\mathbf{x},t\right)$,
and the coin in state $\left\vert \alpha_{\eta}\right\rangle $, the
probability of finding the walker at $\left(\mathbf{x},t\right)$
irrespectively of the coin state is, then, 
\begin{equation}
P_{\mathbf{x},t}=\sum_{\alpha=1}^{N}\sum_{\eta=\pm}\left\vert \psi_{\mathbf{x},t}^{\alpha,\eta}\right\vert ^{2}=\left\langle \psi_{\mathbf{x},t}\right.\left\vert \psi_{\mathbf{x},t}\right\rangle ,\label{prob}
\end{equation}
where we used the fact that $\sum_{\alpha=1}^{N}\sum_{\eta=\pm}\left\vert \alpha_{\eta}\right\rangle \left\langle \alpha_{\eta}\right\vert $
is the identity in $\mathcal{H}_{\mathrm{C}}$. Clearly $\sum_{\mathbf{x}}P_{\mathbf{x},t}=1$
because $\sum_{\mathbf{x}}\left\vert \mathbf{x}\right\rangle \left\langle \mathbf{x}\right\vert $
is the identity in $\mathcal{H}_{\mathrm{P}}$.

The dynamical evolution of the system is ruled by 
\begin{equation}
\left\vert \psi_{t+1}\right\rangle ={\hat{U}}\left\vert \psi_{t}\right\rangle ,\label{map}
\end{equation}
where the unitary operator 
\begin{equation}
\hat{U}=\hat{D}\circ\left(\hat{I}\otimes\hat{C}\right)\label{U}
\end{equation}
is given in terms of the identity operator in $\mathcal{H}_{\mathrm{P}}$,
$\hat{I}$, and two more unitary operators. On the one hand $\hat{C}$
is the so-called coin operator (an operator in $\mathcal{H}_{\mathrm{C}}$),
which can be written in its more general form as 
\begin{equation}
\hat{C}=\sum_{\alpha,\alpha^{\prime}=1}^{N}\,\,\sum_{\eta,\eta^{\prime}=\pm}C_{\alpha^{\prime},\eta^{\prime}}^{\alpha,\eta}\left\vert \alpha_{\eta}\right\rangle \left\langle \alpha_{\eta^{\prime}}^{\prime}\right\vert ,\label{C}
\end{equation}
where the matrix elements $C_{\alpha^{\prime},\eta^{\prime}}^{\alpha,\eta}\equiv\left\langle \alpha_{\eta}\right\vert \hat{C}\left\vert \alpha_{\eta^{\prime}}^{\prime}\right\rangle $
can be arranged as a $2N\times2N$ unitary square matrix $C$. On
the other hand $\hat{D}$ is the conditional displacement operator
in $\mathcal{H}$ 
\begin{equation}
\hat{D}=\sum_{\mathbf{x}}\sum_{\alpha=1}^{N}\sum_{\eta=\pm}\left\vert \mathbf{x}+\eta\mathbf{u}_{\alpha}\right\rangle \left\langle \mathbf{x}\right\vert \otimes\left\vert \alpha_{\eta}\right\rangle \left\langle \alpha_{\eta}\right\vert ,\label{D}
\end{equation}
where $\mathbf{u}_{\alpha}$ is the unit vector along direction $x_{\alpha}$;
note that, depending on the coin state $\left\vert \alpha_{\eta}\right\rangle $,
the walker moves one site to the positive or negative direction of
$x_{\alpha}$ if $\eta=+$ or $\eta=-$, respectively.

Projecting (\ref{map}) onto $\left\langle \mathbf{x}\right\vert $
and using (\ref{proj}) and (\ref{U})--(\ref{D}) we get straightforwardly
\begin{equation}
\left\vert \psi_{\mathbf{x},t+1}\right\rangle =\sum_{\alpha=1}^{N}\sum_{\eta=\pm}\left\vert \alpha_{\eta}\right\rangle \left\langle \alpha_{\eta}\right\vert \hat{C}\left\vert \psi_{\mathbf{x}-\eta\mathbf{u}_{\alpha},t}\right\rangle ,\label{mapket_x}
\end{equation}
which further projected onto $\left\langle \alpha_{\eta}\right\vert $
leads to 
\begin{equation}
\psi_{\mathbf{x},t+1}^{\alpha,\eta}=\sum_{\alpha^{\prime}=1}^{N}\sum_{\eta^{\prime}=\pm}C_{\alpha^{\prime},\eta^{\prime}}^{\alpha,\eta}\psi_{\mathbf{x}-\eta\mathbf{u}_{\alpha},t}^{\alpha^{\prime},\eta^{\prime}}.\label{map_x}
\end{equation}
Equation (\ref{mapket_x}), or equivalently (\ref{map_x}), is the
NDQW map in position representation; it shows that, at each (discrete)
time, the wavefunctions at each point are coherent linear superpositions
of wavefunctions at neighboring points at previous time, the weights
of the superposition being given by the coin operator matrix elements
$C_{\alpha^{\prime},\eta^{\prime}}^{\alpha,\eta}$. Next we proceed
to derive the solution of map (\ref{map_x}).

Given the linearity of the map and the fact that it is space-invariant
($C_{\alpha^{\prime},\eta^{\prime}}^{\alpha,\eta}$ do not depend
on space) a useful technique here is the spatial Discrete Fourier
Transform (DFT), which has been used many times in QW studies (see,
for example,  \cite{Ambainis01}). First we define the DFT pair 
\begin{align}
\left\vert \tilde{\psi}_{\mathbf{k,}t}\right\rangle  & \equiv\sum_{\mathbf{x}}e^{-\mathrm{i}\mathbf{k}\cdot\mathbf{x}}\left\vert \psi_{\mathbf{x},t}\right\rangle ,\label{DFT_k}\\
\left\vert \psi_{\mathbf{x},t}\right\rangle  & \equiv\int\frac{\mathrm{d}^{N}\mathbf{k}}{\left(2\pi\right)^{N}}e^{\mathrm{i}\mathbf{k}\cdot\mathbf{x}}\left\vert \tilde{\psi}_{\mathbf{k},t}\right\rangle ,\label{DFT_x}
\end{align}
where $\mathbf{k}=\left(k_{1},\ldots,k_{N}\right)$ and $k_{\alpha}\in\left[-\pi,\pi\right]$
is the (quasi-)momentum vector \cite{domain-k}. Applying the previous
definitions to the map (\ref{mapket_x}) we readily get 
\begin{equation}
\left\vert \tilde{\psi}_{\mathbf{k},t+1}\right\rangle =\hat{C}_{\mathbf{k}}\left\vert \tilde{\psi}_{\mathbf{k},t}\right\rangle ,\label{mapket_k}
\end{equation}
where we defined a coin operator in the quasi-momentum space 
\begin{equation}
\hat{C}_{\mathbf{k}}\equiv\sum_{\alpha=1}^{N}\sum_{\eta=\pm}\left\vert \alpha_{\eta}\right\rangle \left\langle \alpha_{\eta}\right\vert \hat{C}e^{-\mathrm{i}\eta k_{\alpha}},\label{Ck}
\end{equation}
$k_{\alpha}=\mathbf{k}\cdot\mathbf{u}_{\alpha}$, whose matrix elements
read 
\begin{equation}
\left\langle \alpha_{\eta}\right\vert \hat{C}_{\mathbf{k}}\left\vert \alpha_{\eta^{\prime}}^{\prime}\right\rangle \equiv\left(C_{\mathbf{k}}\right)_{\alpha^{\prime},\eta^{\prime}}^{\alpha,\eta}=e^{-\mathrm{i}\eta k_{\alpha}}C_{\alpha^{\prime},\eta^{\prime}}^{\alpha,\eta}.\label{Ck_elements}
\end{equation}
Projection of (\ref{mapket_k}) onto $\left\langle \alpha_{\eta}\right\vert $
and use of (\ref{Ck},\ref{Ck_elements}) leads to 
\begin{equation}
\tilde{\psi}_{\mathbf{k},t+1}^{\alpha,\eta}=\sum_{\alpha^{\prime}=1}^{N}\sum_{\eta^{\prime}=\pm}e^{-\mathrm{i}\eta\mathbf{k}\cdot\mathbf{u}_{\alpha}}C_{\alpha^{\prime},\eta^{\prime}}^{\alpha,\eta}\tilde{\psi}_{\mathbf{k},t}^{\alpha^{\prime},\eta^{\prime}}.\label{map_k}
\end{equation}
Hence the nonlocal maps (\ref{mapket_x},\ref{map_x}) become local
in the momentum representation (\ref{mapket_k},\ref{map_k}). This
allows solving formally the QW dynamics very easily because map (\ref{mapket_k})
implies 
\begin{equation}
\left\vert \tilde{\psi}_{\mathbf{k},t}\right\rangle =\left(\hat{C}_{\mathbf{k}}\right)^{t}\left\vert \tilde{\psi}_{\mathbf{k},0}\right\rangle ,\label{evol_k}
\end{equation}
and hence the eigensystem of $\hat{C}_{\mathbf{k}}$ (or of $C_{\mathbf{k}}$
in matrix form) is most useful for solving the problem, as we do next.

As the operator $\hat{C}_{\mathbf{k}}\mathcal{\ }$is unitary, its
eigenvalues have all the form $\lambda_{\mathbf{k}}^{\left(s\right)}=\exp\left(-\mathrm{i}\omega_{\mathbf{k}}^{\left(s\right)}\right)$,
$s=1,\ldots,N$, with $\omega_{\mathbf{k}}^{\left(s\right)}$ real.
We will need to know the $\hat{C}_{\mathbf{k}}$ eigenstates too,
$\left\{ \left\vert \phi_{\mathbf{k}}^{\left(s\right)}\right\rangle \right\} _{s=1}^{2N}$.
Once the eigensystem of $\hat{C}_{\mathbf{k}}$ is known, implementing
(\ref{evol_k}) is trivial: Given an initial distribution of the walker
in position representation $\left\vert \psi_{\mathbf{x},0}\right\rangle $
we compute its DFT $\left\vert \tilde{\psi}_{\mathbf{k},0}\right\rangle $
via (\ref{DFT_k}), as well as the projections 
\begin{equation}
\tilde{f}_{\mathbf{k}}^{\left(s\right)}=\left\langle \phi_{\mathbf{k}}^{\left(s\right)}\right.\left\vert \tilde{\psi}_{\mathbf{k},0}\right\rangle ,\label{fsk}
\end{equation}
so that $\left\vert \tilde{\psi}_{\mathbf{k},0}\right\rangle =$ $\sum_{s}\tilde{f}_{\mathbf{k}}^{\left(s\right)}\left\vert \phi_{\mathbf{k}}^{\left(s\right)}\right\rangle $.
Now recalling (\ref{evol_k}) we arrive to 
\begin{equation}
\left\vert \tilde{\psi}_{\mathbf{k},t}\right\rangle =\sum_{s=1}^{2N}e^{-\mathrm{i}\omega_{\mathbf{k}}^{\left(s\right)}t}\tilde{f}_{\mathbf{k}}^{\left(s\right)}\left\vert \phi_{\mathbf{k}}^{\left(s\right)}\right\rangle ,
\end{equation}
where we used $\lambda_{\mathbf{k}}^{\left(s\right)}=\exp\left(-\mathrm{i}\omega_{\mathbf{k}}^{\left(s\right)}\right)$,
while in position representation we get, using (\ref{DFT_x}), 
\begin{align}
\left\vert \psi_{\mathbf{x},t}\right\rangle  & =\sum_{s=1}^{2N}\left\vert \psi_{\mathbf{x},t}^{\left(s\right)}\right\rangle ,\label{evol_x}\\
\left\vert \psi_{\mathbf{x},t}^{\left(s\right)}\right\rangle  & =\int\frac{\mathrm{d}^{N}\mathbf{k}}{\left(2\pi\right)^{N}}e^{\mathrm{i}\left(\mathbf{k}\cdot\mathbf{x-}\omega_{\mathbf{k}}^{\left(s\right)}t\right)}\tilde{f}_{\mathbf{k}}^{\left(s\right)}\left\vert \phi_{\mathbf{k}}^{\left(s\right)}\right\rangle .\label{evol_x_s}
\end{align}
Hence the QW is formally solved: all we need is to compute the eigensystem
of $\hat{C}_{\mathbf{k}}$ and the initial state in reciprocal space
$\left\vert \tilde{\psi}_{\mathbf{k},0}\right\rangle $, which determines
the weight functions $\tilde{f}_{\mathbf{k}}^{\left(s\right)}$ through
(\ref{fsk}).

Equation (\ref{evol_x}) shows that the QW dynamics corresponds to
the superposition of $2N$ independent walks, labeled by $s$. According
to (\ref{evol_x_s}) the $\omega_{\mathbf{k}}^{\left(s\right)}$'s
are the frequencies of the map, each of which defines a dispersion
relation in the system ($2N$ in total). Note as well that what we
have done in the end is to decompose the QW dynamics in terms of plane
waves. In particular, if $\tilde{f}_{\mathbf{k}}^{\left(s\right)}=\delta^{(N)}\left(\mathbf{k}-\mathbf{k}_{0}\right)$,
what means that $\left\vert \tilde{\psi}_{\mathbf{k},0}\right\rangle $
is different from zero only for $\mathbf{k}=\mathbf{k}_{0}$, $\left\vert \psi_{\mathbf{x},t}^{\left(s\right)}\right\rangle =\left(2\pi\right)^{-N}\exp\left[\mathrm{i}\left(\mathbf{k}_{0}\cdot\mathbf{x-}\omega_{\mathbf{k}_{0}}^{\left(s\right)}t\right)\right]\left\vert \phi_{\mathbf{k}_{0}}^{\left(s\right)}\right\rangle $,
which is an unnormalizable plane wave and thus unphysical.

In order to avoid possible confusions, to conclude this initial part
we state that the ordering of the coin base elements we will be using
in the matrix representations of operators and kets is $\left\vert 1_{+}\right\rangle ,\left\vert 1_{-}\right\rangle ,\ldots\left\vert N_{+}\right\rangle ,\left\vert N_{-}\right\rangle $.

\section{Continuous wave equations for spatially extended initial conditions}

In this Section we describe the evolution of spatially extended initial
conditions that are close to the plane waves we have introduced. We
define such spatially extended states as those having a width which
is appreciably larger than the lattice spacing (taken as unity in
this work). This type of initial states are wavepackets which are
easily expressed in reciprocal space, 
\begin{equation}
\left\vert \tilde{\psi}_{\mathbf{k},0}\right\rangle =\sum_{s=1}^{2N}\tilde{F}_{\mathbf{k}-\mathbf{k}_{0}}^{\left(s\right)}\left\vert \phi_{\mathbf{k}_{0}}^{\left(s\right)}\right\rangle ,\label{Inik0}
\end{equation}
where $\mathbf{k}_{0}$ is a reference (carrier) wavevector, chosen
at will, $\left\vert \phi_{\mathbf{k}_{0}}^{\left(s\right)}\right\rangle $
are the associated eigenvectors, and $\tilde{F}_{\mathbf{k}}^{\left(s\right)}$
is a narrow function of $\mathbf{k}$, centered at $\mathbf{k}=0$
($\tilde{F}_{\mathbf{k}-\mathbf{k}_{0}}^{\left(s\right)}$ is centered
at $\mathbf{k}=\mathbf{k}_{0}$) and having a very small width $\Delta k^{\left(s\right)}\ll\pi$
\cite{SD}. Notice that we have chosen an initial coin state which
is independent of $\mathbf{k}$. From here one must distinguish between
\textit{regular points}, where eigenvectors $\left\vert \phi_{\mathbf{k}}^{\left(s\right)}\right\rangle $
have a smooth dependence on $\mathbf{k}$ close to $\mathbf{k}_{0}$,
and \textit{degeneracy points}, where eigenvectors have wild variations
around them, as we will see later.

According to (\ref{DFT_x}) the initial condition (\ref{Inik0}) reads
in position representation 
\begin{equation}
\left\vert \psi_{\mathbf{x},0}\right\rangle =e^{\mathrm{i}\mathbf{k}_{0}\cdot\mathbf{x}}\sum_{s=1}^{2N}F_{\mathbf{x},0}^{\left(s\right)}\left\vert \phi_{\mathbf{k}_{0}}^{\left(s\right)}\right\rangle ,\label{Inix}
\end{equation}
where 
\begin{equation}
F_{\mathbf{x},0}^{\left(s\right)}=\int\frac{\mathrm{d}^{N}\mathbf{k}}{\left(2\pi\right)^{N}}e^{\mathrm{i}\left(\mathbf{k}-\mathbf{k}_{0}\right)\cdot\mathbf{x}}\tilde{F}_{\mathbf{k}-\mathbf{k}_{0}}^{\left(s\right)}
\end{equation}
is a wide and smooth function of $\mathbf{x}$ because $\tilde{F}_{\mathbf{k}}^{\left(s\right)}$
is concentrated around $\mathbf{k}=0$. Hence in real space our initial
condition consists of a coin state $\left\vert \phi_{\mathbf{k}_{0}}^{\left(s\right)}\right\rangle $
equal at all points, multiplied by the carrier $\exp\left[\mathrm{i}\left(\mathbf{k}_{0}\cdot\mathbf{x}\right)\right]$
and by a wide and smooth function of space, $F_{\mathbf{x},0}^{\left(s\right)}$.
We see then that the type of initial conditions we are dealing with
are very close to the plane waves of momentum $\mathbf{k}_{0}$ of
the QW, given by $e^{\mathrm{i}\mathbf{k}_{0}\cdot\mathbf{x}}\left\vert \phi_{\mathbf{k}_{0}}^{\left(s\right)}\right\rangle $.

\subsection{Regular points}

For regular points (i.e., far from degeneracies)\ the initial condition
(\ref{Inik0}) determines the weight functions (\ref{fsk}) as 
\begin{equation}
\tilde{f}_{\mathbf{k}}^{\left(s\right)}=\sum_{s^{\prime}}\tilde{F}_{\mathbf{k}-\mathbf{k}_{0}}^{\left(s^{\prime}\right)}\langle\phi_{\mathbf{k}}^{\left(s\right)}\mid\phi_{\mathbf{k}_{0}}^{\left(s^{\prime}\right)}\rangle=\tilde{F}_{\mathbf{k}-\mathbf{k}_{0}}^{\left(s\right)}+O\left(\Delta k^{\left(s\right)}\right),\label{fks_approx}
\end{equation}
where we took into account that the eigenvectors $\left\vert \phi_{\mathbf{k}_{0}}^{\left(s\right)}\right\rangle $
vary smoothly around $\mathbf{k}=\mathbf{k}_{0}$ and that only for
$\mathbf{k\approx k}_{0}$ the function $\tilde{F}_{\mathbf{k}-\mathbf{k}_{0}}^{\left(s\right)}$
is non-vanishing. Note that we cannot make such an approximation in
the case of degeneracy points because of the strong variations of
the eigenvectors around them.

Now, the system will evolve according to (\ref{evol_x_s}). Approximating
the eigenvector $\left\vert \phi_{\mathbf{k}}^{\left(s\right)}\right\rangle $
appearing in (\ref{evol_x_s}) by $\left\vert \phi_{\mathbf{k}_{0}}^{\left(s\right)}\right\rangle $
using the same arguments as before, the partial waves $\left\vert \psi_{\mathbf{x},t}^{\left(s\right)}\right\rangle $
can be written as 
\begin{align}
\left\vert \psi_{\mathbf{x},t}^{\left(s\right)}\right\rangle  & =F_{\mathbf{x},t}^{\left(s\right)}\exp\left[\mathrm{i}\left(\mathbf{k}_{0}\cdot\mathbf{x-}\omega_{\mathbf{k}_{0}}^{\left(s\right)}t\right)\right]\left\vert \phi_{\mathbf{k}_{0}}^{\left(s\right)}\right\rangle +O\left(\Delta k^{\left(s\right)}\right),\label{psis-Fs}\\
F_{\mathbf{x},t}^{\left(s\right)} & =\int\frac{\mathrm{d}^{N}\mathbf{k}}{\left(2\pi\right)^{N}}\exp\left(\mathrm{i}\left[\left(\mathbf{k}-\mathbf{k}_{0}\right)\cdot\mathbf{x-}\left(\omega_{\mathbf{k}}^{\left(s\right)}-\omega_{\mathbf{k}_{0}}^{\left(s\right)}\right)t\right]\right)\nonumber \\
\times & \tilde{F}_{\mathbf{k}-\mathbf{k}_{0}}^{\left(s\right)},\label{F_discrete}
\end{align}
where we have defined new functions $\left\{ F_{\mathbf{x},t}^{\left(s\right)}\right\} _{s=1}^{2N}$.
Note that, at $t=0$, $\left\vert \psi_{\mathbf{x},0}^{\left(s\right)}\right\rangle =F_{\mathbf{x},0}^{\left(s\right)}\exp\left[\mathrm{i}\left(\mathbf{k}_{0}\cdot\mathbf{x}\right)\right]\left\vert \phi_{\mathbf{k}_{0}}^{\left(s\right)}\right\rangle $
, in agreement with (\ref{Inix}).

The problem is then solved if functions $\left\{ F_{\mathbf{x},t}^{\left(s\right)}\right\} _{s=1}^{2N}$
are determined. Instead of doing so by brute force, i.e. by integrating
--maybe numerically-- Eq. (\ref{F_discrete}), what we do now is to
look for a continuous wave equation that, by comparison with other
known cases, sheds light onto the expected behavior of the QW. In
order to test the quality of our analysis based on our continuous
equations and the knowledge of the dispersion relations, we will present
some plots that are obtained directly by iteration of Eq. (\ref{map_x})
without any approximation.

To avoid confusion, we introduce a function $F^{\left(s\right)}\left(\mathbf{x},t\right)$
of continuous real arguments exactly in the same way as in (\ref{F_discrete})
as there is nothing forbidding such a definition. We have then $F_{\mathbf{x},t}^{\left(s\right)}=F^{\left(s\right)}\left(\mathbf{x},t\right)$
for $\mathbf{x\in\mathbb{Z}}^{N}$ and $t\in\mathbb{N}$. In order
to simplify the derivation we rewrite Eq. (\ref{F_discrete}) by making
the variable change $\mathbf{k}-\mathbf{k}_{0}\rightarrow\mathbf{k}$.
Finally, as for the limits of integration (originally from $-\pi$
to $+\pi$ for each dimension), we extend them from $-\infty$ to
$+\infty$ in agreement with the continuous limit. We then have 
\begin{align}
F^{\left(s\right)}\left(\mathbf{x},t\right) & \equiv\int\frac{\mathrm{d}^{N}\mathbf{k}}{\left(2\pi\right)^{N}}\exp\left[\mathrm{i}\left(\mathbf{k}\cdot\mathbf{x-}\left[\omega_{\mathbf{k}_{0}+\mathbf{k}}^{\left(s\right)}-\omega_{\mathbf{k}_{0}}^{\left(s\right)}\right]t\right)\right]\nonumber \\
\times & \tilde{F}_{\mathbf{k}}^{\left(s\right)}.\label{F_cont}
\end{align}

Let us obtain the (approximate) wave equation. First we take the time
derivative of (\ref{F_cont}) and get 
\begin{align}
\mathrm{i}\partial_{t}F^{\left(s\right)}\left(\mathbf{x},t\right) & =\int\frac{\mathrm{d}^{N}\mathbf{k}}{\left(2\pi\right)^{N}}\exp\left[\mathrm{i}\left(\mathbf{k}\cdot\mathbf{x-}\left[\omega_{\mathbf{k}_{0}+\mathbf{k}}^{\left(s\right)}-\omega_{\mathbf{k}_{0}}^{\left(s\right)}\right]t\right)\right]\nonumber \\
 & \times\left[\omega_{\mathbf{k}_{0}+\mathbf{k}}^{\left(s\right)}-\omega_{\mathbf{k}_{0}}^{\left(s\right)}\right]\tilde{F}_{\mathbf{k}}^{\left(s\right)}.\label{dFdtaux}
\end{align}
Then we Taylor expand the function $\left[\omega_{\mathbf{k}_{0}+\mathbf{k}}^{\left(s\right)}-\omega_{\mathbf{k}_{0}}^{\left(s\right)}\right]$
(except in the exponent: otherwise large errors at long times would
be introduced) around $\mathbf{k}=0$, 
\begin{equation}
\omega_{\mathbf{k}_{0}+\mathbf{k}}^{\left(s\right)}-\omega_{\mathbf{k}_{0}}^{\left(s\right)}=\sum_{i=1}^{N}\varpi_{i}^{\left(s\right)}k_{i}+\frac{1}{2!}\sum_{i,j=1}^{N}\varpi_{ij}^{\left(s\right)}k_{i}k_{j}+\cdots,
\end{equation}
where 
\begin{align}
\varpi_{i}^{\left(s\right)} & =\left.\partial\omega_{\mathbf{k}_{0}+\mathbf{k}}^{\left(s\right)}/\partial k_{i}\right\vert _{\mathbf{k}=0}\nonumber \\
 & =\left.\partial\omega_{\mathbf{k}}^{\left(s\right)}/\partial k_{i}\right\vert _{\mathbf{k}=\mathbf{k}_{0}}=\left[\mathbf{v}_{\mathrm{g}}^{\left(s\right)}\left(\mathbf{k}_{0}\right)\right]_{i},\\
\ \varpi_{ij}^{\left(s\right)} & =\left.\partial^{2}\omega_{\mathbf{k}_{0}+\mathbf{k}}^{\left(s\right)}/\partial k_{i}\partial k_{j}\right\vert _{\mathbf{k}=0}\nonumber \\
 & =\left.\partial^{2}\omega_{\mathbf{k}}^{\left(s\right)}/\partial k_{i}\partial k_{j}\right\vert _{\mathbf{k}=\mathbf{k}_{0}},
\end{align}
etc. In the latter equation, $\mathbf{v}_{\mathrm{g}}^{\left(s\right)}\left(\mathbf{k}_{0}\right)=\left.\nabla_{\mathbf{k}}\omega_{\mathbf{k}}^{\left(s\right)}\right\vert _{\mathbf{k}=\mathbf{k}_{0}}$
is the group velocity at the point $\mathbf{k}=\mathbf{k}_{0}$ (see
discussion below). Now it is trivial to transform the so-obtained
right hand side of Eq. (\ref{dFdtaux}) as a sum of spatial derivatives
of $F^{\left(s\right)}\left(\mathbf{x},t\right)$, leading to the
result 
\begin{align}
\mathrm{i}\frac{\partial F^{\left(s\right)}\left(\mathbf{x},t\right)}{\partial t} & =-\mathrm{i}\mathbf{v}_{\mathrm{g}}^{\left(s\right)}\left(\mathbf{k}_{0}\right)\cdot\nabla F^{\left(s\right)}\nonumber \\
 & -\frac{1}{2!}\sum_{i,j=1}^{N}\varpi_{ij}^{\left(s\right)}\frac{\partial^{2}F^{\left(s\right)}}{\partial x_{i}\partial x_{j}}+\cdots,\label{WE}
\end{align}
which is the sought continuous wave equation. One should understand
that, in general, few terms are necessary on the right hand side of
Eq. (\ref{WE}) because $F^{\left(s\right)}\left(\mathbf{x},t\right)$
is a slowly varying function of space, as commented.

The first term on the right hand side of the wave equation is an advection
term, implying that the initial condition $F_{0}^{\left(s\right)}\left(\mathbf{x}\right)\equiv F^{\left(s\right)}\left(\mathbf{x},0\right)$
is shifted in time as $F^{\left(s\right)}\left(\mathbf{x},t\right)=F_{0}^{\left(s\right)}\left(\mathbf{x}-\mathbf{v}_{\mathrm{g}}^{\left(s\right)}\left(\mathbf{k}_{0}\right)t\right)$
without distortion, to the leading order. In fact we can get rid of
this term by defining a moving reference frame $\mathbf{X}$ such
that 
\begin{equation}
\mathbf{X}=\mathbf{x}-\mathbf{v}_{\mathrm{g}}^{\left(s\right)}\left(\mathbf{k}_{0}\right)t,\ \ \ A^{\left(s\right)}\left(\mathbf{X},t\right)=F^{\left(s\right)}\left(\mathbf{x},t\right),
\end{equation}
and then Eq. (\ref{WE}) becomes 
\begin{equation}
\mathrm{i}\frac{\partial A^{\left(s\right)}\left(\mathbf{X},t\right)}{\partial t}=-\frac{1}{2}\sum_{i,j=1}^{N}\varpi_{ij}^{\left(s\right)}\frac{\partial^{2}A^{\left(s\right)}}{\partial X_{i}\partial X_{j}}+\cdots,\label{Schr}
\end{equation}
which is an $N$-dimensional Schrödinger-like equation, to the leading
order. This means that the evolution of the wave packet consists of
the advection of the initial wave packet at the corresponding group
velocity and, on top of that, the wavefunction itself evolves according
to (\ref{Schr}). Equation (\ref{Schr}) is a main result of this
article as it governs the nontrivial dynamics of the QW when spatially
extended initial conditions, as we have defined them, are considered.
It evidences the role played by the dispersion relations as anticipated:
For distributions whose DFT is centered around some $\mathbf{k}_{0}$,
the local variations of $\omega$ around $\mathbf{k}_{0}$ determine
the type of wave equation controlling the QW dynamics.

Just two additional remarks: First, if the initial condition only
projects onto one of the sheets of the dispersion relation (hence
the sum in $s$ in (\ref{Inik0}) reduces to a single element), all
$F^{\left(s\right)}\left(\mathbf{X},t\right)$ will be zero, except
the one corresponding to the chosen eigenvector. This means that the
coin state is preserved along the evolution, to the leading order.
Second, if the initial state projects onto several eigenvectors the
probability of finding the walker at $\left(\mathbf{x},t\right)$,
irrespectively of the coin state, follows from (\ref{prob}), (\ref{evol_x})
and (\ref{psis-Fs}), and reads 
\begin{equation}
P_{\mathbf{x},t}=\sum_{s}\left\vert F_{\mathbf{x},t}^{\left(s\right)}\right\vert ^{2}=\sum_{s}\left\vert A^{\left(s\right)}\left(\mathbf{X},t\right)\right\vert ^{2}\boldsymbol{,}
\end{equation}
to the leading order, where we used $\langle\phi_{\mathbf{k}_{0}}^{\left(s\right)}\mid\phi_{\mathbf{k}_{0}}^{\left(s^{\prime}\right)}\rangle=\delta_{s,s^{\prime}}$.
Hence $P_{\mathbf{x},t}$ is just the sum of partial probabilities:
there is not interference among the different sub-QW's.

\subsection{Degeneracy points}

This case requires special care because there are eigenvectors of
the QW that vary strongly around the degeneracy, forbidding the very
initial approximation taken in the case of regular points, namely
Eq. (\ref{fks_approx}). Given the singular nature of the problem
we will try to give a rather general (but not fully general) theory
here: We will present a treatment that covers the case of (hyper-)
conical intersections. These appear in the 2D Grover walk (as we will
see in the next section), in the 3D Alternate QW \cite{Roldan12}
(that will not be treated here) and, very likely, in some other other
cases. The basic assumptions are: (i) There are just two (hyper-)sheets
in the dispersion relation that display a conical intersection at
$\mathbf{k}=\mathbf{k}_{\mathrm{D}}$ (diabolical point) and we label
them by $s=1,2$ for the sake of definiteness; and (ii) There are
other sheets, degenerate with the diabolical ones at $\mathbf{k}=\mathbf{k}_{\mathrm{D}}$,
whose associated frequencies $\omega_{\mathbf{k}}^{\left(s\right)}$
are constant around the diabolical point.

In order to alleviate the notation we will denote by $\omega_{_{\mathrm{D}}}$
the value of the degenerate frequencies at $\mathbf{k}=\mathbf{k}_{\mathrm{D}}$
\begin{equation}
\omega_{_{\mathrm{D}}}=\omega_{\mathbf{k}_{\mathrm{D}}}^{\left(1\right)}=\omega_{\mathbf{k}_{\mathrm{D}}}^{\left(2\right)}.
\end{equation}
Notice that there can be other sheets with $\omega_{\mathbf{k}_{\mathrm{D}}}^{\left(s\neq1,2\right)}=\omega_{_{\mathrm{D}}}$
(as it happens with $\omega_{\mathbf{k}_{\mathrm{D}}}^{\left(3\right)}$
in the 2D Grover map).

We consider an initial wave packet defined in reciprocal space by
\begin{equation}
\left\vert \tilde{\psi}_{\mathbf{k},0}\right\rangle =\tilde{F}_{\mathbf{k}-\mathbf{k}_{\mathrm{D}}}\left\vert \Xi\right\rangle ,\label{InikD}
\end{equation}
where $\left\vert \Xi\right\rangle $\ is an eigenstate of the coin
operator $\hat{C}_{\mathbf{k}}$ at $\mathbf{k}=\mathbf{k}_{\mathrm{D}}$,
and $\tilde{F}_{\mathbf{k}-\mathbf{k}_{\mathrm{D}}}$ is again a narrow
function centered at $\mathbf{k}=\mathbf{k}_{\mathrm{D}}$. In the
position representation this initial condition reads, see (\ref{DFT_x}),
\begin{equation}
\mid\psi_{\mathbf{x},0}\rangle=e^{\mathrm{i}\mathbf{k}_{\mathrm{D}}\cdot\mathbf{x}}F_{\mathbf{x},0}\left\vert \Xi\right\rangle ,\label{InixD}
\end{equation}
where $F_{\mathbf{x},0}=\int\frac{\mathrm{d}^{N}\mathbf{k}}{\left(2\pi\right)^{N}}e^{\mathrm{i}\left(\mathbf{k-k}_{\mathrm{D}}\right)\cdot\mathbf{x}}\tilde{F}_{\mathbf{k}-\mathbf{k}_{\mathrm{D}}}$
is the DFT of $\tilde{F}_{\mathbf{k}}$. Hence, as in the regular
case, $\left\vert \Xi\right\rangle $ represents the state of the
coin at the initial condition. Clearly $\left\vert \Xi\right\rangle $
is not unique because of the degeneracy. We do not use the notation
$\left\vert \phi_{\mathbf{k}_{\mathrm{D}}}^{\left(s\right)}\right\rangle $
but instead $\left\vert \Xi\right\rangle $ because the former are
ill defined, as we have seen in the example of the 2D Grover walk.

The initial condition (\ref{InikD}) determines the weight functions
(\ref{fsk}) as 
\begin{equation}
\tilde{f}_{\mathbf{k}}^{\left(s\right)}=\tilde{F}_{\mathbf{k}-\mathbf{k}_{\mathrm{D}}}\langle\phi_{\mathbf{k}}^{\left(s\right)}\mid\Xi\rangle.\label{fksD}
\end{equation}
We will assume that $\left\vert \Xi\right\rangle $ does not project
on regular sheets (those not degenerate with the conical intersection);
otherwise those projections will evolve as described in the previous
case of regular points. This is equivalent to saying that, in the
remainder of this subsection, all sums on $s$ will be restricted
to sheets that are degenerate at the conical intersection.

Definition (\ref{fksD}) poses no problem but for $\mathbf{k}=\mathbf{k}_{\mathrm{D}}$.
However, as this is a single point (a set of null measure) its influence
on the final result, given by integrals in $\mathbf{k}$, is null
\cite{null measure}.

A main difference with the regular case is seen at this stage because
there is not any choice of the initial coin state $\left\vert \Xi\right\rangle $
making that only one value of $s$ be populated (remind that $\left\vert \phi_{\mathbf{k}}^{\left(s\right)}\right\rangle $
vary strongly around $\mathbf{k}=\mathbf{k}_{\mathrm{D}}$). This
has consequences as we see next.

Now the system will evolve according to (\ref{evol_x_s}). The partial
waves $\left\vert \psi_{\mathbf{x},t}^{\left(s\right)}\right\rangle $
read now 
\begin{align}
\left\vert \psi_{\mathbf{x},t}^{\left(s\right)}\right\rangle  & =e^{\mathrm{i}\left(\mathbf{k}_{\mathrm{D}}\cdot\mathbf{x-}\omega_{_{\mathrm{D}}}t\right)}\left\vert \mathbf{F}_{\mathbf{x},t}^{\left(s\right)}\right\rangle \label{psiD_aux}\\
\left\vert \mathbf{F}_{\mathbf{x},t}^{\left(s\right)}\right\rangle  & =\int\frac{\mathrm{d}^{N}\mathbf{k}}{\left(2\pi\right)^{N}}\exp\left[\mathrm{i}\left(\mathbf{k}\cdot\mathbf{x-}\left[\omega_{\mathbf{k}_{\mathrm{D}}+\mathbf{k}}^{\left(s\right)}-\omega_{_{\mathrm{D}}}\right]t\right)\right]\nonumber \\
 & \times\tilde{F}_{\mathbf{k}}\,\,\langle\phi_{\mathbf{k}_{\mathrm{D}}+\mathbf{k}}^{\left(s\right)}\mid\Xi\rangle\mid\phi_{\mathbf{k}_{\mathrm{D}}+\mathbf{k}}^{\left(s\right)}\rangle,\label{FsD}
\end{align}
where we made the variable change $\mathbf{k}-\mathbf{k}_{\mathrm{D}}\rightarrow\mathbf{k}$,
and extended the limits of integration to infinity as corresponding
to the continuous limit. Note that $\omega_{\mathbf{k}_{\mathrm{D}}}^{\left(s\neq1,2\right)}-\omega_{_{\mathrm{D}}}=0$
according to assumption (ii) above. As in the regular case, the evolution
has a fast part, given by the carrier wave $e^{\mathrm{i}\left(\mathbf{k}_{\mathrm{D}}\cdot\mathbf{x-}\omega_{_{\mathrm{D}}}t\right)}$,
and a slow part (both in time and in space) given by $\left\vert \mathbf{F}_{\mathbf{x},t}^{\left(s\right)}\right\rangle $.

Note that the previous approximations reproduce the correct result
at $t=0$, Eq. (\ref{InixD}), upon using that $\sum_{s}\left\vert \phi_{\mathbf{k}_{\mathrm{D}}+\mathbf{k}}^{\left(s\right)}\right\rangle \left\langle \phi_{\mathbf{k}_{\mathrm{D}}+\mathbf{k}}^{\left(s\right)}\right\vert $
is very approximately the unit operator in the \emph{degenerate} subspace,
to which $\left\vert \Xi\right\rangle $ belongs.

It is important to understand that the vector (coin) part of the state
evolves in time in this diabolical point situation, because more than
one of the sheets that become degenerate at the conical intersection
become necessarily populated, i.e. there are at least two $s$ values
for which $\langle\phi_{\mathbf{k}_{\mathrm{D}}+\mathbf{k}}^{\left(s\right)}\mid\Xi\rangle\neq0$,
unlike the regular case.

Finding a wave equation in this case is by far more complicated than
in the regular case, as the vector part of $\left\vert \mathbf{F}_{\mathbf{x},t}^{\left(s\right)}\right\rangle $
depends on time (because $\left\vert \phi_{\mathbf{k}_{\mathrm{D}}+\mathbf{k}}^{\left(s\right)}\right\rangle $
cannot be approximated by $\left\vert \phi_{\mathbf{k}_{\mathrm{D}}}^{\left(s\right)}\right\rangle $),
unlike the regular case where it can be approximated by the constant
vector  $\left\vert \phi_{\mathbf{k}_{0}}^{\left(s\right)}\right\rangle $,
see (\ref{psis-Fs}). We will not try deriving such a wave equation
but conform ourselves with trying to understand the evolution of the
system under the assumed conditions.

What we can say is that the frequency offsets $\left[\omega_{\mathbf{k}_{\mathrm{D}}+\mathbf{k}}^{\left(s=1,2\right)}-\omega_{_{\mathrm{D}}}\right]$
in (\ref{FsD}) are conical for small $\mathbf{k}$ (i.e. in the vicinity
of the diabolical point, which is the region selected by $\tilde{F}_{\mathbf{k}}$);
in other words, $\left[\omega_{\mathbf{k}_{\mathrm{D}}+\mathbf{k}}^{\left(s=1,2\right)}-\omega_{_{\mathrm{D}}}\right]\approx\pm ck$,
where $c$ is the group speed (the modulus of the group velocity)
and $k=\left\vert \mathbf{k}\right\vert $; see the 2D Grover walk
case in Eq. (\ref{Omega-diabolo}), where $c=1/\sqrt{2}$. For the
rest of sheets we have assumed $\omega_{\mathbf{k}_{\mathrm{D}}}^{\left(s\neq1,2\right)}-\omega_{_{\mathrm{D}}}=0$.
Hence we can write, from (\ref{FsD}) 
\begin{align}
\left\vert \mathbf{F}_{\mathbf{x},t}^{\left(s=1,2\right)}\right\rangle  & =\int\frac{\mathrm{d}^{N}\mathbf{k}}{\left(2\pi\right)^{N}}\exp\left[\mathrm{i}\left(\mathbf{k}\cdot\mathbf{x}\mp ckt\right)\right]\nonumber \\
 & \times\langle\phi_{\mathbf{k}_{\mathrm{D}}+\mathbf{k}}^{\left(s\right)}\mid\Xi\rangle\mid\phi_{\mathbf{k}_{\mathrm{D}}+\mathbf{k}}^{\left(s\right)}\rangle,\label{Fs12}
\end{align}
while the rest of partial waves verify 
\begin{align}
\left\vert \mathbf{F}_{\mathbf{x},t}^{\left(s\neq1,2\right)}\right\rangle  & =\int\frac{\mathrm{d}^{N}\mathbf{k}}{\left(2\pi\right)^{N}}\exp\left[\mathrm{i}\left(\mathbf{k}\cdot\mathbf{x}\right)\right]\nonumber \\
 & \times\tilde{F}_{\mathbf{k}}\langle\phi_{\mathbf{k}_{\mathrm{D}}+\mathbf{k}}^{\left(s\right)}\mid\Xi\rangle\mid\phi_{\mathbf{k}_{\mathrm{D}}+\mathbf{k}}^{\left(s\right)}\rangle=\left\vert \mathbf{F}_{\mathbf{x},0}^{\left(s\neq1,2\right)}\right\rangle ,
\end{align}
which is a constant, equal to its initial value. Hence, the projection
of the initial state onto the sheets of constant frequency does not
evolve, as expected, leading to a possible localization of a part
of the initial wave packet. We thus consider in the following that
$\langle\phi_{\mathbf{k}_{\mathrm{D}}+\mathbf{k}}^{\left(s\neq1,2\right)}\mid\Xi\rangle=0$.

Regarding the sheets $s=1,2$ we cannot progress unless we assume
some property of the eigenvectors $\left\vert \phi_{\mathbf{k}_{\mathrm{D}}+\mathbf{k}}^{\left(s\right)}\right\rangle $.
We will assume that $\left\vert \phi_{\mathbf{k}_{\mathrm{D}}+\mathbf{k}}^{\left(s=1,2\right)}\right\rangle $
(for $\mathbf{k}$ close to zero, i.e. around the diabolical point)
depend just on the angular part of $\mathbf{k}$ (let us call it $\Omega$)
but not on its modulus $k$. We also restrict ourselves to cases in
which $\tilde{F}_{\mathbf{k}}$ only depends on $k$ (spherical symmetry),
i.e. $\tilde{F}_{\mathbf{k}}=\tilde{F}_{k}$. Hence we write the integral
in (\ref{Fs12}) in ($N$-dimensional) spherical coordinates as 
\begin{align}
\left\vert \mathbf{F}_{\mathbf{x},t}^{\left(s=1,2\right)}\right\rangle  & =\left(2\pi\right)^{-N}\int\mathrm{d}k\exp\left(\mp\mathrm{i}ckt\right)k^{N-1}\tilde{F}_{k}\left\vert \mathbf{E}_{\mathbf{x}}^{\left(s\right)}\left(k\right)\right\rangle ,\label{FD}\\
\left\vert \mathbf{E}_{\mathbf{x}}^{\left(s\right)}\left(k\right)\right\rangle  & =\int\mathrm{d}^{N}\Omega\exp\left(\mathrm{i}\mathbf{k}\cdot\mathbf{x}\right)\langle\phi_{\mathbf{k}_{\mathrm{D}}+\mathbf{k}}^{\left(s\right)}\mid\Xi\rangle\mid\phi_{\mathbf{k}_{\mathrm{D}}+\mathbf{k}}^{\left(s\right)}\rangle,
\end{align}
where we wrote $\mathrm{d}^{N}\mathbf{k}=k^{N-1}\mathrm{d}k\mathrm{d}^{N}\Omega$
and $\mathrm{d}^{N}\Omega$ is the $N$-dimensional solid angle element.

Up to here the theory is somehow general. However we cannot progress
unless we particularize to some special case. We shall do this below
for the 2D Grover walk.

\section{Application to the 2D Grover QW}

So far we have derived general expressions for the $N$--dimensional
QW. In this section we consider the special case of the two--dimensional
QW using the Grover coin operator.

\subsection{Diagonalization of the 2D Grover map: Dispersion relations and diabolical
points}

The so-called Grover coin of dimension $2N$ has matrix elements $C_{\alpha^{\prime},\eta^{\prime}}^{\alpha,\eta}=1/N-\delta_{\alpha,\alpha^{\prime}}\delta_{\eta,\eta^{\prime}}$.
In the present case ($N=2$), the corresponding matrix $C_{\mathbf{k}}$
(\ref{Ck_elements}), with $\mathbf{k}=(k_{1},k_{2})$, has the form
\begin{equation}
C_{\mathbf{k}}=\frac{1}{2}\begin{pmatrix}-e^{-\mathrm{i}k_{1}} & e^{-\mathrm{i}k_{1}} & e^{-\mathrm{i}k_{1}} & e^{-\mathrm{i}k_{1}}\\
e^{\mathrm{i}k_{1}} & -e^{\mathrm{i}k_{1}} & e^{\mathrm{i}k_{1}} & e^{\mathrm{i}k_{1}}\\
e^{-\mathrm{i}k_{2}} & e^{-\mathrm{i}k_{2}} & -e^{-\mathrm{i}k_{2}} & e^{-\mathrm{i}k_{2}}\\
e^{\mathrm{i}k_{2}} & e^{\mathrm{i}k_{2}} & e^{\mathrm{i}k_{2}} & -e^{\mathrm{i}k_{2}}
\end{pmatrix},\label{G_k}
\end{equation}
whose diagonalization yields the eigenvalues $\lambda_{\mathbf{k}}^{\left(s\right)}=\exp\left(-\mathrm{i}\omega_{\mathbf{k}}^{\left(s\right)}\right)$
with 
\begin{equation}
\omega_{\mathbf{k}}^{\left(1\right)}=\pi+\Omega_{\mathbf{k}},\text{ }\omega_{\mathbf{k}}^{\left(2\right)}=\pi-\Omega_{\mathbf{k}},\ \omega_{\mathbf{k}}^{\left(3\right)}=\pi,\text{ }\omega_{\mathbf{k}}^{\left(4\right)}=0,\label{Lambdas}
\end{equation}
where 
\begin{equation}
\Omega_{\mathbf{k}}=\arccos\left[\frac{1}{2}\left(\cos{k_{1}}+\cos{k_{2}}\right)\right]\in\left[0,\pi\right].\label{disp}
\end{equation}
Note that adding a multiple integer of $2\pi$ to any of the $\omega$'s
does not change anything because time is discrete and runs in steps
of $1$, see (\ref{evol_x_s}).

We see, according to (\ref{Lambdas}), that the last two eigenvalues
$\lambda_{\mathbf{k}}^{\left(3,4\right)}=\exp\left(-\mathrm{i}\omega_{\mathbf{k}}^{\left(3,4\right)}\right)=-1,1$
do not depend on $\mathbf{k}$, and the first two eigenvalues $\lambda_{\mathbf{k}}^{\left(1,2\right)}=\exp\left(-\mathrm{i}\omega_{\mathbf{k}}^{\left(1,2\right)}\right)=-\exp\left(\mp\mathrm{i}\Omega_{\mathbf{k}}\right)$
are complex-conjugate of each other, hence we can choose $\Omega_{\mathbf{k}}\in\left[0,\pi\right]$
without loss of generality. A plot of the dispersion relations (\ref{Lambdas})
is given in Fig. 1, where five degeneracy points (the origin and the
four corners, see also Fig. \ref{figcontour}) are observed with 3-fold
degeneracies. Moreover the frequencies $\pm\Omega_{\mathbf{k}}$ have
a conical form (diabolo-like) at those degeneracies and this is why
we call them \textit{diabolical points}, in analogy to other diabolos
found in different systems \cite{graphene1,graphene2,Mead79,Cederbaum03,Berry1,Berry2,Berry3,Zhang,Torrent12}.
At the central diabolical point the degeneracy is between $\omega_{\mathbf{k}}^{\left(s=1,2,4\right)}$,
while at the corners it is between $\omega_{\mathbf{k}}^{\left(s=1,2,3\right)}$.

\begin{figure}
\begin{minipage}[t]{1\columnwidth}%
\includegraphics[width=9cm]{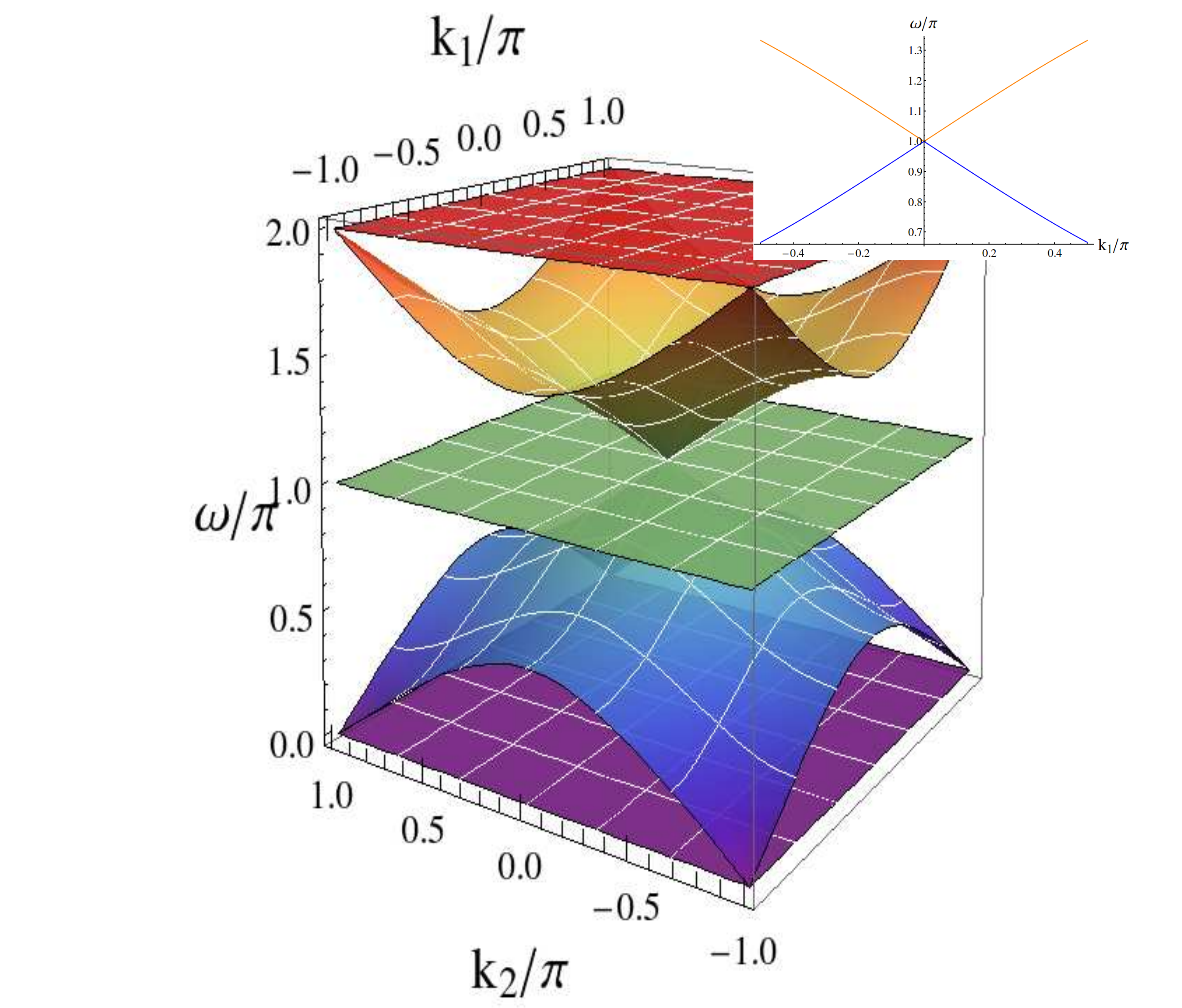}%
\end{minipage}

\caption{(Main Figure): Dispersion relation sketch around the diabolic point.
The value of $\omega/\pi$ is plotted for the different sheets (i.e.,
different eigenvalues). The middle (flat) surface corresponds to $\omega_{k}^{(3)}=\pi$,
the surface with values ranging from $1$ to $2$ corresponds to $\omega_{k}^{(1)}$.
The bottom plane is associated with $\omega_{k}^{(4)}$ and the surface
with values ranging from $0$ to $1$ corresponds to $\omega_{k}^{(2)}$.
(Inset): Detail of $\omega_{k}^{(1)}$ and $\omega_{k}^{(2)}$ around
the origin (orange and blue curves, respectively) for $k_{2}=0$ as
a function of $k_{1}$, showing the intersection at the diabolical
point.}

\label{figbranches}
\end{figure}

For the sake of later use we note that the frequency $\Omega_{\mathbf{k}}$
reads, close to the diabolical point $\mathbf{k}=0$, 
\begin{equation}
\Omega_{\mathbf{k}}\simeq\frac{k}{\sqrt{2}}-\frac{k^{3}\cos\left(2\theta\right)}{48\sqrt{2}}+O\left(k^{5}\right),\label{Omega-diabolo}
\end{equation}
where $\mathbf{k}=k\left(\cos\theta,\sin\theta\right)$. This means
that, very close to the diabolical point, the frequency $\Omega_{\mathbf{k}}$
actually has a conical dependence on the wavenumber. Later we will
understand the consequences of the existence of these points.

The fact that $\omega_{\mathbf{k}}^{\left(s=3,4\right)}$ are constant
causes that $\left\vert \psi_{\mathbf{x},t}^{\left(s=3,4\right)}\right\rangle $,
see (\ref{evol_x_s}), do not have a wave character; only $\left\vert \psi_{\mathbf{x},t}^{\left(s=1,2\right)}\right\rangle $
are true waves. Hence propagation in the 2D Grover walk occurs only
if the initial condition projects onto subspaces $1$ or $2$; otherwise
the walker remains localized. This explains why strong localization
in the 2D Grover map is usually observed, as first noted in Ref. \cite{Inui04}.
All this can be put more formally in terms of the group velocity $\mathbf{v}_{\mathrm{g}}\left(\mathbf{k}\right)$\ of
the waves in the system, given by the gradient (in $\mathbf{k}$)
of the wave frequency. The group velocity, as in any linear wave system,
has the meaning of velocity at which an extended wave packet, centered
in $\mathbf{k}$-space around some value $\mathbf{k}_{0}$, moves
(there are other effects affecting extended wavepackets which we analyze
below). In our case $\omega_{\mathbf{k}}^{\left(s=3,4\right)}$ are
constant, hence their gradient is null: these eigenvalues entail no
motion. On the contrary $\omega_{\mathbf{k}}^{\left(s=1,2\right)}$
depend on $\mathbf{k}$ (\ref{Lambdas},\ref{disp}) and then define
non null group velocities: 
\begin{equation}
\mathbf{v}_{\mathrm{g}}^{\left(1,2\right)}\left(\mathbf{k}\right)=\pm\nabla_{\mathbf{k}}\Omega_{\mathbf{k}},\label{veg}
\end{equation}
where the plus and minus signs stand for $s=1,2$, respectively. Using
(\ref{disp}) these velocities become 
\begin{equation}
\mathbf{v}_{\mathrm{g}}^{\left(1,2\right)}\left(\mathbf{k}\right)=\pm\frac{(\sin{k_{1}},\sin{k_{2}})}{\sqrt{4-(\cos{k_{1}}+\cos{k_{2}})^{2}}}.\label{vg}
\end{equation}

\begin{figure}
\includegraphics[width=8cm]{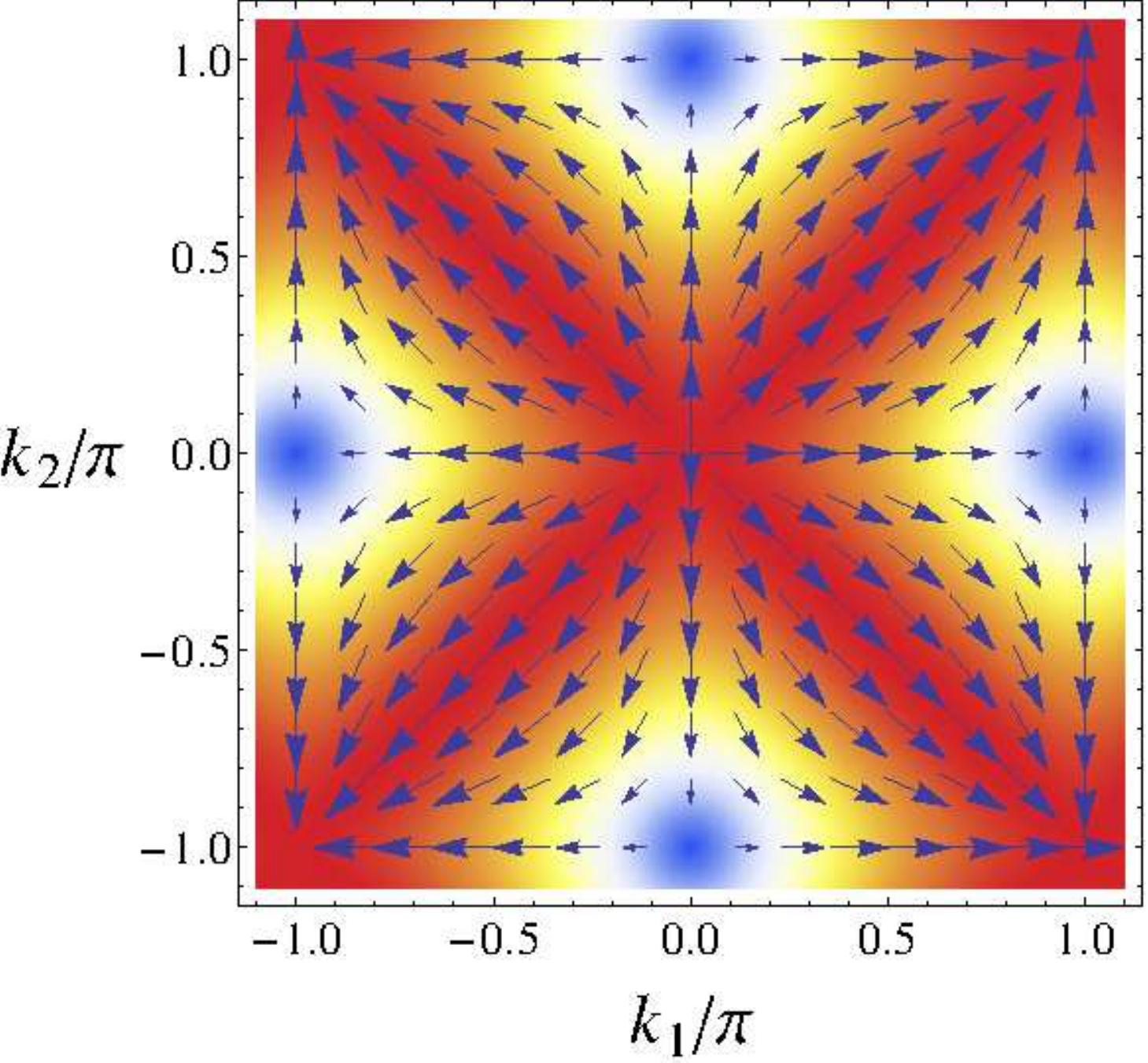}\caption{Vector plot of the group velocity (shown as arrows), superimposed
to a density plot of the velocity modulus Eq. (\ref{disp}).}

\label{figcontour}
\end{figure}

Hence the maximum group velocity modulus in the 2D Grover walk is
$\frac{1}{\sqrt{2}}$, as can be easily checked and is obtained, in
particular, along the diagonals $k_{2}=\pm k_{1}$. This is the maximum
speed at which any feature of the QW can propagate. Figure 2 displays
a vector plot of $\mathbf{v}_{\mathrm{g}}^{\left(1,2\right)}\left(\mathbf{k}\right)$
where a number of interesting points (in $\mathbf{k}$ space) is revealed:
On the one hand there are points of null group velocity, located at
$\mathbf{k}=\left(0,\pm\pi\right)$ and at $\mathbf{k}=\left(\pm\pi,0\right)$,
in correspondence with the saddle points observed in Fig. 1. On the
other hand there are five \textit{singular} points at $\mathbf{k}=\left(0,0\right)$
as well as at the four corners. These points are singular because
the group velocity is undefined at them (they look like sources or
sinks). We note that these points coincide with the diabolical points
in Fig. 1.

We find it worth mentioning that in two--dimensional QWs with coin
operators other that the Grover, diabolical points are also found,
as we have checked by using the DFT coin \cite{Mackay} that displays
several conical intersections. In this case of the DFT coin none of
the four leaves of the dispersion relation is constant. We mention
too that a behavior very similar to that of the 2D Grover walk is
found in the Alternate QW \cite{Roldan12}, except for the fact that
in this last case there are no constant surfaces in the dispersion
relation (and hence no localization phenomena).

\subsection{Eigenvectors}

The eigenvectors of the 2D Grover walk matrix (\ref{G_k})\ are given
by \cite{Watabe}. Use of these eigenvalues in (\ref{evol_x_s}) allows
finally to compute the state of the system at any time.

Whenever $\mathbf{k}$ is not close to a diabolical point these eigenvectors
vary smoothly around $\mathbf{k}$. Here we just want to study the
behavior of the eigenvectors close to the diabolical point at $\mathbf{k}=\mathbf{k}_{\mathrm{D}}\equiv\left(0,0\right)$
--at the corners the conclusions are similar. We find it convenient
to use polar coordinates $\left(k_{1},k_{2}\right)=\left(k\cos\theta,k\sin\theta\right)$.
Performing the limit for $k\rightarrow0$ we find 
\begin{align}
\phi_{\mathbf{k}}^{\left(1\right)} & =\frac{1}{2\sqrt{2}}\left(\begin{array}{c}
1+\sqrt{2}\cos\theta\\
1-\sqrt{2}\cos\theta\\
-1-\sqrt{2}\sin\theta\\
-1+\sqrt{2}\sin\theta
\end{array}\right),\nonumber \\
\phi_{\mathbf{k}}^{\left(2\right)} & =\frac{1}{2\sqrt{2}}\left(\begin{array}{c}
1-\sqrt{2}\cos\theta\\
1+\sqrt{2}\cos\theta\\
-1+\sqrt{2}\sin\theta\\
-1-\sqrt{2}\sin\theta
\end{array}\right),\nonumber \\
\phi_{\mathbf{k}}^{\left(3\right)} & =\frac{1}{\sqrt{2}}\left(\begin{array}{c}
\sin\theta\\
-\sin\theta\\
\cos\theta\\
-\cos\theta
\end{array}\right),\ \ \phi_{\mathbf{k}}^{\left(4\right)}=\frac{1}{2}\left(\begin{array}{c}
1\\
1\\
1\\
1
\end{array}\right).\label{eigen_close}
\end{align}

Hence, close to the diabolical point $\mathbf{k}_{\mathrm{D}}$, there
are three eigenvectors, corresponding to $s=1,2,3$, displaying a
strong azimuthal dependence, while $\phi_{\mathbf{k}}^{\left(4\right)}$
is constant around $\mathbf{k}_{\mathrm{D}}$ (its variations are
smooth and hence tend to zero as $\mathbf{k}\rightarrow\mathbf{k}_{\mathrm{D}}$).
Thus, even if only $\phi_{\mathbf{k}}^{\left(1\right)}$ and $\phi_{\mathbf{k}}^{\left(2\right)}$
participate in the conical intersection and define the diabolical
point, $\phi_{\mathbf{k}}^{\left(3\right)}$ is also affected by this
feature because it is associated with eigenvalue $\lambda_{\mathbf{k}}^{\left(3\right)}=-1$,
which is degenerate with $\lambda_{\mathbf{k}_{\mathrm{D}}}^{\left(s=1,2\right)}$.
The fact that the diabolical point is singular is easy to understand:
depending on the direction we approach to it the eigenvectors of the
problem are different.

Just at the diabolical point $\lambda_{\mathbf{k}}^{\left(s=1,2,3\right)}=-1$
and hence there is a 3D eigensubspace formed by all vectors orthogonal
to the fourth eigenvector, namely $\phi_{\mathbf{k}_{\mathrm{D}}}^{\left(4\right)}=\frac{1}{2}\operatorname{col}\left(1,1,1,1\right)$,
which is equal to $\phi_{\mathbf{k}}^{\left(4\right)}$: see (\ref{eigen_close}).
A sensible choice for these eigenvectors follows from noting that
$\phi_{\mathbf{k}}^{\left(1\right)}+\phi_{\mathbf{k}}^{\left(2\right)}$
in (\ref{eigen_close}) is independent of the azimuth and is orthogonal
to $\phi_{\mathbf{k}}^{\left(4\right)}$. Hence one of the basis elements
for this 3D degenerate subspace can chosen as 
\begin{equation}
\phi_{_{\mathrm{D}}}=\frac{\phi_{\mathbf{k}}^{\left(1\right)}+\phi_{\mathbf{k}}^{\left(2\right)}}{\sqrt{2}}=\frac{1}{2}\operatorname{col}\left(1,1,-1,-1\right),\label{fi_D}
\end{equation}
which has the outstanding property of being the single vector that
projects, close to the diabolical point $\mathbf{k}_{\mathrm{D}}$,
only onto $\phi_{\mathbf{k}}^{\left(s=1,2\right)}$, i.e. onto the
eigenspaces of the diabolo. As for the other two eigenvectors we just
impose their orthogonality with $\phi_{_{\mathrm{D}}}$ and $\phi_{\mathbf{k}_{\mathrm{D}}}^{\left(4\right)}$
and we choose them arbitrarily as 
\begin{equation}
\phi_{_{\mathrm{D}}}^{\prime}=\frac{1}{\sqrt{2}}\operatorname{col}\left(1,-1,0,0\right),\ \ \phi_{_{\mathrm{D}}}^{\prime\prime}=\frac{1}{\sqrt{2}}\operatorname{col}\left(0,0,-1,1\right).
\end{equation}
Note that we are using a ``primed'' notation instead of keeping
the notation with the label $s$, except for $s=4$, because none
of these eigenvectors keeps being an eigenvector for $\mathbf{k}\neq\mathbf{k}_{\mathrm{D}}$
(only for special directions $\theta$) and hence there is no sense
in attaching any of these eigenvectors to a particular sheet of the
dispersion relation. To conclude we list the projections of these
eigenvectors onto the eigenvectors (\ref{eigen_close}) in the neighborhood
of $\mathbf{k}_{\mathrm{D}}$: 
\begin{equation}
\begin{array}{ll}
\langle\phi_{_{\mathrm{D}}}\mid\phi_{\mathbf{k}}^{\left(1\right)}\rangle=\langle\phi_{_{\mathrm{D}}}\mid\phi_{\mathbf{k}}^{\left(2\right)}\rangle=\frac{1}{\sqrt{2}}, & \langle\phi_{_{\mathrm{D}}}\mid\phi_{\mathbf{k}}^{\left(3\right)}\rangle=0,\\
\langle\phi_{_{\mathrm{D}}}^{\prime}\mid\phi_{\mathbf{k}}^{\left(1\right)}\rangle=-\langle\phi_{_{\mathrm{D}}}^{\prime}\mid\phi_{\mathbf{k}}^{\left(2\right)}\rangle=\frac{\cos\theta}{\sqrt{2}}, & \langle\phi_{_{\mathrm{D}}}^{\prime}\mid\phi_{\mathbf{k}}^{\left(3\right)}\rangle=-\sin\theta,\\
\langle\phi_{_{\mathrm{D}}}^{\prime\prime}\mid\phi_{\mathbf{k}}^{\left(1\right)}\rangle=-\langle\phi_{_{\mathrm{D}}}^{\prime\prime}\mid\phi_{\mathbf{k}}^{\left(2\right)}\rangle=\frac{\sin\theta}{\sqrt{2}}, & \langle\phi_{_{\mathrm{D}}}^{\prime\prime}\mid\phi_{\mathbf{k}}^{\left(3\right)}\rangle=\cos\theta,
\end{array}\label{proj_close}
\end{equation}
and $\left\langle \phi_{_{\mathrm{D}}}\right.\left\vert \phi_{\mathbf{k}}^{\left(4\right)}\right\rangle =\left\langle \phi_{_{\mathrm{D}}}^{\prime}\right.\left\vert \phi_{\mathbf{k}}^{\left(4\right)}\right\rangle =\left\langle \phi_{_{\mathrm{D}}}^{\prime\prime}\right.\left\vert \phi_{\mathbf{k}}^{\left(4\right)}\right\rangle =0$.
All this has strong consequences on the QW dynamics near $\mathbf{k}_{\mathrm{D}}$
as we will show below.

\subsection{Continuous wave equations}

In order to illustrate how the evolution of the QW can be controlled
via an appropriate choice of the initial conditions, we will show
some results obtained numerically from the evolution of the 2D quantum
walk with the Grover operator. Unless otherwise specified, the initial
profile in position space is a Gaussian centered at the origin with
cylindrical symmetry (independent of $\alpha$ or $\eta$ ):
\begin{equation}
\psi_{\mathbf{x},0}=\mathcal{N}e^{i\mathbf{k}_{0}\cdot\mathbf{x}}e^{-\frac{x_{1}^{2}+x_{2}^{2}}{2\sigma^{2}}},\label{initpsi}
\end{equation}
with $\mathbf{x}=(x_{1},x_{2})$, which implies that in the continuous
limit we also have a Gaussian, in momentum space, centered at $\mathbf{k}_{0}=(k_{1},k_{2})$.
Here, $\mathcal{N}$ is a constant that guarantees the normalization
of the state. In the numerics, we have taken a sufficiently large
value $\sigma=50$ for the Gaussian, so as to make it possible the
connection with the continuous limit, although we will also discuss
the situation for smaller values of $\sigma$ in order to show the
robustness of the results.

\begin{figure}
\begin{minipage}[t]{1\columnwidth}%
\includegraphics[width=8cm]{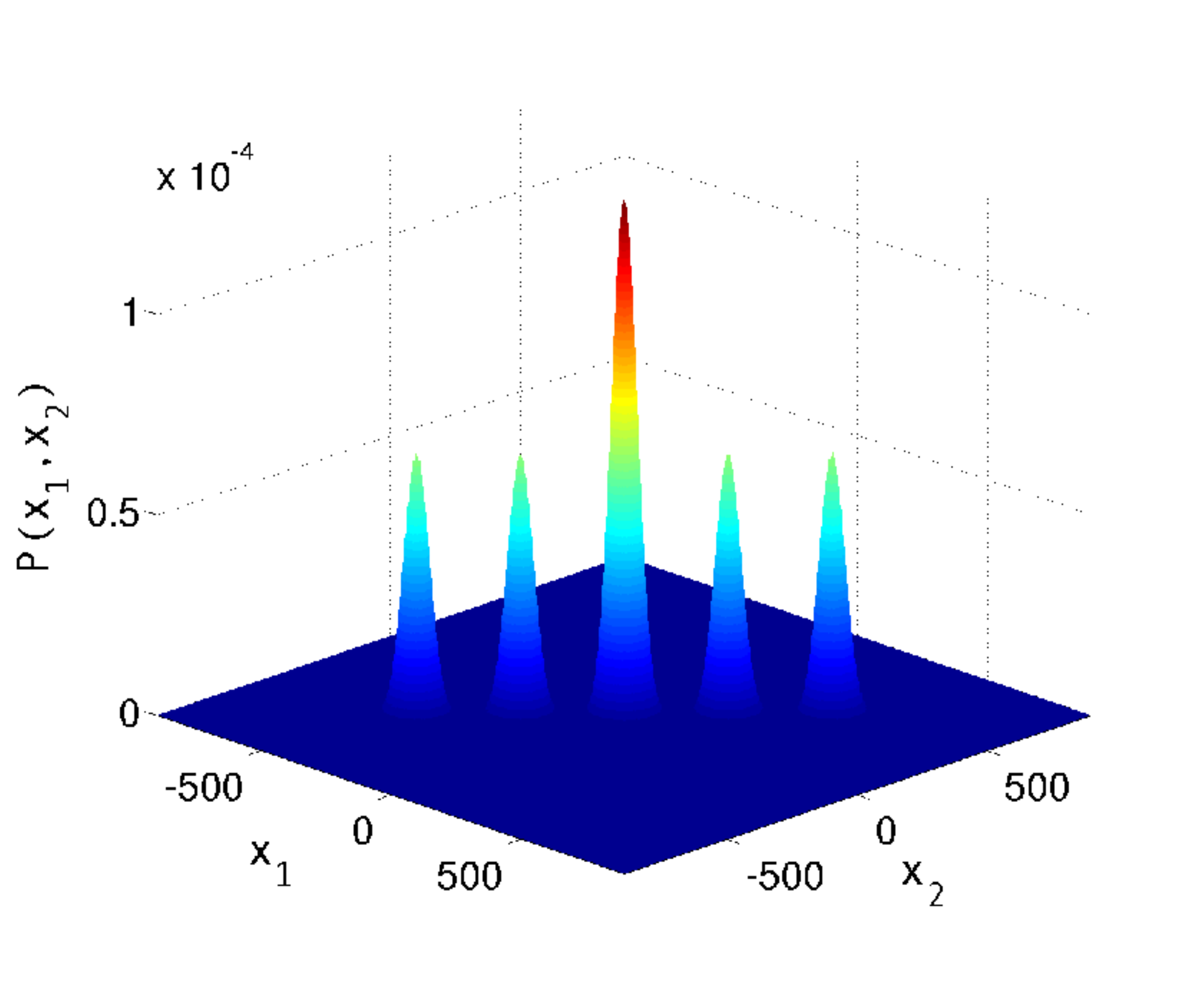}

\includegraphics[width=4cm]{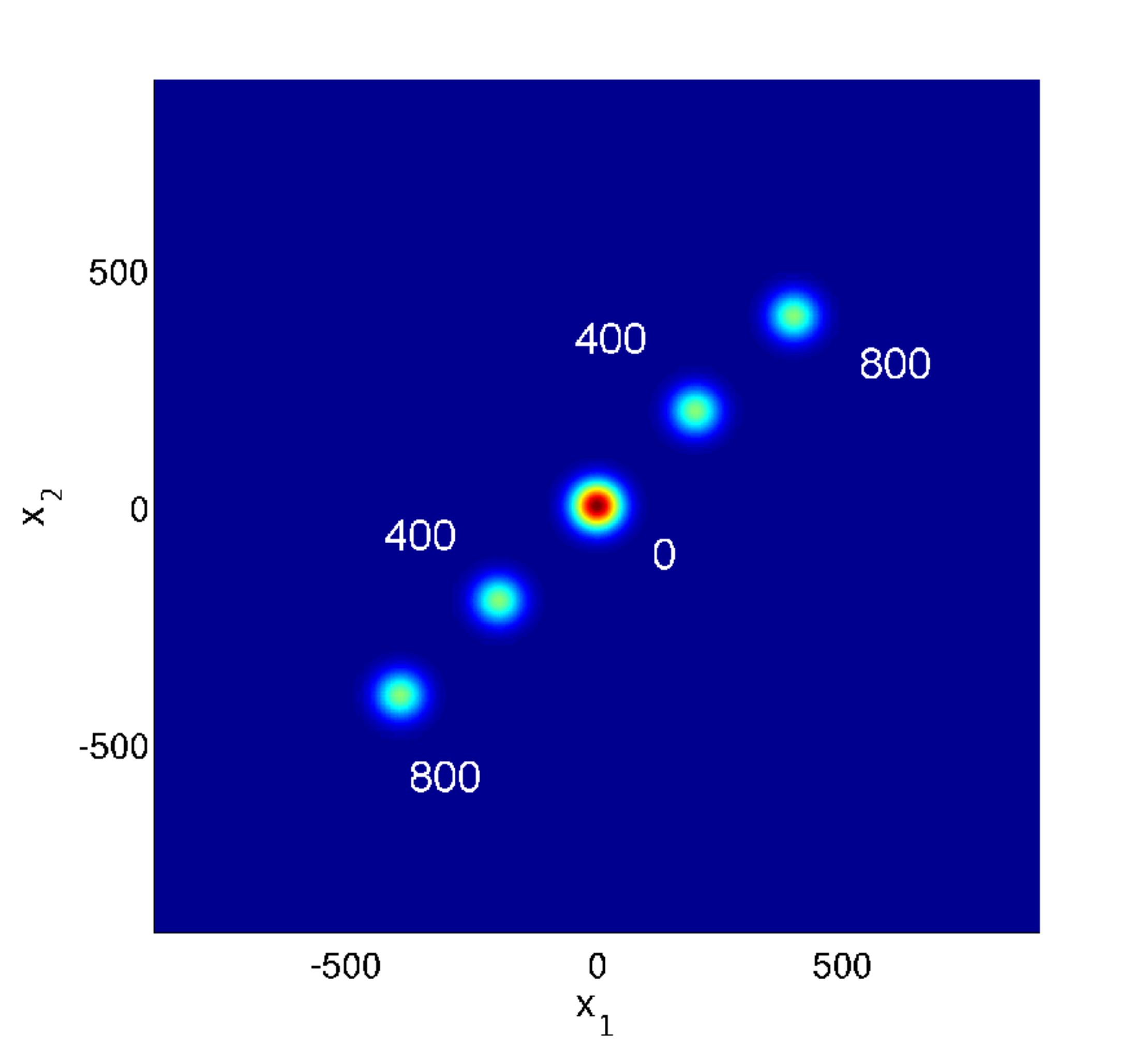}\includegraphics[width=4cm]{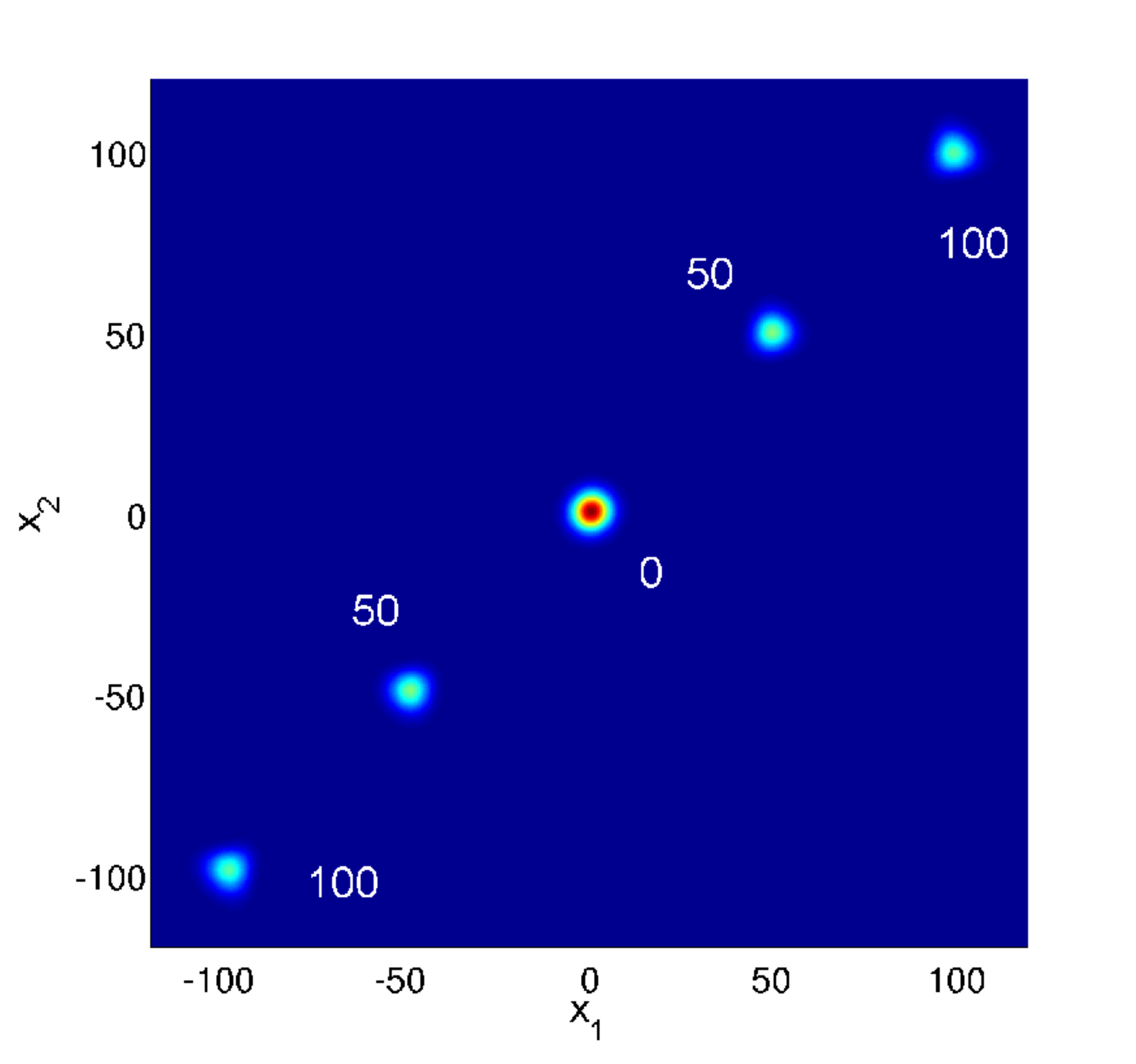}%
\end{minipage}

\caption{(Top panel): Probability distribution as a function of the dimensionless
$(x_{1},x_{2})$ position. The initial condition is given by $\frac{1}{\sqrt{2}}(\mid\phi_{\mathbf{k}_{0}}^{(1)}\rangle+\mid\phi_{\mathbf{k}_{0}}^{(2)}\rangle)$
and $k_{1}=k_{2}=\pi/2$.. The distribution is shown for three times
t=0, 400, 800 respectively. (Bottom panel): Top view of the previous
figure, both for $\sigma=50$ (left) and $\sigma=5$ (right). The
numbers indicate, in each case, the value of time for different snapshots.}

\label{2gaussnodist}
\end{figure}

Let us first consider a simple case having $k_{1}=k_{2}=\pi/2$, with
an initial coin state $\frac{1}{\sqrt{2}}(\mid\phi_{\mathbf{k}_{0}}^{(1)}\rangle+\mid\phi_{\mathbf{k}_{0}}^{(2)}\rangle)$,
where $\mid\phi_{\mathbf{k}_{0}}^{(1)}\rangle=\frac{1}{\sqrt{2}}\operatorname{col}\left(1,0,-1,0\right)$
and $\mid\phi_{\mathbf{k}_{0}}^{(2)}\rangle=\frac{1}{\sqrt{2}}\operatorname{col}\left(0,1,0,-1\right)$
that projects on the $s=1$ and $s=2$ branches with equal weights.
The group velocity is given by $\mathbf{v}_{\mathrm{g}}^{\left(s\right)}\left(\mathbf{k}_{0}\right)=\pm(\frac{1}{2},\frac{1}{2})$,
respectively for $s=1,2$, implying that the original Gaussian will
split into two pieces that will move along the diagonal of the lattice
in opposite directions, as shown in Fig. \ref{2gaussnodist}. It must
be noticed that, since the second derivatives vanish at this point
($\varpi_{ij}=0)$, the Gaussian distribution moves with no appreciable
distortion. On the right bottom panel of this figure, we show the
result for $\sigma=5$ and we observe that, in this case, the results
obtained for large $\sigma$ are very robust even for such a low value
of $\sigma$.

\begin{figure}
\begin{minipage}[t]{1\columnwidth}%
\includegraphics[width=8cm]{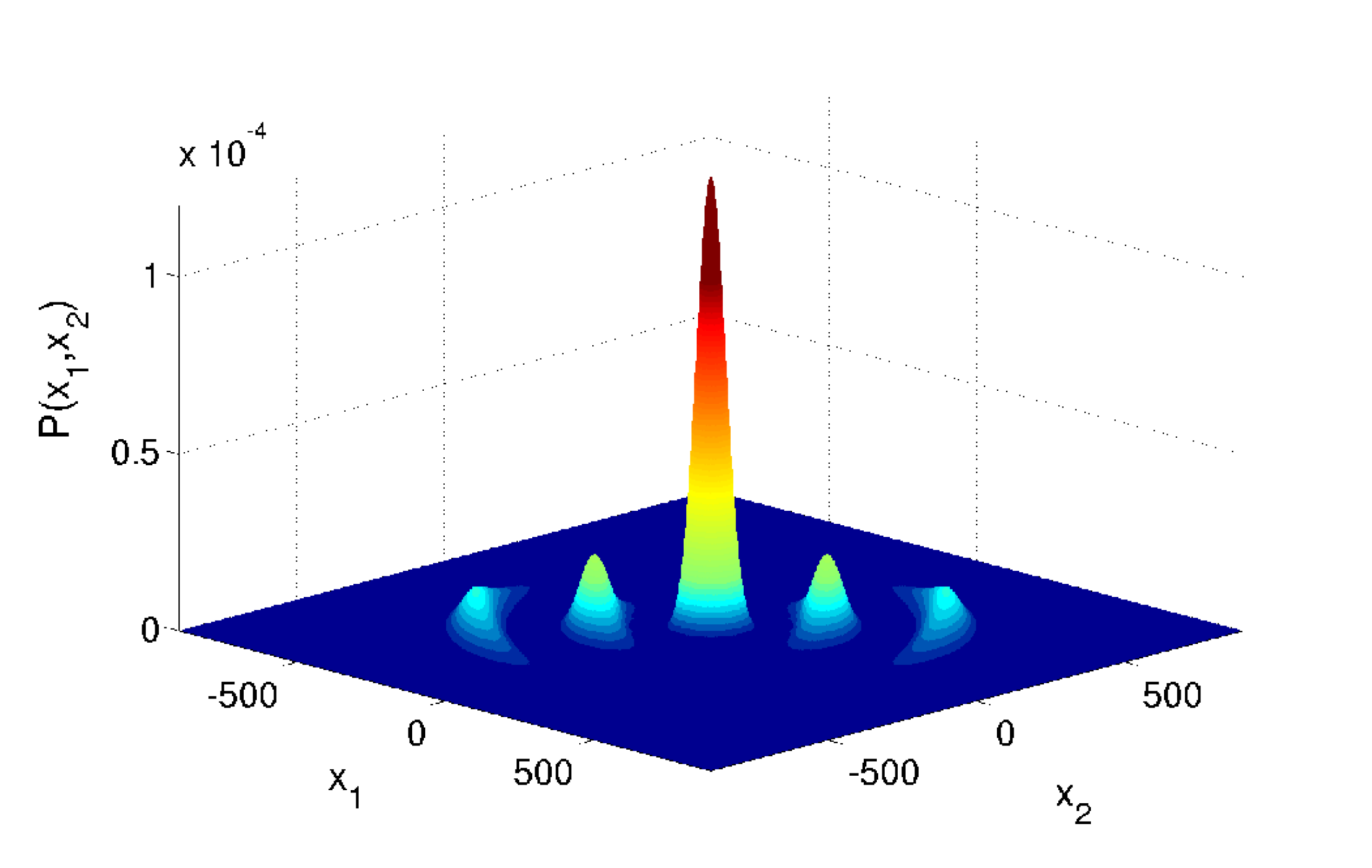}

\includegraphics[width=8cm]{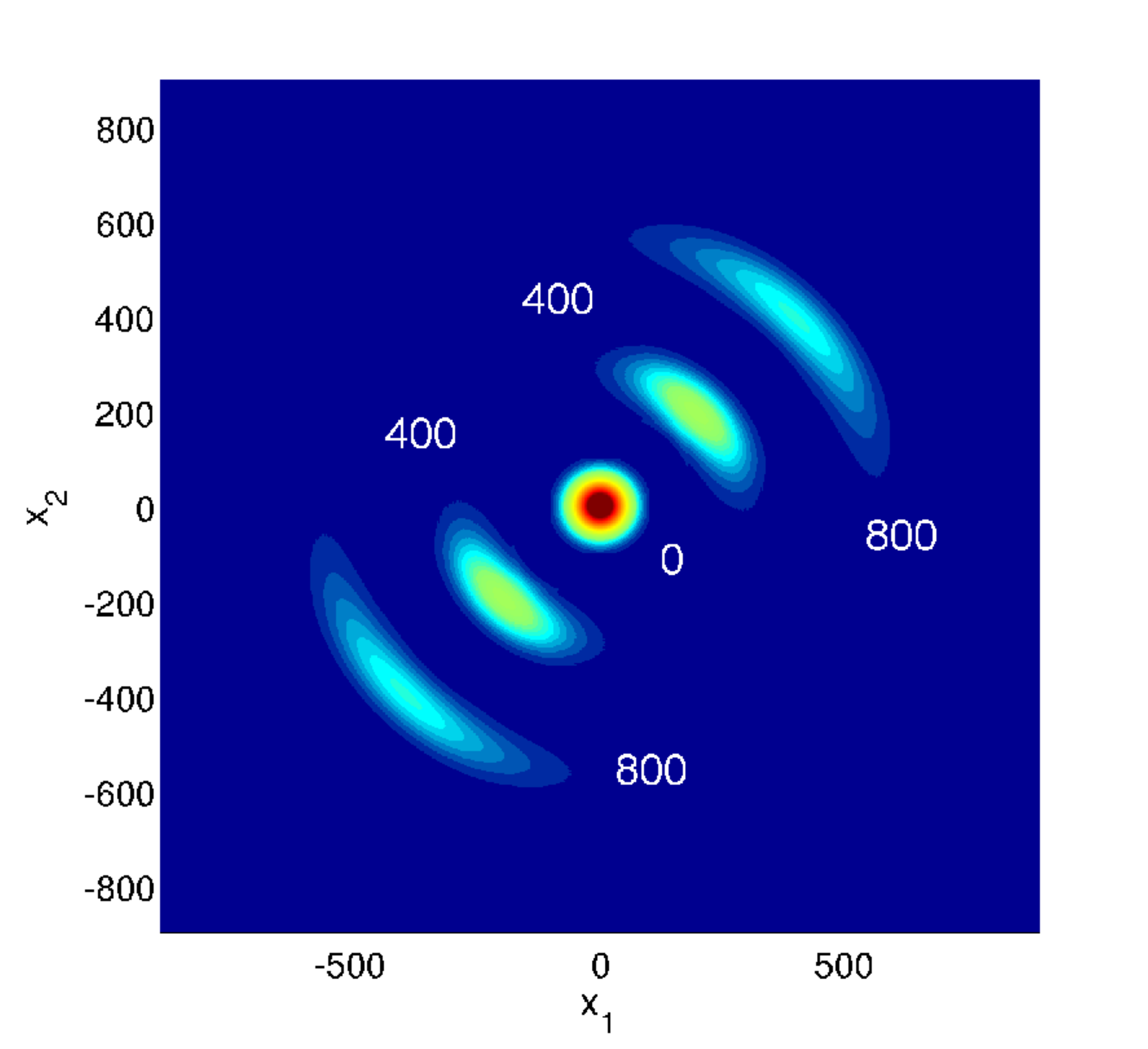}%
\end{minipage}

\caption{(Top panel): Probability distribution as a function of the dimensionless
$(x_{1},x_{2})$ position. The initial condition is given by $\frac{1}{\sqrt{2}}(\mid\phi_{\mathbf{k}_{0}}^{(1)}\rangle+\mid\phi_{\mathbf{k}_{0}}^{(2)}\rangle)$
and $k_{1}=k_{2}=0.01\pi$. The distribution is shown for three times
t=0, 400, 800 respectively. (Bottom panel): Top view of the previous
figure, showing the value of the corresponding time steps.}

\label{2gausswithdist}
\end{figure}

However, if we choose an initial value $\mathbf{k}_{0}$ close to
the origin, then the second derivatives take a large value, with the
consequence that now the two pieces experience a considerable distortion
along the line perpendicular to the line of motion, as can be seen
from Fig. \ref{2gausswithdist}, where $k_{1}=k_{2}=0.01\pi$, implying
a curvature $\varpi_{ij}\simeq7.96$ in the perpendicular direction.

\begin{figure}
\begin{minipage}[t]{1\columnwidth}%
\includegraphics[width=4cm]{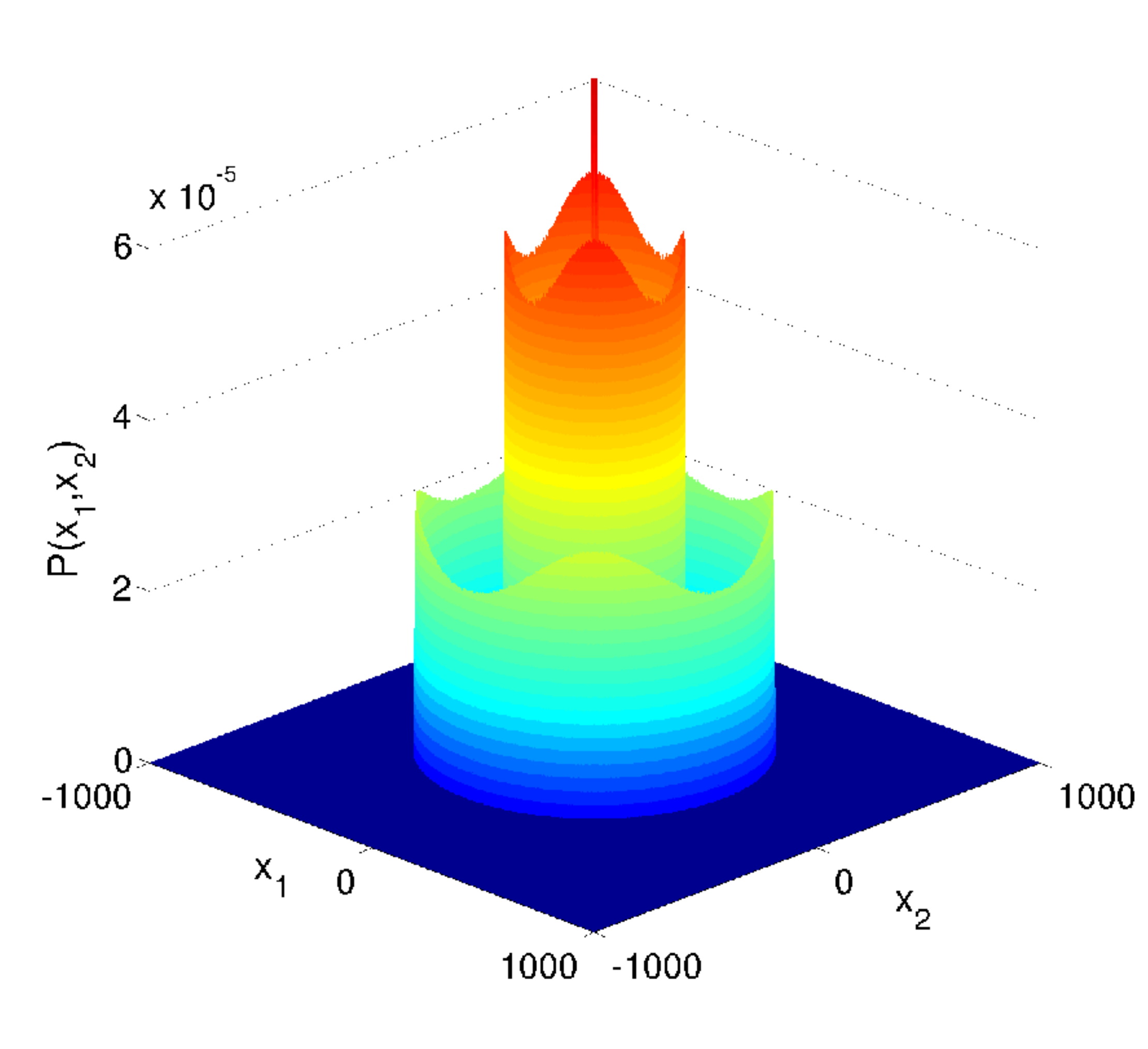}\includegraphics[width=4cm]{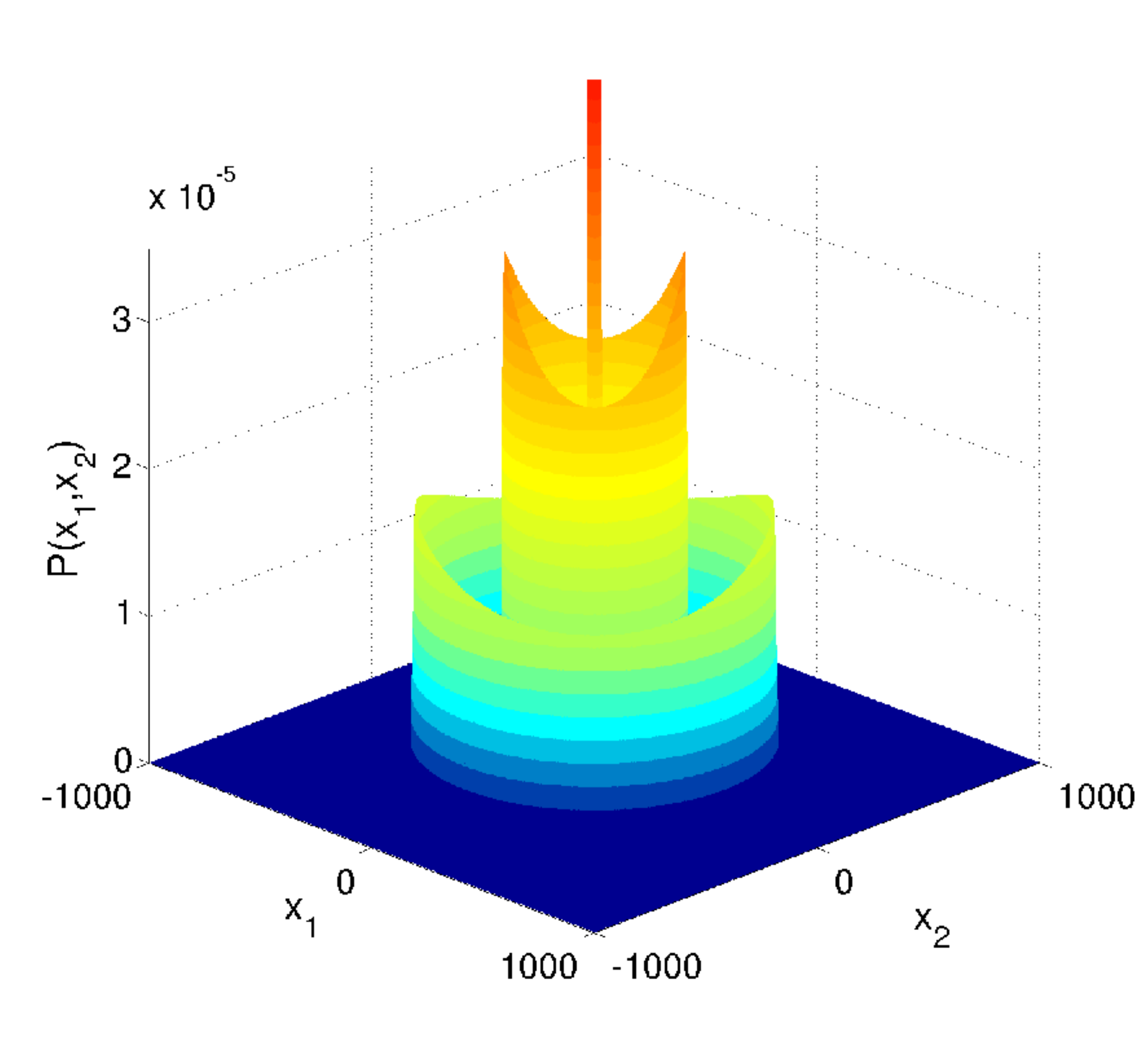}

\includegraphics[width=4cm]{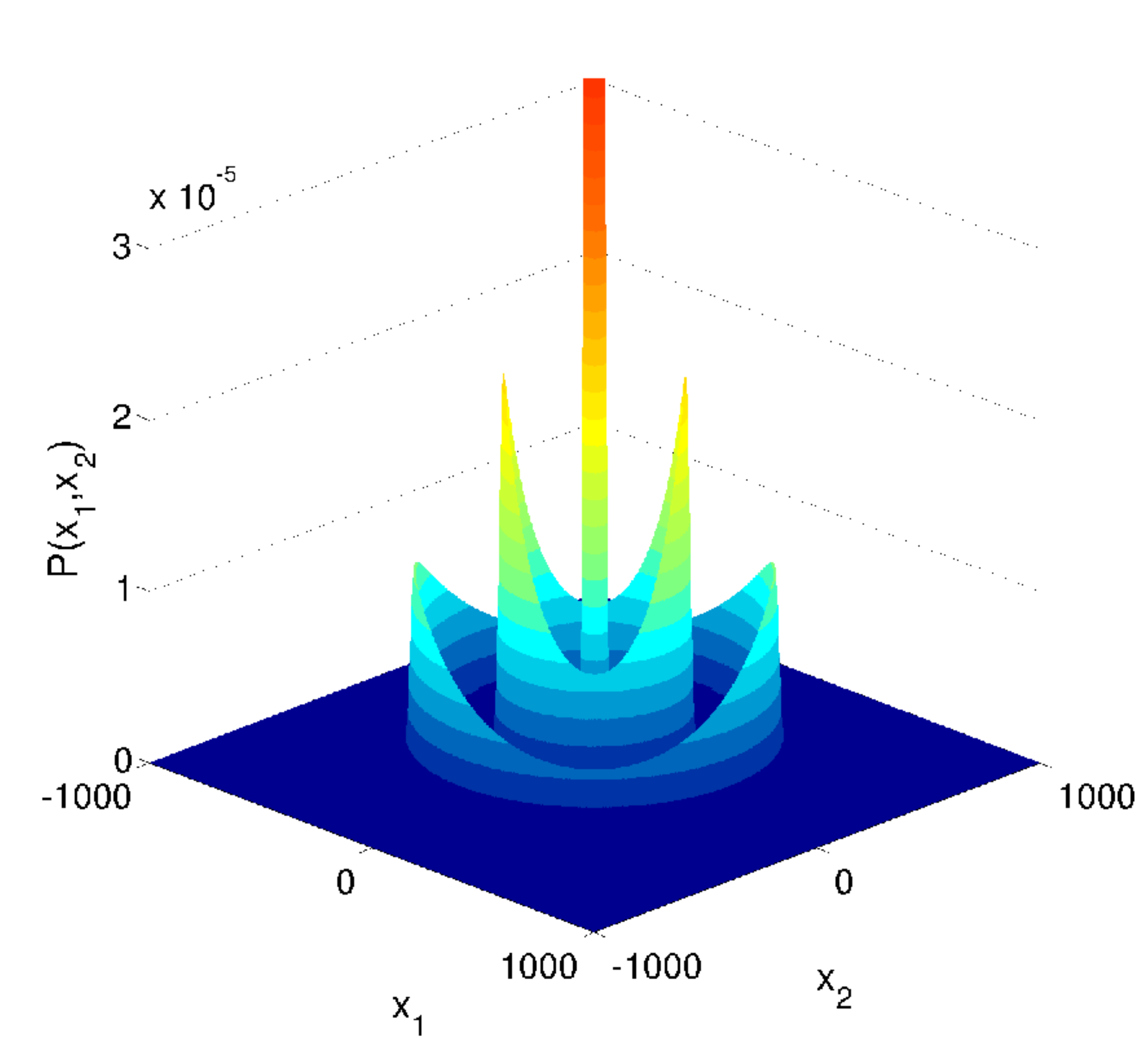}\includegraphics[width=4cm]{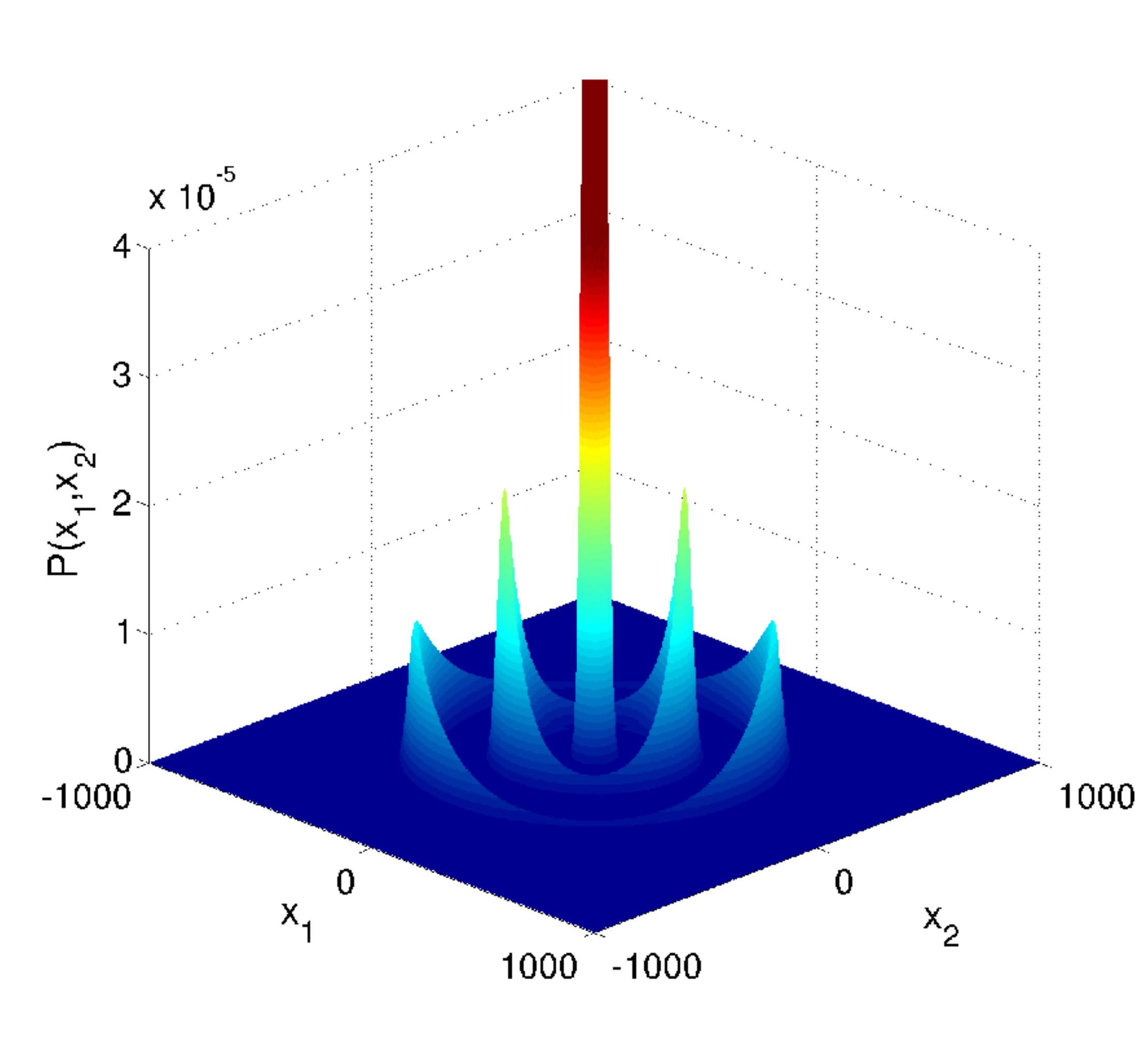}%
\end{minipage}

\caption{Same as Fig. \ref{2gausswithdist}, corresponding to four different
values of $\sigma$. From left to right and from top to bottom: $\sigma=5$,
$\sigma=10$, $\sigma=20$ and $\sigma=30$, respectively. The initial
condition is given by $\frac{1}{\sqrt{2}}(\mid\phi_{\mathbf{k}_{0}}^{(1)}\rangle+\mid\phi_{\mathbf{k}_{0}}^{(2)}\rangle)$
and $k_{1}=k_{2}=0.01\pi$.}

\label{k001varyingsigma}
\end{figure}

In this case, at variance with the case considered in Fig. \ref{2gaussnodist},
we have to face a peculiar situation, as the distribution is now peaked
around a value close to the diabolical point $\mathbf{k}_{D}=\left(0,0\right)$.
As far as the initial distribution does not contain this point with
a significant probability (i.e., for a large value of $\sigma$),
the behavior will be as described above. However, if $\sigma$ decreases
below a certain limit, the evolution is dominated by the singular
behavior of the diabolic point (to be studied in detail in the next
subsection). This transition is clearly observed in Fig. \ref{k001varyingsigma},
where the resulting evolution is depicted as $\sigma$ is increased
from $\sigma=5$ to $\sigma=30$, the latter subpanel already showing
a close resemblance with the result in Fig. \ref{2gausswithdist}.
Distributions initially centered around non-singular points, however,
are much more robust against a smaller $\sigma$, as previously shown
in Fig. \ref{2gaussnodist}.

\begin{figure}
\includegraphics[width=8cm]{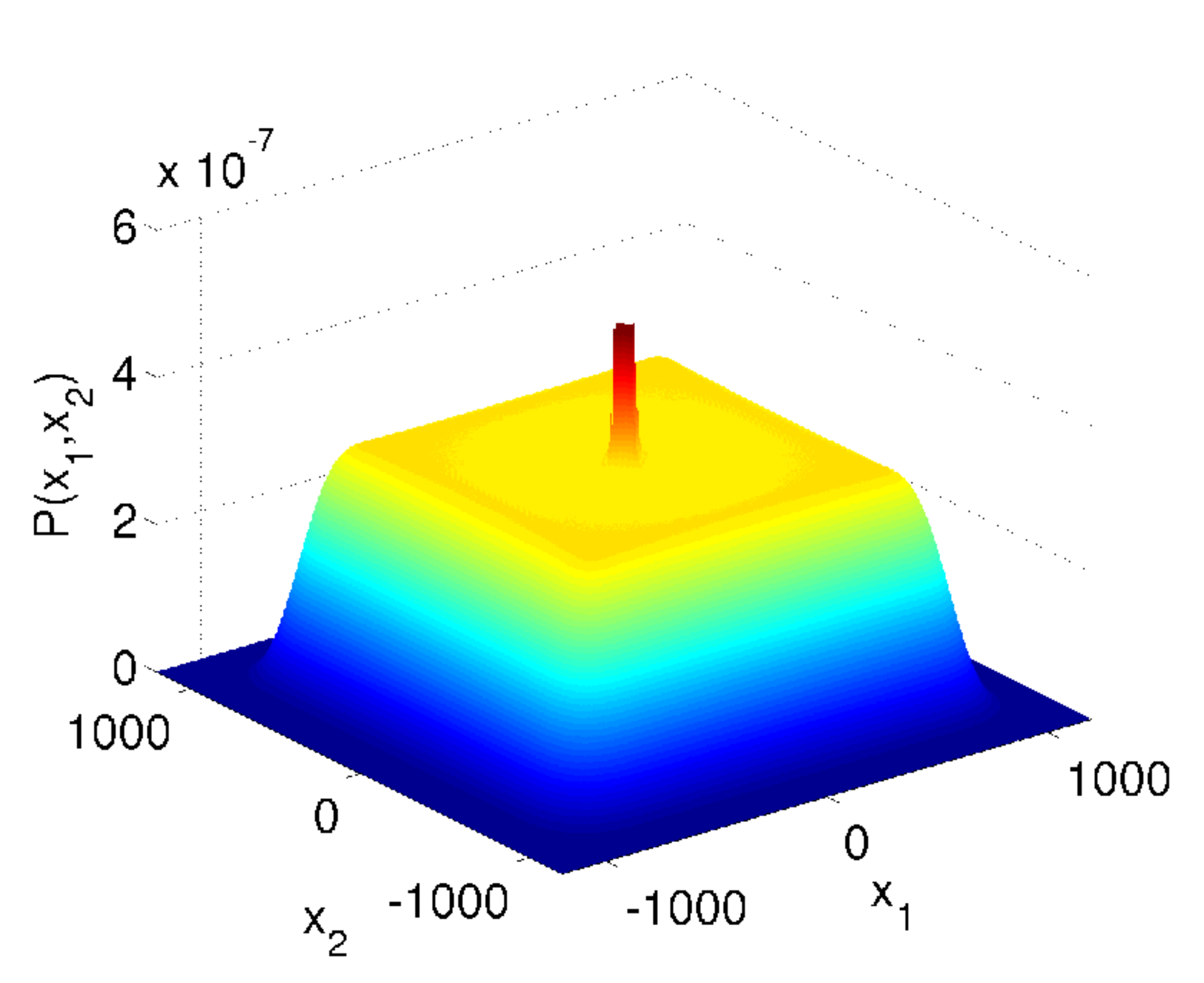}

\caption{Probability distribution as a function of the dimensionless $(x_{1},x_{2})$
position at time t=9000, for initial conditions at the saddle point
$\mathbf{k}_{0}=(0,\pi)$. The coin initial state is $\mid\phi_{\mathbf{k}_{0}}^{(1)}\rangle=\frac{1}{2}\operatorname{col}\left(1+i,1+i,1-i,1-i\right)$,
and the starting space distribution as given by Eq. (\ref{saddlesinc}).
We have taken a value $\sigma_{0}=15$ for this simulation. The central
peak has been cut for a better display of the rest of the features.}

\label{figsaddle}
\end{figure}

A completely different situation arises when $\mathbf{k}_{0}$ corresponds
to one of the saddle points (such as the one located at $\mathbf{k}_{sp}=\left(0,\pi\right)$).
As the group velocity vanishes at that point, i.e., $\mathbf{v}_{g}^{\left(1,2\right)}\left(\mathbf{k}_{0}\right)=\left(0,0\right)$,
the center of the probability distribution is expected to remain at
rest but will spread in time analogously to diffracting optical waves,
as Eq. (\ref{Schr}) reduces to the paraxial wave equation of optical
diffraction \cite{Goodman}. In our case, by considering $s=1$ as
the initial coin state, one finds $\varpi_{11}=-1/2,\varpi_{12}=\varpi_{21}=0,\varpi_{22}=1/2.$
Therefore, Eq. (\ref{Schr}) becomes 
\begin{equation}
\mathrm{i}\frac{\partial A^{\left(1\right)}\left(\mathbf{X},t\right)}{\partial t}=-\frac{1}{4}\frac{\partial^{2}A^{\left(1\right)}}{\partial X_{1}^{2}}+\frac{1}{4}\frac{\partial^{2}A^{\left(1\right)}}{\partial X_{2}^{2}}.\label{eqparax}
\end{equation}
This wave equation is similar to that of paraxial optical diffraction
\cite{Goodman}, differing from it in that the sign of spatial derivatives
are different for the two spatial directions. Hence this hyperbolic
equation describes a situation in which there is diffraction in one
direction and anti-diffraction in the other direction, as it happens
in certain optical systems \cite{Kolpakov}. This causes that the
cylindrical symmetry proper of diffraction is replaced by a rectangular
symmetry. Basing on the analogy with diffraction, it is possible to
tailor the initial distribution in order to reach a desired asymptotic
distribution as we already demonstrated in the one-dimensional QW
\cite{German}. Hence if we want to reach a final homogeneous (top
hat) distribution, we must use an initial condition of the form \cite{Goodman,German}
\begin{equation}
\psi_{\mathbf{x},0}=\mathcal{N}e^{i\mathbf{k}_{0}\cdot\mathbf{x}}e^{-\frac{x_{1}^{2}+x_{2}^{2}}{2\sigma^{2}}}sinc\left(x_{1}/\sigma_{0}\right)\,\, sinc\left(x_{2}/\sigma_{0}\right),\label{saddlesinc}
\end{equation}
with $sinc\left(x\right)=\sin\left(\pi x\right)/\pi x$ and $\sigma_{0}$
a constant that accounts for the width of the distribution. In Fig.
\ref{figsaddle} we observe the evolution of the Grover walk with
the above initial condition and the initial coin $\left\vert \phi_{\mathbf{k_{0}},0}^{(1)}\right\rangle =\frac{1}{2}col\left(1+i,1+i,1-i,1-i\right)$,
equal for all populated sites. The figure shows a final distribution
that is quite homogeneous along most of its support, except in its
outer borders, where it shows a smooth but rapid fall out (the ratio
$\sigma/\sigma_{0}$ has been chosen to optimize the result \cite{German}).
This is the expected result, but Fig. \ref{figsaddle} also shows
a central peak that has been cut for a better display, the height
of this central peak being quite large (around $1.5\cdot10^{-5}$
) as compared to the plateau (around $4\cdot10^{-7}$). The existence
of that central peak is a consequence of the common initial coin $\left\vert \phi_{\mathbf{k_{0}},0}\right\rangle $.
This coin state guarantees that the initial distribution projects
just onto the relevant branch of the dispersion relation at its central
spatial frequency but, as the width of the distribution is finite,
for the larger values of $\left\vert \mathbf{k}-\mathbf{k}_{0}\right\vert $
it will provide a non--negligible projection onto the static branches
of the dispersion relation and the localized part of these projections
are what we see as a central peak in Fig. \ref{figsaddle}.

\subsection{Dynamics at the diabolical points}

In this case $N=2$, $\mathrm{d}^{N}\Omega=\mathrm{d}\theta$, with
$\theta$ the azimuth, and 
\begin{equation}
\left\vert \mathbf{E}_{\mathbf{x}}^{\left(s\right)}\left(k\right)\right\rangle =\int_{0}^{2\pi}\mathrm{d}\theta\exp\left(\mathrm{i}\mathbf{k}\cdot\mathbf{x}\right)\langle\phi_{\mathbf{k}_{\mathrm{D}}+\mathbf{k}}^{\left(s\right)}\mid\Xi\rangle\mid\phi_{\mathbf{k}_{\mathrm{D}}+\mathbf{k}}^{\left(s\right)}\rangle.\label{Exs}
\end{equation}
We remind that here $\left\vert \Xi\right\rangle $ is assumed to
project just onto $\left\vert \phi_{\mathbf{k}_{\mathrm{D}}+\mathbf{k}}^{\left(s=1,2\right)}\right\rangle $.
This means that $\left\vert \Xi\right\rangle $ coincides with $\left\vert \phi_{\mathrm{D}}\right\rangle $,
see (\ref{fi_D}). According to the projections (\ref{proj_close})
and to the vectors (\ref{eigen_close}) we can write, in matrix form,
\begin{equation}
\mathbf{E}_{\mathbf{x}}^{\left(s\right)}\left(k\right)=\frac{1}{4}\int_{0}^{2\pi}\mathrm{d}\theta\exp\left(\mathrm{i}\mathbf{k}\cdot\mathbf{x}\right)\left(\begin{array}{c}
1\pm\sqrt{2}\cos\theta\\
1\mp\sqrt{2}\cos\theta\\
-1\mp\sqrt{2}\sin\theta\\
-1\pm\sqrt{2}\sin\theta
\end{array}\right),
\end{equation}
where the upper (lower) sign in $\pm$ or $\mp$ corresponds to $s=1(2)$,
respectively, and we remind that $\mathbf{k}=\left(k\cos\theta,k\sin\theta\right)$.
Upon writing $\mathbf{k}\cdot\mathbf{x}=kx\cos\left(\theta-\varphi\right)$,
where $\mathbf{x}=x\left(\cos\varphi,\sin\varphi\right)$ and performing
the integral we arrive to 
\begin{equation}
\mathbf{E}_{\mathbf{x}}^{\left(s\right)}\left(k\right)=\frac{\pi}{2}\left(\begin{array}{c}
J_{0}\left(kx\right)\pm\mathrm{i}\sqrt{2}J_{1}\left(kx\right)\cos\varphi\\
J_{0}\left(kx\right)\mp\mathrm{i}\sqrt{2}J_{1}\left(kx\right)\cos\varphi\\
-J_{0}\left(kx\right)\mp\mathrm{i}\sqrt{2}J_{1}\left(kx\right)\sin\varphi\\
-J_{0}\left(kx\right)\pm\mathrm{i}\sqrt{2}J_{1}\left(kx\right)\sin\varphi
\end{array}\right),\label{Exk_diab}
\end{equation}
where $J_{n}$ are the Bessel functions of the first kind of order
$n$. Using this result in (\ref{FD}) we get, in matrix form, 
\begin{equation}
\mathbf{F}_{\mathbf{x},t}^{\left(s\right)}=\left(2\pi\right)^{-2}\int\mathrm{d}k\exp\left(\mp\mathrm{i}kct\right)k\tilde{F}_{k}\mathbf{E}_{\mathbf{x}}^{\left(s\right)}\left(k\right).\label{Fxt_diab}
\end{equation}

Let us compute the probability of finding the walker at some point,
regardless the coin state, as given by (\ref{prob}), (\ref{evol_x}),
(\ref{psiD_aux}) and (\ref{Fxt_diab}). After simple algebra we obtain
\begin{equation}
P_{\mathbf{x},t}=\sum_{s,s^{\prime}}\left[\mathbf{F}_{\mathbf{x},t}^{\left(s\right)}\right]^{\dag}\cdot\mathbf{F}_{\mathbf{x},t}^{\left(s^{\prime}\right)}=\left[p_{\mathbf{x},t}^{\left(0\right)}\right]^{2}+\left[p_{\mathbf{x},t}^{\left(1\right)}\right]^{2},\label{PD}
\end{equation}
where 
\begin{align}
p_{\mathbf{x},t}^{\left(0\right)} & =\left(2\pi\right)^{-1}\int_{0}^{\infty}\mathrm{d}k\cos\left(kct\right)J_{0}\left(kx\right)k\tilde{F}_{k},\label{p0}\\
p_{\mathbf{x},t}^{\left(1\right)} & =\left(2\pi\right)^{-1}\int_{0}^{\infty}\mathrm{d}k\sin\left(kct\right)J_{1}\left(kx\right)k\tilde{F}_{k},\label{p1}
\end{align}
are amplitude probabilities. We see that the spatial dependence of
the probability amplitude on the azimuth $\varphi$ in (\ref{Fxt_diab},\ref{Exk_diab})
disappears when looking at the full probability $P_{\mathbf{x},t}$.
This means that a coin-sensitive measurement of the probability should
display angular features.

As a summary of what we know up to here about the neighborhood of
the diabolical point, we can say that when the initial coin $\left\vert \Xi\right\rangle $
coincides with $\left\vert \phi_{\mathrm{D}}\right\rangle $, see
(\ref{fi_D}), we predict (i) no localization and, (ii) when the initial
wave packet $F_{\mathbf{x},0}$ is radially symmetric (it only depends
on $x=\left\vert \mathbf{x}\right\vert $), an evolution of the probability
$P_{\mathbf{x},t}$ that keeps the radial symmetry along time, according
to (\ref{PD},\ref{p0},\ref{p1}).

Equations (\ref{p0}) and (\ref{p1}) describe the evolution of the
probability of an initial wave packet centered at a diabolical point
and with radial symmetry. The actual probability $P_{\mathbf{x},t}$
can be computed from them numerically, but little can be said in general.
In order to gain some insight we consider next the relevant long time
limit, which turns out to be analytically tractable. First of all
it is convenient to scale the radial wavenumber $k$ to the (loosely
defined) width of the initial state in real space, $F_{\mathbf{x},0}$,
which we denote by $\sigma$. Accordingly we define $\kappa=\sigma k$
so that (\ref{p0}) and (\ref{p1}) become 
\begin{align}
p_{\mathbf{x},t}^{\left(0\right)} & =\left(2\pi\sigma^{2}\right)^{-1}\int_{0}^{\infty}\mathrm{d}\kappa\cos\left(\frac{ct}{\sigma}\kappa\right)J_{0}\left(\frac{x}{\sigma}\kappa\right)\kappa\tilde{f}_{\kappa},\label{p0k}\\
p_{\mathbf{x},t}^{\left(1\right)} & =\left(2\pi\sigma^{2}\right)^{-1}\int_{0}^{\infty}\mathrm{d}\kappa\sin\left(\frac{ct}{\sigma}\kappa\right)J_{1}\left(\frac{x}{\sigma}\kappa\right)\kappa\tilde{f}_{\kappa},\label{p1k}
\end{align}
where $\tilde{f}_{\kappa}=\tilde{F}_{k=\kappa/\sigma}$ is non null
only for $\kappa\lesssim1$. Hence when $ct\gg\sigma$, $\cos\left(\frac{ct}{\sigma}\kappa\right)$
and $\sin\left(\frac{ct}{\sigma}\kappa\right)$ are strongly oscillating
functions of $\kappa$, around zero, what would make $p_{\mathbf{x},t}^{\left(0,1\right)}$
to vanish. However in the integrals defining such amplitude probabilities
other oscillating functions, $J_{0}\left(\frac{x}{\sigma}\kappa\right)$
and $J_{1}\left(\frac{x}{\sigma}\kappa\right)$, appear. Thus it can
be expected that when the oscillations of the latter and the ones
of the former are partially in phase, a non null value of the integrals
is got. According to the asymptotic behavior of $J_{n}\left(z\right)$
at large $z$, $J_{n}\left(z\right)\approx\sqrt{2/\pi z}\cos\left(z-n\pi/2-\pi/4\right)$
for $z\gg\left\vert n^{2}-1/4\right\vert $ \cite{Bessel}, this partial
phase matching occurs when $x\approx ct$, which is expected from
physical considerations. Hence we are led to consider the limit $x=ct+\sigma\xi$,
with $ct\gg\sigma$, where $\xi$ is the scaled radial offset with
respect to $ct$, in which case (\ref{p0k}) and (\ref{p1k}), after
expressing the products of trigonometric functions as sums, become
\begin{align}
p_{\mathbf{x},t}^{\left(0\right)} & =p_{\mathbf{x},t}^{\left(1\right)}\nonumber \\
 & \approx\left(2\pi\sigma^{2}\right)^{-1}\sqrt{\frac{\sigma}{2\pi ct}}\int_{0}^{\infty}\mathrm{d}\kappa\sin\left(\kappa\xi+\pi/4\right)\sqrt{\kappa}\tilde{f}_{\kappa}\nonumber \\
 & =\left(2\pi\right)^{-1}/\sqrt{2\pi ct}\int_{0}^{\infty}\mathrm{d}k\sin\left(k\sigma\xi+\pi/4\right)\sqrt{k}\tilde{F}_{k},\label{eqp0p1}
\end{align}
to the leading order, where the approximation follows from disregarding
highly oscillating terms. As an application of this result we consider
next the Gaussian case $\tilde{F}_{k}=2\sigma\sqrt{\pi}\exp\left(-\frac{1}{2}\sigma^{2}k^{2}\right)$,
where $\sigma$ is the standard deviation of the initial probability
in real space, Eq. (\ref{initpsi}). We readily obtain 
\begin{align}
p_{\mathbf{x},t}^{\left(0\right)} & =p_{\mathbf{x},t}^{\left(1\right)}=\frac{1}{8\sqrt{\sigma ct}}e^{-\xi^{2}/4}\nonumber \\
 & \times\left\{ \xi\sqrt{\left\vert \xi\right\vert }\left[I_{-1/4}\left(\xi^{2}/4\right)-I_{3/4}\left(\xi^{2}/4\right)\right]\right.\nonumber \\
 & +\left.\frac{\xi^{2}I_{5/4}\left(\xi^{2}/4\right)-\left(\xi^{2}-2\right)I_{1/4}\left(\xi^{2}/4\right)}{\sqrt{\left\vert \xi\right\vert }}\right\} ,\label{pogen0}
\end{align}
where $\xi=\frac{x-ct}{\sigma}$ and $I_{n}$ is the modified Bessel
function of the first kind of order $n$. The total probability $P_{\mathbf{x},t}\equiv P(\xi)$
follows from using (\ref{pogen0}) in (\ref{PD}), from which we observe
that the initial width $\sigma$ acts only as a scale factor for the
height and shape of the probability at long times ($ct\gg\sigma$).
A plot of the probability can be seen in Fig. \ref{cutvaso}, where
two unequal spikes, whose maxima are located at $\xi=-1.74623$ and
at $\xi=0.550855$, separated by a zero at $\xi=-0.765951$, are apparent.
We remind that $x=ct+\sigma\xi$ is a radial coordinate, hence these
maxima correspond to two concentric rings separated by a dark ring.
The situation is fully analogous to the so-called ``Poggendorff rings\textquotedblright{}\ appearing
in the paraxial propagation of a beam incident along an optic axis
of a biaxial crystal, a situation in which a conical intersection
(a diabolical point) is present. Remarkably the result we have obtained
for $P_{\mathbf{x},t}$ fully coincides with that described in \cite{Berry2}
whose Fig. 7 is to be compared to our Fig. \ref{cutvaso}.

\begin{figure}
\begin{minipage}[t]{1\columnwidth}%
\includegraphics[width=8cm]{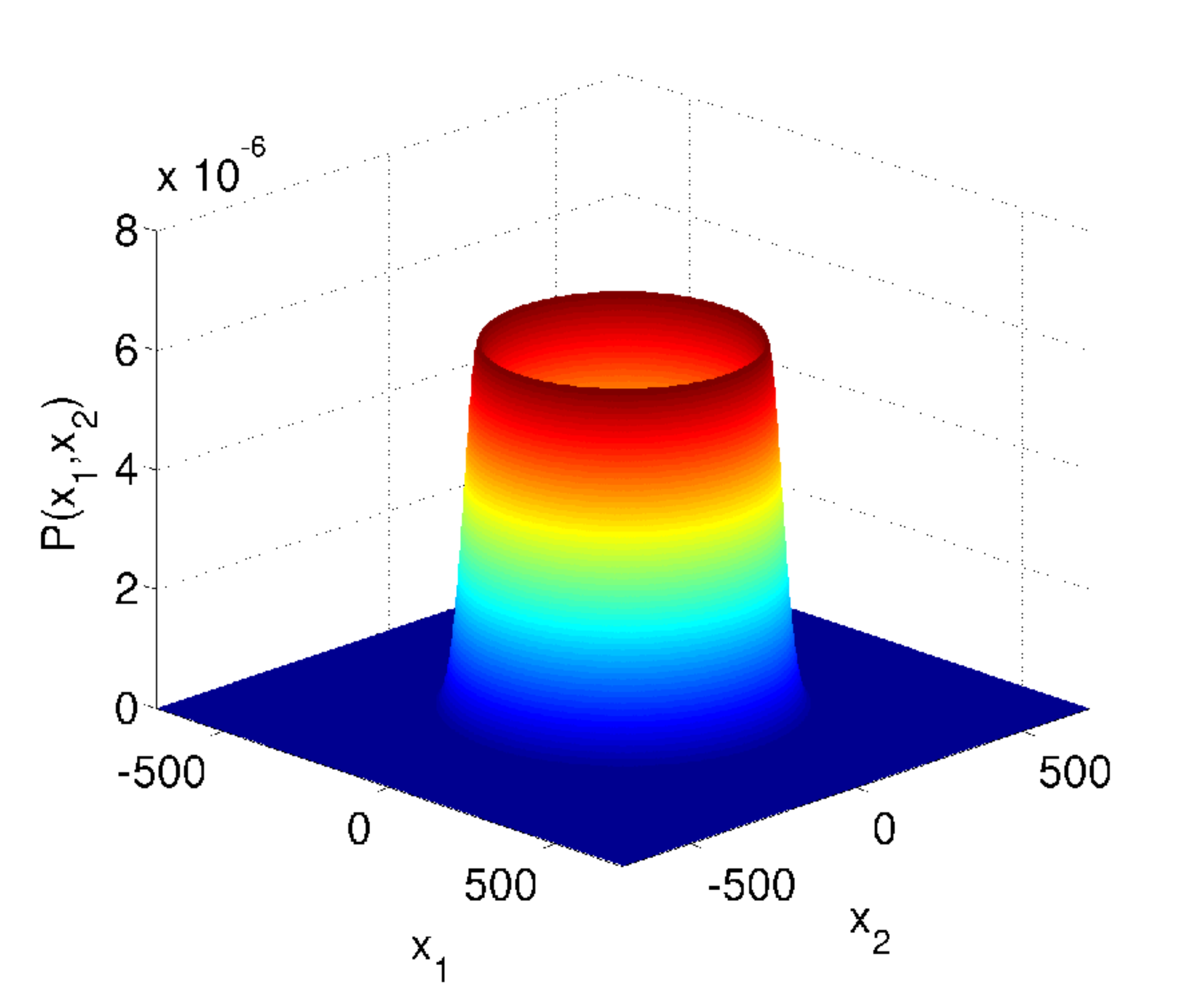}

\includegraphics[width=4cm]{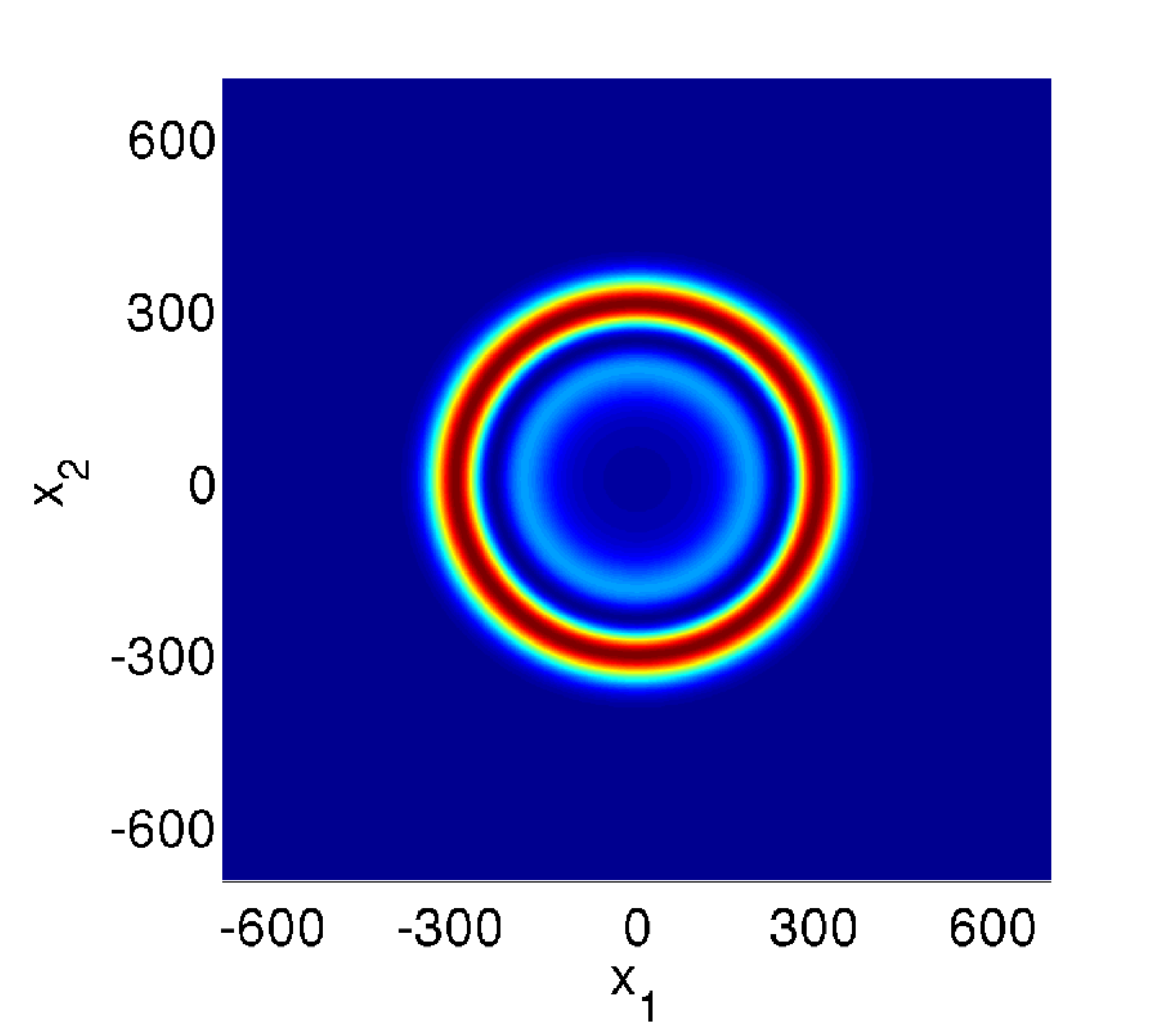}\includegraphics[width=4cm]{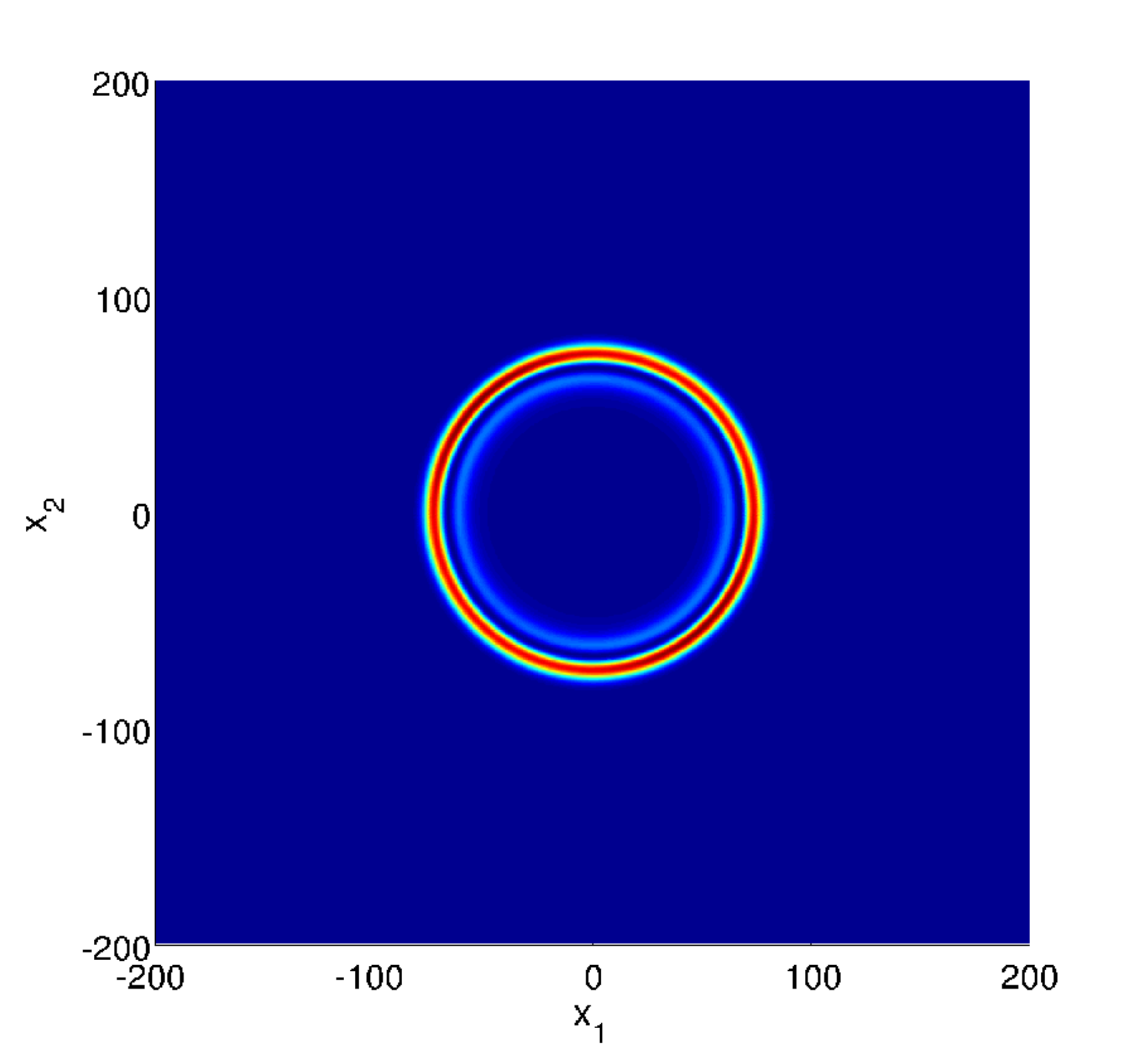}%
\end{minipage}

\caption{(Top panel): Probability distribution as a function of the dimensionless
$(x_{1},x_{2})$ position, at time $t=400$. The initial condition
is given by Eq. (\ref{fi_D}) around the diabolic point $\mathbf{k}_{\mathrm{D}}=(0,0)$,
with $\sigma=50$. (Bottom panel): Top view of the previous figure
with $\sigma=50$ (left) and $\sigma=5$ (right), but now at time
$t=100$.}

\label{vaso}
\end{figure}

\begin{figure}
\includegraphics[width=8cm]{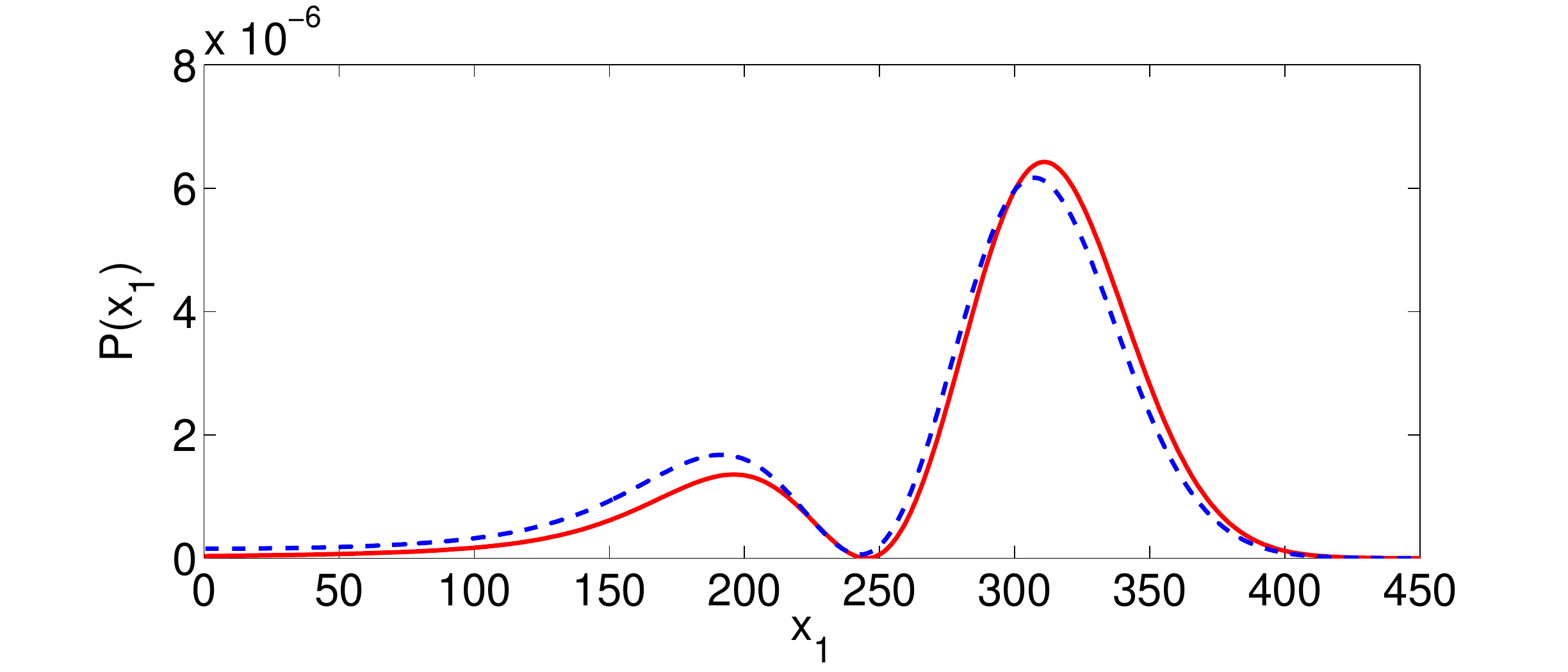}

\caption{Radial cut, with $x_{2}=0$, for the top panel of Fig. \ref{vaso}
(blue, dotted curve), showing the Poggendorff rings (see the text
for explanation). Also shown for comparison is our analytical result,
as obtained form Eqs. (\ref{PD}) and (\ref{pogen0}) (solid, red
curve).}

\label{cutvaso}
\end{figure}

As an example that illustrates the above features, we have studied
the evolution of a state of the form Eq. (\ref{initpsi}) for the
diabolic point $\mathbf{k}_{\mathrm{D}}=(0,0)$, and a coin state
as defined in Eq. (\ref{fi_D}). As seen from Fig. \ref{vaso}, the
probability distribution keeps its cylindrical symmetry while it expands
in position. It can also be seen from this figure that the general
features remain valid for a wider distribution (in $k$ space) corresponding
to $\sigma=5$, even if the distribution in spatial coordinates is
narrower for a given time step. The Poggendorff rings can be appreciated
in Fig. \ref{cutvaso}, which shows a radial cut of the previous figure.
We also plot for comparison our analytical result, as obtained form
Eqs. (\ref{PD}) and (\ref{pogen0}) , which shows a good agreement
with the (exact) numerical result, as can be seen from this figure.

\begin{figure}
\includegraphics[width=8cm]{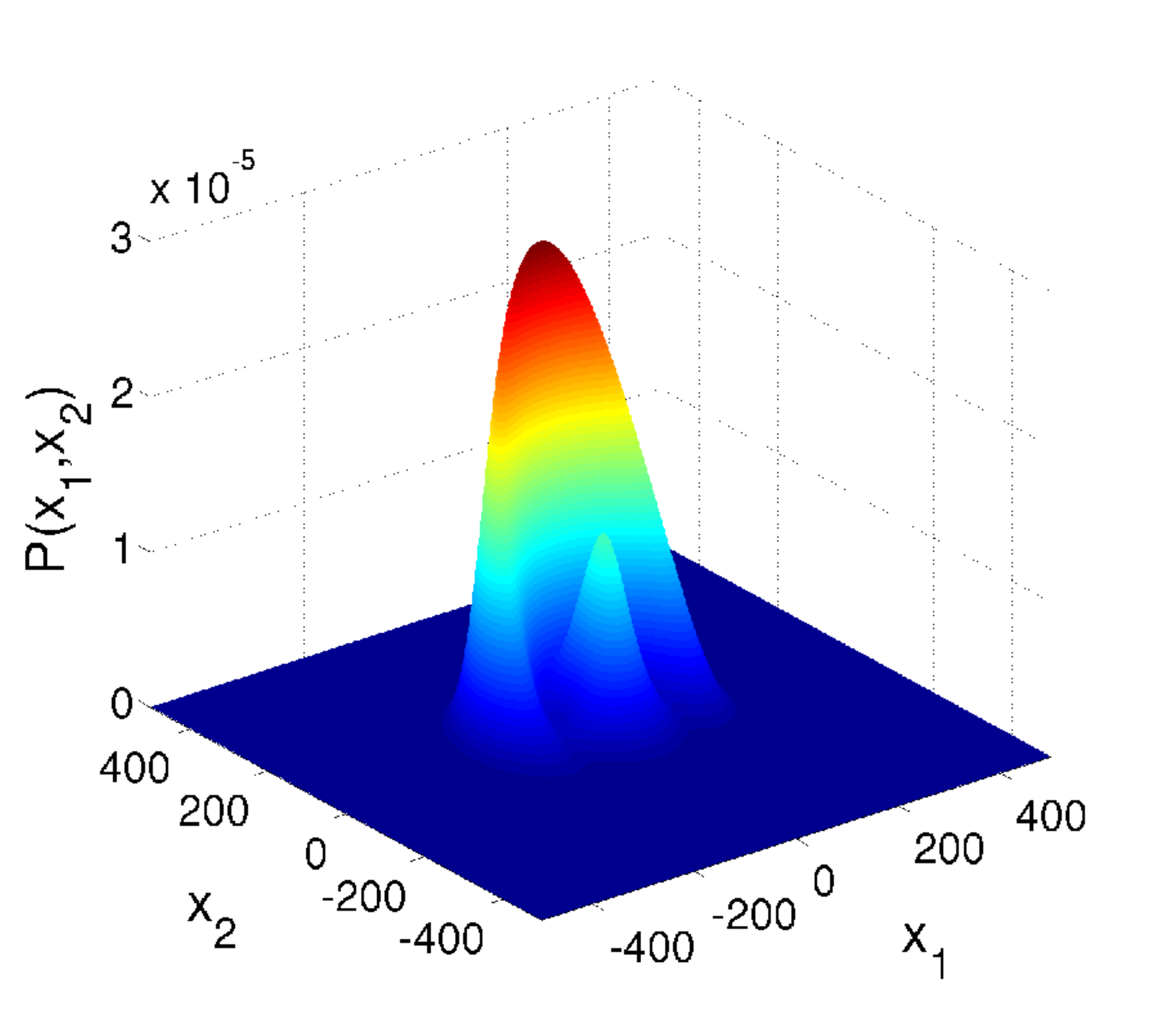}

\caption{Probability distribution as a function of the dimensionless $(x_{1},x_{2})$
position, at time $t=200$. The initial condition is given by $\left\vert \phi_{\mathbf{k}_{\mathrm{D}}}^{\left(2\right)}\right\rangle $
around the diabolic point with $\theta=\pi/2$ .}

\label{diabphi2pis2}
\end{figure}

All these features crucially depend on the choice of the coin initial
conditions, as clearly observed when these conditions are chosen differently.
As an example, we show in Fig. \ref{diabphi2pis2} the evolution after
200 time steps, of the probability distribution when we start from
a state, also centered around the diabolic point, but the vector coin
is now $\left\vert \phi_{\mathbf{k}_{\mathrm{D}}}^{\left(2\right)}\right\rangle $
with $\theta=\pi/2$. Most of the probability is directed along the
positive $x_{2}$ axis, which corresponds to the choice of $\theta$.
In fact, by changing the value of this parameter one can, at will,
obtain a chosen direction for propagation.

\section{Application to the 3D Grover QW}

In this section we briefly analyze the three--dimensional Grover QW
in order to illustrate similarities and differences with the two--dimensional
case. In 3D, the Grover-coin operator reads
\begin{equation}
C=\frac{1}{3}\left(\begin{array}{cccccc}
-2 & 1 & 1 & 1 & 1 & 1\\
1 & -2 & 1 & 1 & 1 & 1\\
1 & 1 & -2 & 1 & 1 & 1\\
1 & 1 & 1 & -2 & 1 & 1\\
1 & 1 & 1 & 1 & -2 & 1\\
1 & 1 & 1 & 1 & 1 & -2
\end{array}\right),\label{Grovercoin3D}
\end{equation}
and the diagonalization of the corresponding matrix $C_{\mathbf{k}}$
(\ref{Ck_elements}), with $\mathbf{k}=(k_{1},k_{2},k_{3})$, yields
six eigenvalues $\lambda_{\mathbf{k}}^{\left(s\right)}=\exp\left(-\mathrm{i}\omega_{\mathbf{k}}^{\left(s\right)}\right)$
with 
\begin{equation}
\omega_{\mathbf{k}}^{\left(1,2\right)}=\pm\Omega_{\mathbf{k}}^{\left(+\right)},\text{ }\omega_{\mathbf{k}}^{\left(3,4\right)}=\pm\Omega_{\mathbf{k}}^{\left(-\right)},\ \omega_{\mathbf{k}}^{\left(5\right)}=0,\text{ }\omega_{\mathbf{k}}^{\left(6\right)}=\pi,\label{Lambdas3D}
\end{equation}
where 
\begin{align}
\cos\Omega_{\mathbf{k}}^{\left(\pm\right)} & =-\frac{1}{3}\left[\sum_{i=1,2,3}\cos{k_{i}\pm}\right.\nonumber \\
 & \left.\sqrt{\sum_{i=1,2,3}^{j>i}\left(\cos^{2}{k_{i}}-\cos{k_{i}\cos}k_{j}\right)}\right].\label{disp3D}
\end{align}
Remember that adding a multiple integer of $2\pi$ to any of the $\omega$'s
does not change anything because time is discrete and runs in steps
of $1$. This implies that $\omega=-\pi$ and $\omega=\pi$ represent
the same frequency.

\begin{figure}
\begin{minipage}[t]{1\columnwidth}%
\includegraphics[width=3cm]{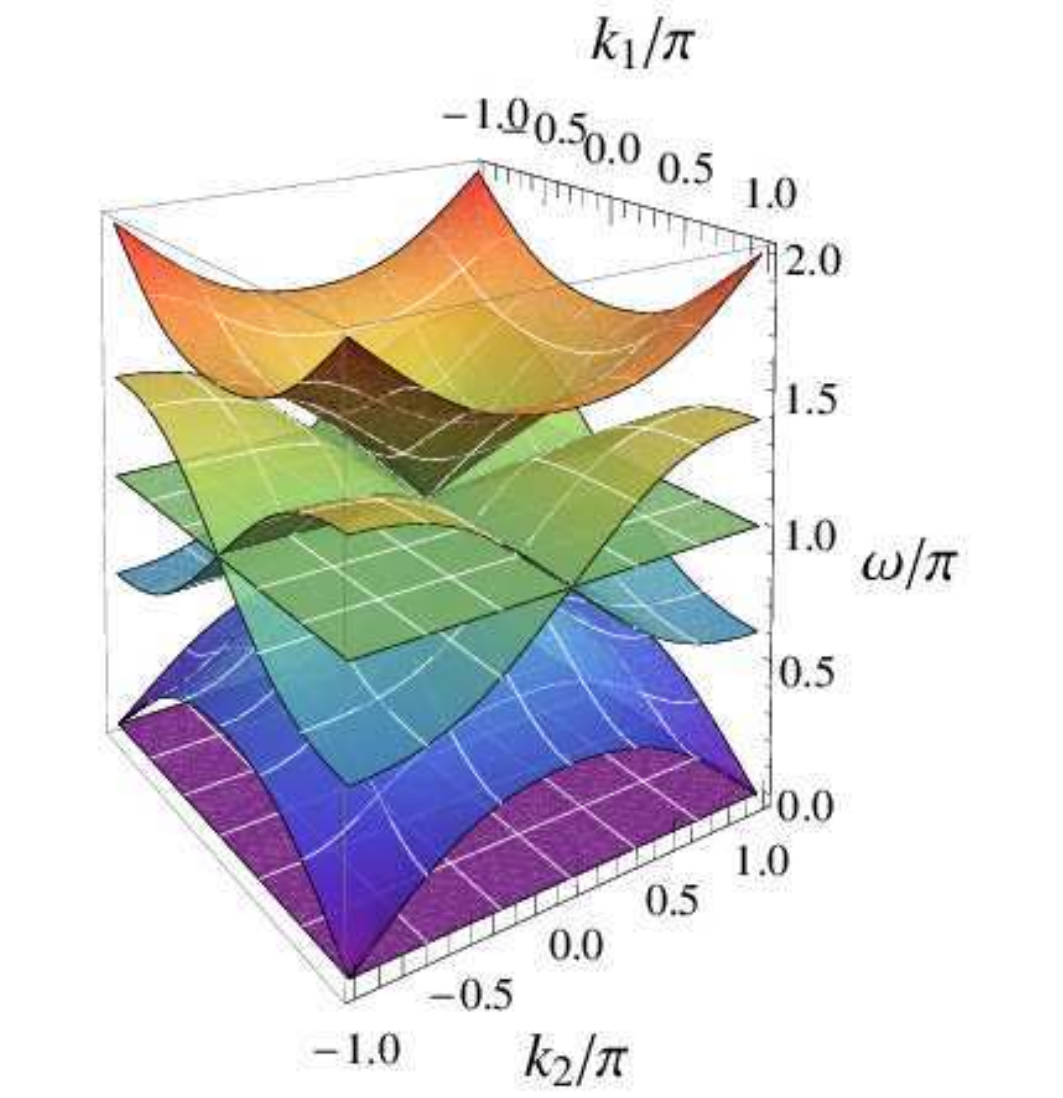}\includegraphics[width=3cm]{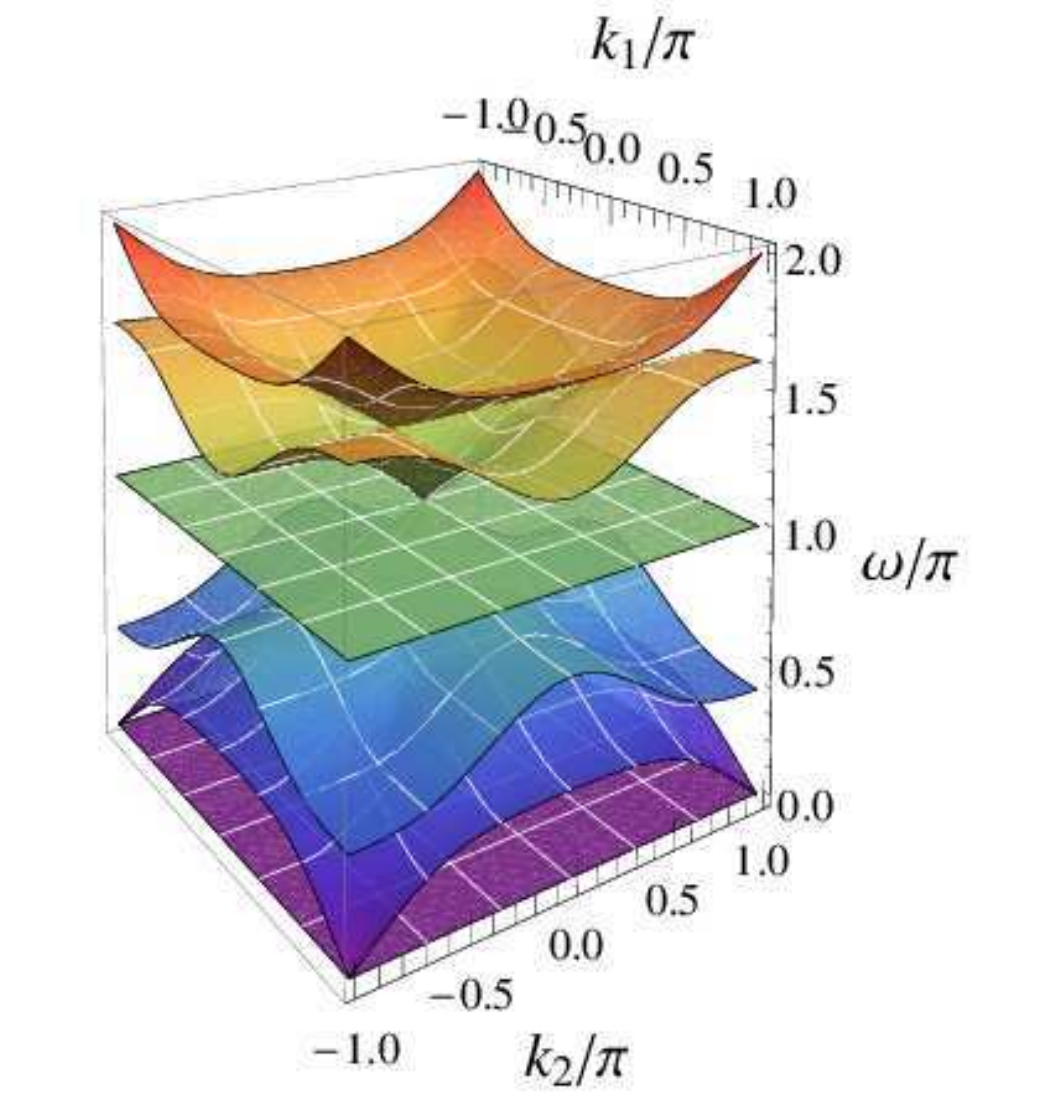}\includegraphics[width=3cm]{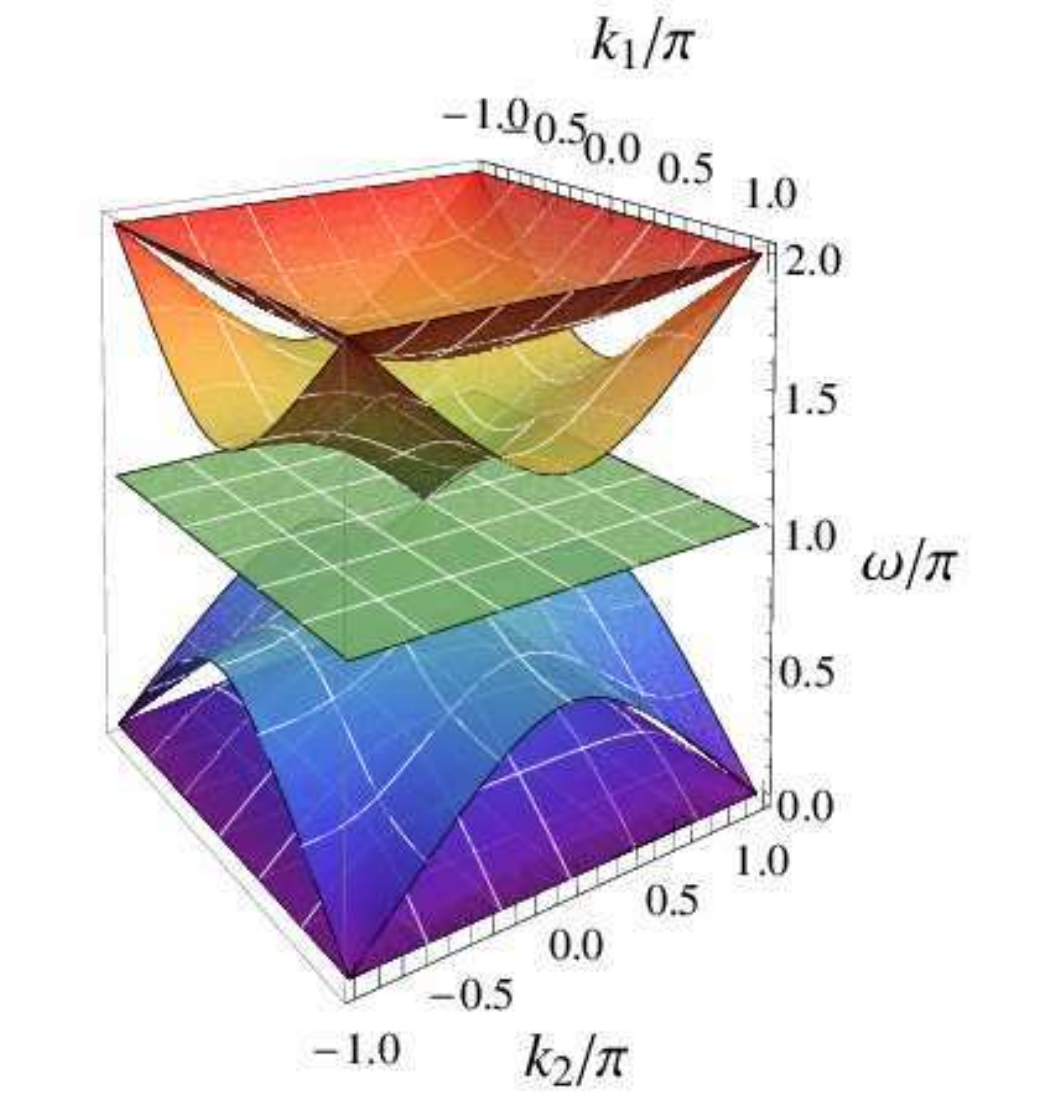}%
\end{minipage}

\caption{Representations of the Grover 3D six dispersion relation surfaces
for $k_{3}=0$ (left), $\pi/2$ (center), and $\pi$ (right).}

\label{disper3d}
\end{figure}

\begin{figure}
\begin{minipage}[t]{1\columnwidth}%
\includegraphics[width=3cm]{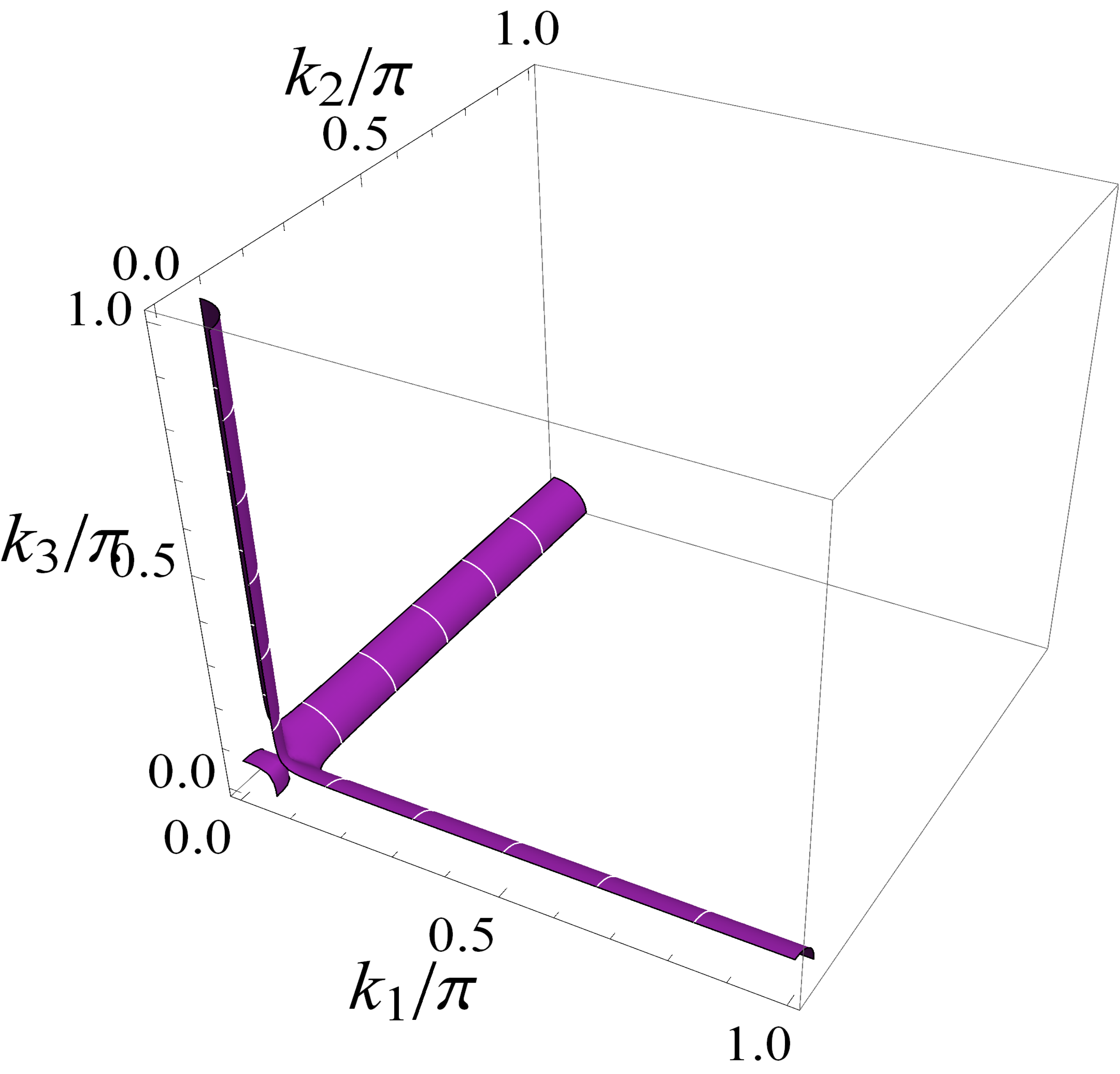}\includegraphics[width=3cm]{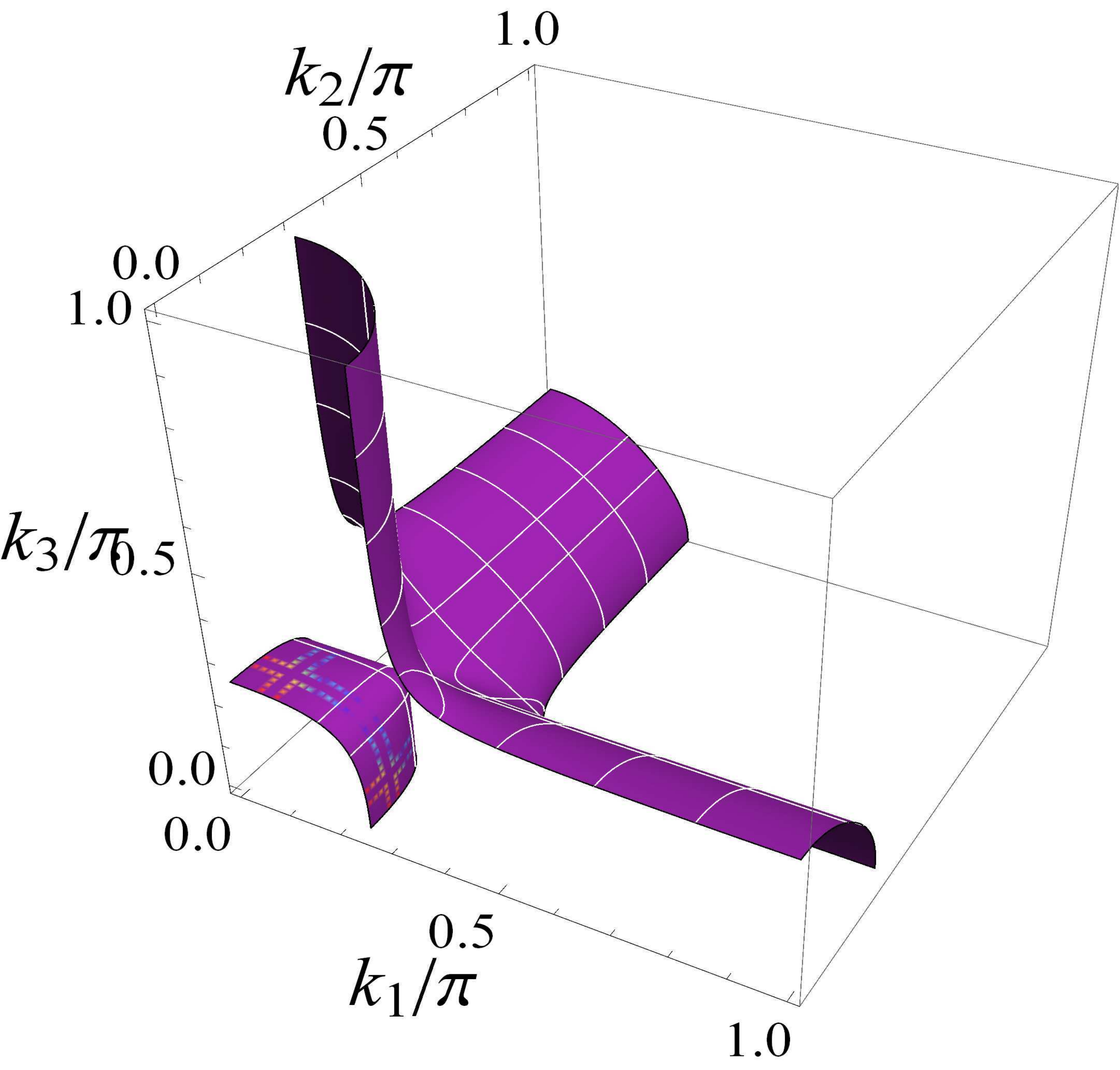}\includegraphics[width=3cm]{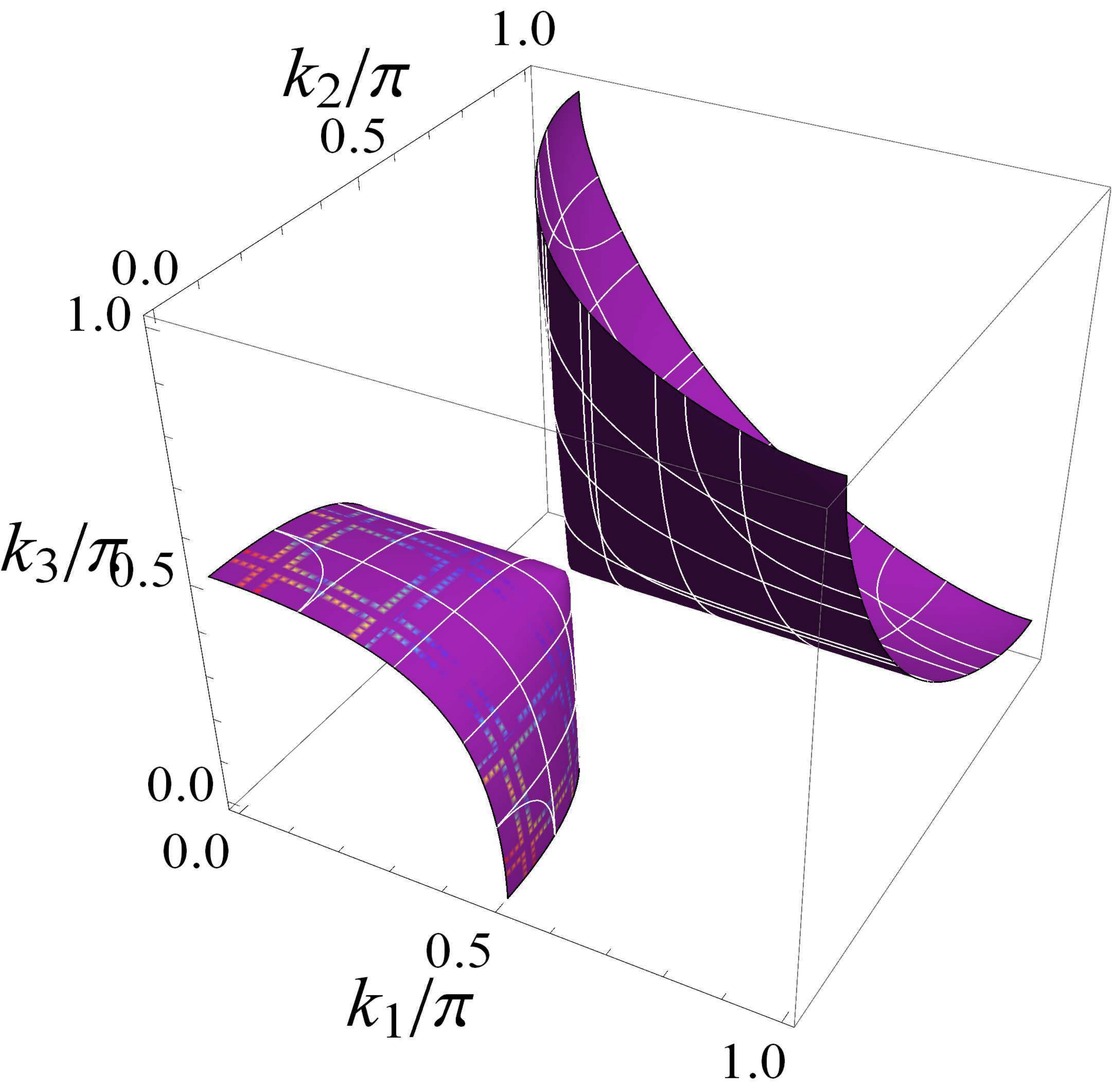}

\includegraphics[width=3cm]{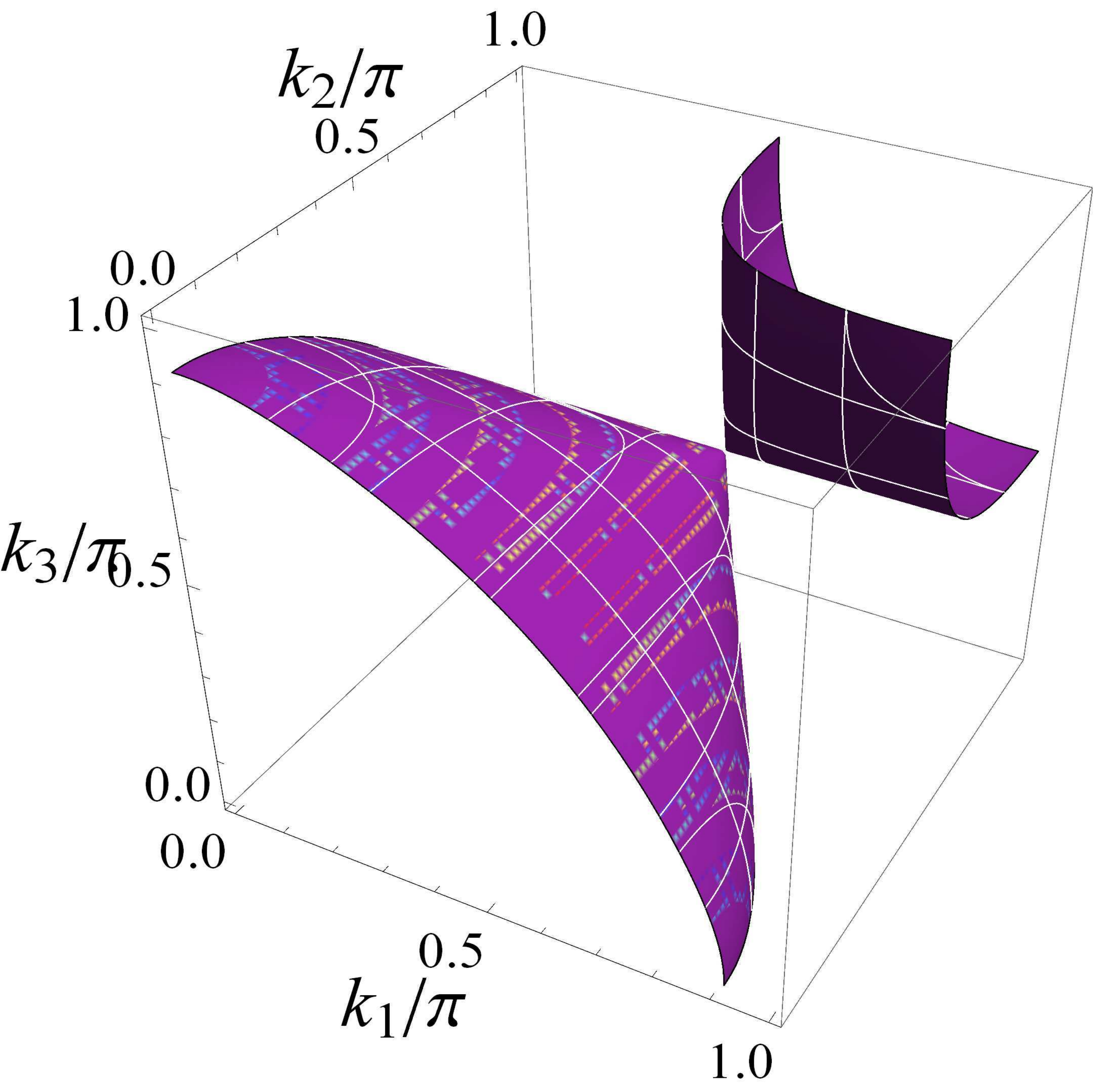}\includegraphics[width=3cm]{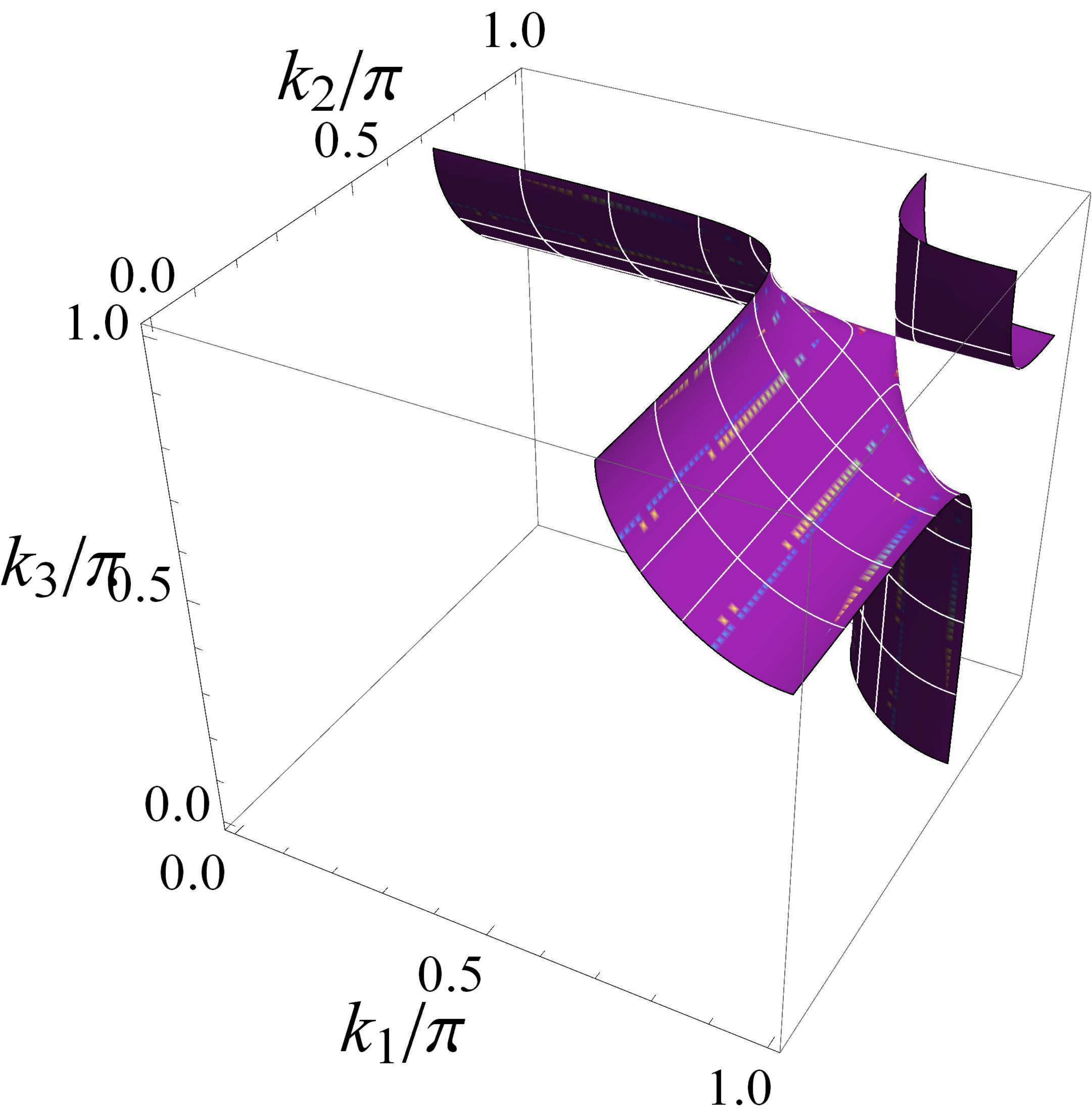}\includegraphics[width=3cm]{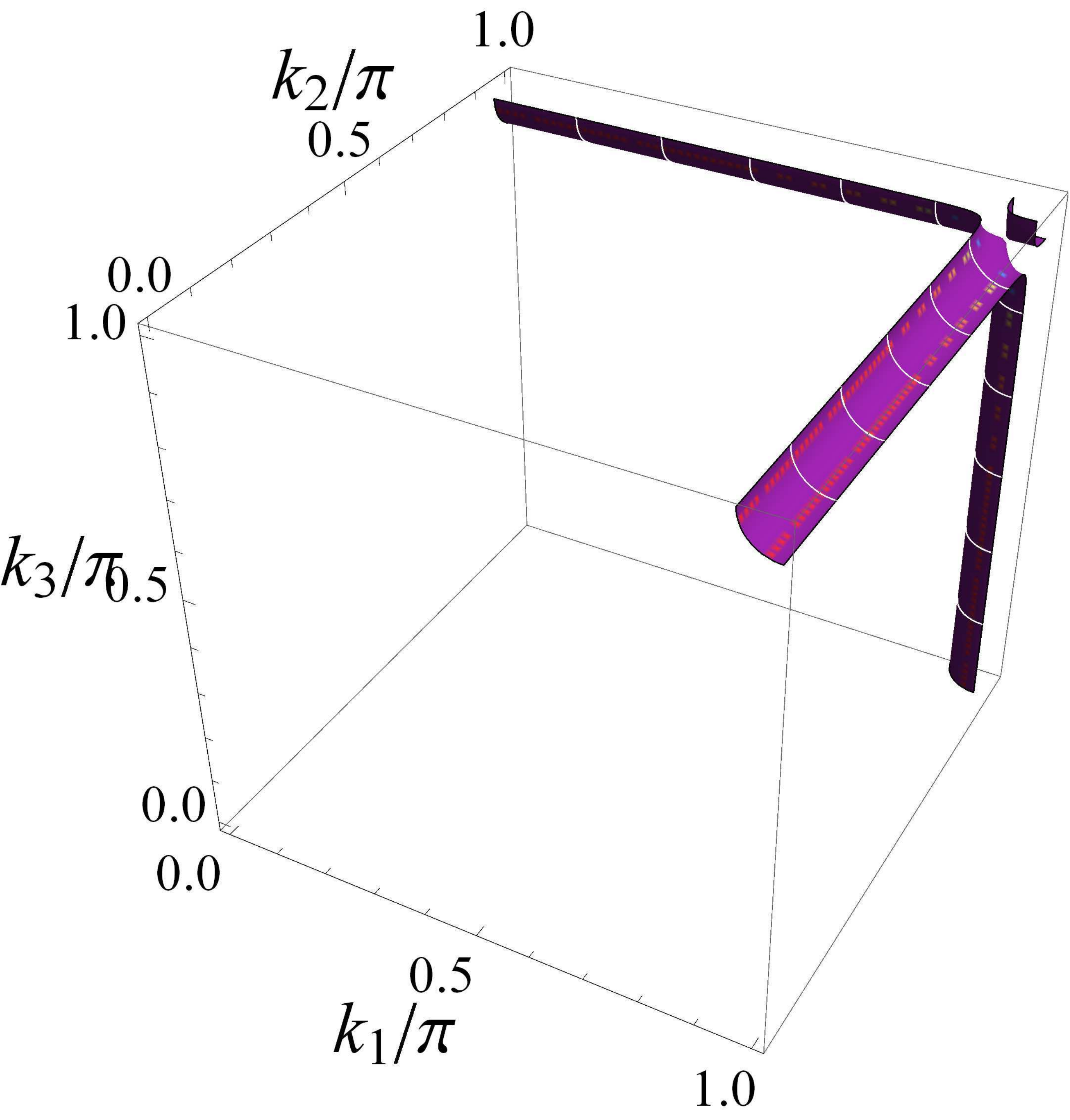}%
\end{minipage}

\caption{Representations of the Grover 3D isofrequency surfaces, as obtained
from $\Omega_{\mathbf{k}}^{\left(+\right)}$ in the first octant of
the Brillouin zone for $\omega=0.95\pi$ (a), $0.8\pi$ (b), $0.6\pi$
(c), $0.4\pi$ (d), $0.2\pi$ (e), and $0.05\pi$ (f). For $\omega=\pi$
the two dispersion relation coincides with the cube axes and the cube
origin, whilst for $\omega=0$ they have migrated to the opposite
side of the cube.}

\label{movindeg3d}
\end{figure}

The graphical representation is more complicated in the 3D case: Plots
of the dispersion relations (\ref{Lambdas3D}) for particular values
of $k_{3}$ and for particular values of $\omega$ are given in Figs.
\ref{disper3d} and \ref{movindeg3d}, respectively. 

A large variety of propagation properties can be expected depending
on the particular region in $k$-space that the initial distribution
occupies. Notice the existence of two constant dispersion relations,
$\omega_{\mathbf{k}}^{\left(5,6\right)}$, which will give rise to
localization phenomena, as already discussed in the 2D case.

There exist several degeneracies in the above dispersion relations,
appearing in the crossings of several dispersion curves. For $\omega=\pi$,
there is a 5-fold degeneracy at $\mathbf{k}=\mathbf{0}$, where $\omega_{\mathbf{k}}^{\left(j\right)}=\pi$
for $j=1,2,3,4,6$. There is also a 3-fold degeneracy occurring when
$\mathbf{k}$ takes the form $\mathbf{k}=\mathbf{k}_{||}\equiv k\mathbf{e}_{i}$,
where $\mathbf{e}_{i}$ is one of the vectors of the canonical basis
in $\mathbb{R}^{3}$(i.e., $\mathbf{k}_{||}$ has two null components
and an arbitrary third component), where $\omega_{\mathbf{k}}^{\left(j\right)}=\pi$
for $j=1,2,6$. For $\omega=0$ these degeneracies migrate to the
corners and borders, respectively, of the first Brillouin zone, see
Fig. \ref{movindeg3d}. The frequencies $\Omega_{\mathbf{k}}^{\left(\pm\right)}$,
close to the degeneracies for $\omega=\pi$, read 
\begin{align}
\left.\cos\Omega_{\mathbf{k}}^{\left(\pm\right)}\right\vert _{\mathbf{k}=0} & \simeq-1+\frac{1}{6}\left\vert \mathbf{k}\right\vert ^{2}\mp\nonumber \\
 & \frac{1}{6}\sqrt{\left\vert \mathbf{k}\right\vert ^{4}-3\left(k_{x}^{2}k_{y}^{2}+k_{x}^{2}k_{z}^{2}+k_{y}^{2}k_{z}^{2}\right)}+O\left(k^{4}\right).
\end{align}
In the case that $\mathbf{k}$ takes the form $\mathbf{k}=\mathbf{k}_{||}\equiv k\mathbf{e}_{i}$,
we obtain

\begin{equation}
\cos\Omega_{\mathbf{k}}^{\left(+\right)}=-1
\end{equation}

\begin{equation}
\cos\Omega_{\mathbf{k}}^{\left(-\right)}=-\frac{1}{3}(1+2\cos k)
\end{equation}
which shows the above-mentioned degeneracy. 

We see that the point degeneracy occurring at $\mathbf{k}=0$, $\cos\Omega_{\mathbf{k}}^{\left(+\right)}=-1$,
is not a true diabolical point because, for fixed frequency, the dispersion
relation has not spherical symmetry (hence the dispersion relation
surface is not a hyperdiabolo). This implies that the theory developed
in Sec. III.B cannot be applied to this case. (We notice, however,
that it can be applied when coins different to the Grover one are
used as it occurs, for example, in the Alternate 3D QW, where true
3D diabolical points exist \cite{Roldan12}.) 

Its not difficult to see that the group velocities 
\begin{equation}
\mathbf{v}_{\mathrm{g}}^{\left(1,2,3,4\right)}\left(\mathbf{k}\right)=\pm\nabla_{\mathbf{k}}\Omega_{\mathbf{k}}^{\left(\pm\right)},\label{veg3D}
\end{equation}
have now the expression 
\begin{align}
\left(\mathbf{v}_{\mathrm{g}}^{\left(1,2\right)}\left(\mathbf{k}\right)\right)_{i} & =\mp\frac{(2\cos\Omega_{\mathbf{k}}^{\left(+\right)}+\sum_{j\neq i}\cos k_{j})\sin k_{i}}{2\left(3\cos\Omega_{\mathbf{k}}^{\left(+\right)}+\sum_{j}\cos k_{j}\right)\sin\Omega_{\mathbf{k}}^{\left(+\right)}},\label{veg3D2}
\end{align}
\begin{align}
\left(\mathbf{v}_{\mathrm{g}}^{\left(3,4\right)}\left(\mathbf{k}\right)\right) & _{i}=\mp\frac{(2\cos\Omega_{\mathbf{k}}^{\left(-\right)}+\sum_{j\neq i}\cos k_{j})\sin k_{i}}{2\left(3\cos\Omega_{\mathbf{k}}^{\left(-\right)}+\sum_{j}\cos k_{j}\right)\sin\Omega_{\mathbf{k}}^{\left(-\right)}},\label{veg3D2-1}
\end{align}
 and $\mathbf{v}_{\mathrm{g}}^{\left(5,6\right)}\left(\mathbf{k}\right)=0$.

We see that the extrema of the dispersion relations for branches 1,2,3
and 4 giving points of null velocity (those for which a Schrödinger
type equation is a priori well suited for extended initial conditions)
occur, in particular, when all $\sin k_{i}=0$. At some of these points
we find the just discussed degeneracies, hence at them a Schrödinger
type equation is not appropriate. However, some of the points of null
velocity do not correspond to degeneracies, in particular those $\mathbf{k}$
with two null components and the third one equal to $\pm\pi$. Consider,
for example, the point $\mathbf{k}_{0}=\left(0,0,\pm\pi\right)$ at
which $\Omega_{\mathbf{k}}^{\left(+\right)}$ presents degeneracies
but not $\Omega_{\mathbf{k}}^{\left(-\right)}$, which takes the value
$\Omega_{\mathbf{k}}^{\left(-\right)}=\arccos1/3$. In order to particularize
the wave equation (\ref{Sch}) to this case we must calculate the
nine coefficients 
\begin{equation}
\ \varpi_{ij}^{\left(3,4\right)}=\pm\left.\partial^{2}\Omega_{\mathbf{k}}^{\left(-\right)}/\partial k_{i}\partial k_{j}\right\vert _{\mathbf{k}=\mathbf{k}_{0}},
\end{equation}
which turn out to be $\varpi_{11}^{\left(3,4\right)}=\varpi_{22}^{\left(3,4\right)}=-\varpi_{33}^{\left(3,4\right)}/4=\mp1/4\sqrt{2}$
the rest being zero. Hence the continuous description in this case
is given by 
\begin{equation}
\mathrm{i}\frac{\partial A^{\left(3,4\right)}}{\partial t}=\pm\frac{1}{8\sqrt{2}}\left(\frac{\partial^{2}}{\partial X_{1}^{2}}+\frac{\partial^{2}}{\partial X_{2}^{2}}-4\frac{\partial^{2}}{\partial X_{3}^{2}}\right)A^{\left(3,4\right)},\label{Schr3D}
\end{equation}
which exhibits an anisotropic diffraction (diffraction in the $\left(X_{1},X_{2}\right)$
plane and anti-diffraction, with a different coefficient, in the $X_{3}$
direction). For $\mathbf{k}_{0}=\left(\pi,0,0\right)$ and $\mathbf{k}_{0}=\left(0,\pi,0\right)$
the result is the same after making the changes $X_{1}\rightarrow X_{3}$
and $X_{2}\rightarrow X_{3}$, respectively. Eq. (\ref{Schr3D}) can
be solved via a Fourier transform method. For a starting Gaussian
profile as given below, the solution will factorize in three one dimensional
functions of the form 

\begin{equation}
\frac{e^{-\frac{ix^{2}}{2t\varpi+2i\sigma^{2}}}}{\sqrt{\sigma^{2}-it\varpi}}
\end{equation}
 (except for an overall constant), where $\varpi$ is any of the coefficients
$\varpi_{ij}$ that appear in Eq. (\ref{Schr3D}), and $x$ represents
one of the coordinates $X_{i}$ with $i=1,2,3$. This implies a characteristic
time scale $t\sim\sigma^{2}/|\varpi|$ for the Gaussian to broaden. 

We now show two examples that illustrate the above results. As in
the 2D cases, we start with an initial condition (\ref{InixD}) defined
by a Gaussian shape 

\begin{equation}
\psi_{\mathbf{x},0}=\mathcal{N}e^{i\mathbf{k}_{0}\cdot\mathbf{x}}e^{-\frac{x_{1}^{2}+x_{2}^{2}+x_{3}^{2}}{2\sigma^{2}}},\label{initpsi3d}
\end{equation}
with $\mathcal{N}$ an appropriate constant defining the normalization,
and $\left\vert \Xi\right\rangle $ a constant (position-independent)
vector in the coin space, which is chosen to be one of the eigenvectors
of the matrix $C_{\mathbf{\mathbf{k}}}$ (\ref{Ck_elements}) at the
point $\mathbf{\mathbf{k}=k}_{0}$. We have not explored those cases
where $\mathbf{k}_{0}$ lies at, or very close to, any of the degeneracies
discussed above. The problem at these points seems to be much more
involved than in the 2D case. For example, the extension of the eigenvectors
obtained in \cite{Watabe} to this case encounters a singular behavior
at the axis (which are degeneracy lines). One needs to carefully take
care of the appropriate limit as one approaches these lines, and perform
a systematic study in order to find the relevant linear combinations
providing a sensible time evolution. Such a study is beyond the scope
of this paper.

\begin{figure}
\begin{minipage}[t]{1\columnwidth}%
\includegraphics[width=4cm,height=3.5cm]{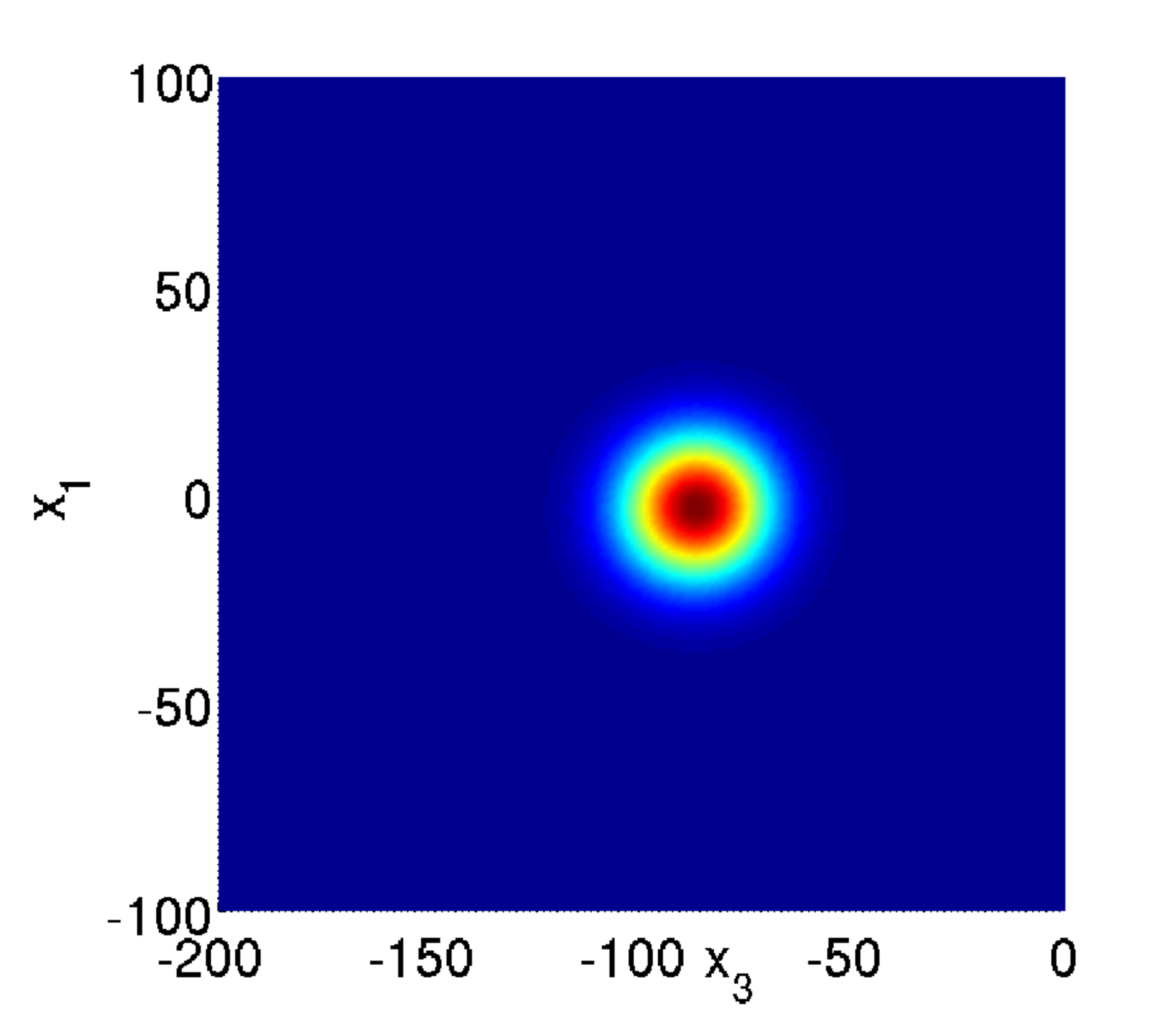}\includegraphics[width=4cm,height=3.5cm]{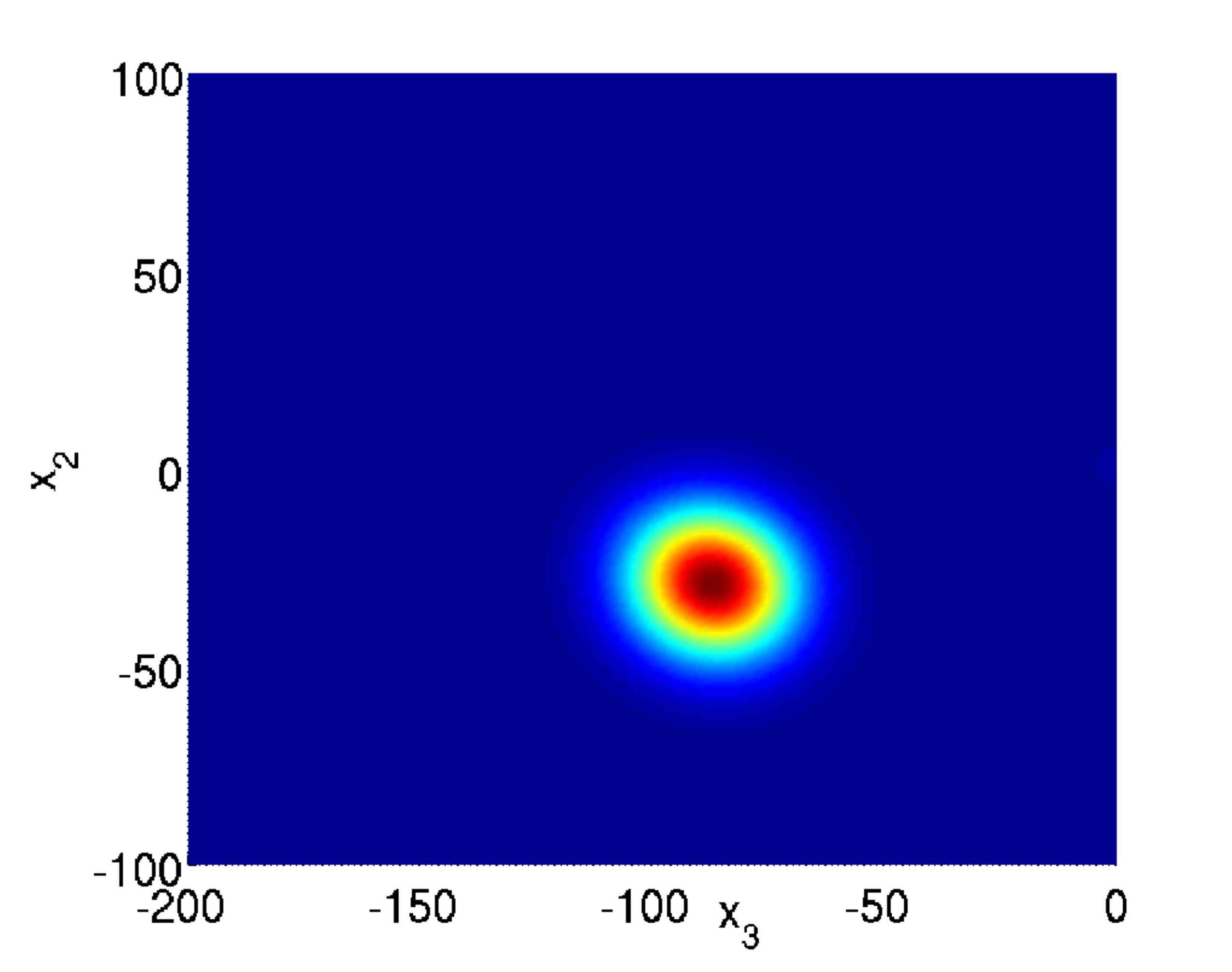}%
\end{minipage}

\caption{Top view of the probability distribution in position space on the
$x_{2}=0$ plane (left) and on the $x_{1}=0$ plane (right) for a
time step $t=125$. The initial distribution is given by Eqs. (\ref{InixD})
and (\ref{initpsi3d}), with $\mathbf{k}_{0}=(0.1,0.2,0.3)\pi$ and
$\left\vert \Xi\right\rangle $ the eigenvector of the coin operator
corresponding to $s=3$. }

\label{Fig0102033D}
\end{figure}

\begin{figure}
\includegraphics[width=8cm]{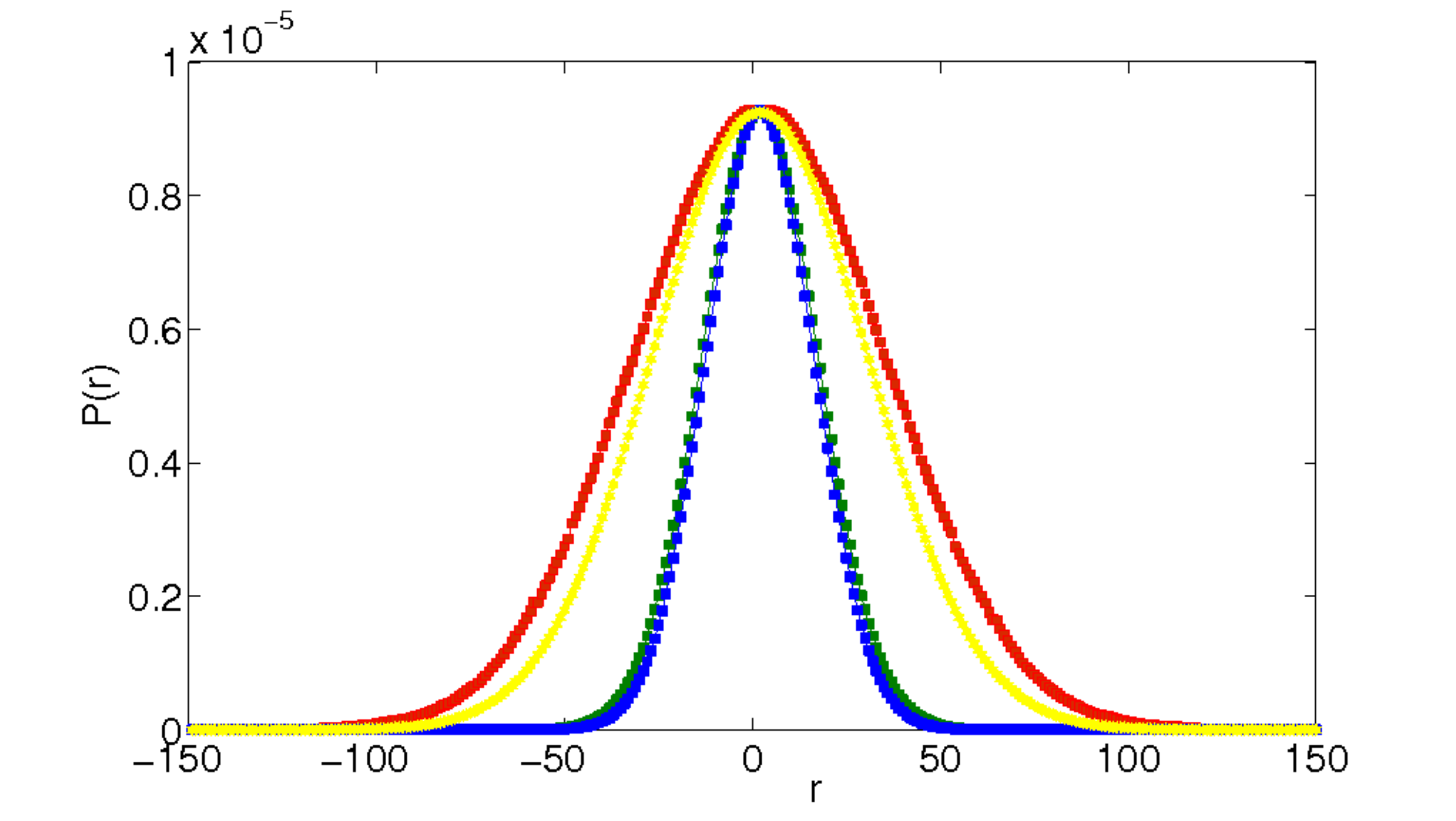}

\caption{Two different radial cuts of the evolved probability distribution,
represented by $P(r)$, where $r$ is the coordinate along the cut,
at $t=1000$ when the initial distribution is a spherically symmetric
Gaussian function with $\sigma=20$, with $\mathbf{k}_{0}=(0,0,\pi)$
and $\left\vert \Xi\right\rangle $ the eigenvector of the coin operator
corresponding to $s=3$. The figure also shows for comparison the
analytical result as obtained from Eq. (\ref{Schr3D}). The curve
with red symbols corresponds to a cut along the $x_{3}$ axis, to
be compared with the analytical approximation (yellow curve). The
curve with blue symbols is the plot for a cut along the $x_{3}$ axis,
to be compared with the analytical approximation (green). }

\label{Fig00Pi3D}
\end{figure}

We first start with a value $\mathbf{k}_{0}=(0.1,0.2,0.3)\pi$ that
lies far from any degeneracy or points with zero group velocity. For
$s=3$, the group velocity is $\mathbf{v}_{\mathrm{g}}^{\left(3\right)}\left(\mathbf{k}_{0}\right)=(-0.028,-0.232,-0.704)$.
In Fig. \ref{Fig0102033D} we show two different views corresponding
to the evolved probability distribution. 
 Within the timescale displayed on this Figure, the initial Gaussian
has not appreciably broaden, and we can instead observe the motion
of the center of the wave packet according to what is expected from
the group velocity. A somehow opposite behavior is observed if we
choose a point giving a zero velocity, such as $\mathbf{k}_{0}=(0,0,\pi)$.
As described above, we expect no motion of the center and a broadening
which depends on the axis we choose. This broadening can be approximately
described by Eq. (\ref{Schr3D}) and, for the values of $\sigma$
we are assuming here, one needs to follow the time evolution during
a considerably larger time scale. As a consequence, higher order effects
neglected in deriving Eq. (\ref{Schr3D}) may accumulate. In Fig.
\ref{Fig00Pi3D} we show two different radial cuts of the 3D probability
distribution. 
As observed from this Figure, the accumulated discrepancy between
the exact numerical evolution and the approximated analytical Eq.
(\ref{Schr3D}) may depend on the spatial direction. However, given
all these considerations, we see that our derived analytical result
gives a reasonable description for the main features of the time evolution,
even in 3D.

\section{Conclusions}

In this article we have formally solved the standard multidimensional
QW, taking into account extended initial conditions of arbitrary shape
and width. In the limit of large width, the continuous limit is particularly
well suited and clearly reveals the wave essence of the walk. The
use of the dispersion relations is the central argument behind this
view as its analysis allows to get much insight into the propagation
properties of the walk, even allowing for the tailoring of the initial
distribution in order to reach a desired asymptotic distribution,
as we have demonstrated. We must insist here in that our goal was
not to obtain approximate solutions to the QW, thus quantifying their
degree of accuracy, but rather to get qualitative insight into the
long term solutions. The equations we have derived will be quite accurate
for wide initial conditions, but the interesting point is that they
provide a good qualitative description for relatively narrow initial
distributions, specially far from degeneracies.

We have also shown that the two dimensional Grover QW exhibits a diabolical
point in its dispersion relations, and have analyzed in detail the
dynamics around this point, closely connecting the Grover walk with
the phenomenon of conical refraction \cite{Berry2,Berry3}. In the
three--dimensional Grover walk we have found other types of degeneracies
whose influence on the dynamics we have not illustrated. The detailed
study of the construction of the eigenvectors providing a clear qualitative
interpretation turns out to be much more complicated in the 3D than
in the 2D case, and we leave this extensive study for a future work. 

We want to insist in that the asymptotic distributions found in the
continuous limit, valid for large initial distributions, can be qualitatively
reached for not very broad distributions, say $\sigma\sim5$, and
this should be easy to implement in systems such as the optical interferometers
of Ref. \cite{Schreiber,Schreiber12}. An exception to this general
rule are those distributions which are peaked close to the diabolic
point, since in this case a broad distribution will be dominated by
the singular nature of that point.

As a final comment we would like to say, on the one hand, that the
multidimensional QW can be viewed as a simulator of a large variety
of linear differential equations depending on the particular region
of the dispersion relation that governs the evolution of the initial
wave--packet. On the other hand, we stress that continuous approximations
to the QW must follow the dispersion relation if they are to be taken
as good approximations.

\section*{Acknowledgments}

This work was supported by Projects FIS2011-26960, FPA2011-23897 of
the Spanish Government\emph{, Generalitat Valenciana} grant PROMETEO/2009/128
and FEDER. A.R. acknowledges financial support from PEDECIBA, ANII,
Universitat de València and Generalitat Valenciana.


\begin{thebibliography}{10}
\bibitem{SD}Obviosly this width will be different, in general, along
different directions in $\mathbf{k}$-space. However we are just interested
in an order-of-magnitude estimate for $\Delta k^{\left(s\right)}$.
To be conservative this width can be taken as the largest standard
deviation of $\left\vert \tilde{F}_{\mathbf{k}}^{\left(s\right)}\right\vert ^{2}$
in any direction or, more formally, as the largest eigenvalue of the
covariance matrix $\sigma$, such that $\left(\sigma^{2}\right)_{ij}=\int\mathrm{d}^{N}\mathbf{k}\left(k_{i}-\bar{k}_{i}\right)\left(k_{j}-\bar{k}_{j}\right)\rho_{\mathbf{k}}^{\left(s\right)}$
with $\rho_{\mathbf{k}}^{\left(s\right)}=\left\vert \tilde{F}_{\mathbf{k}}^{\left(s\right)}\right\vert ^{2}/\int\mathrm{d}^{N}\mathbf{k}\left\vert \tilde{F}_{\mathbf{k}}^{\left(s\right)}\right\vert ^{2}$
and $\bar{k}_{i}=\int\mathrm{d}^{N}\mathbf{k}k_{i}\rho_{\mathbf{k}}^{\left(s\right)}$,
$i,j\in\left\{ 1,\ldots,N\right\} $. In our case $\bar{k}_{i}=0$
as $\tilde{F}_{\mathbf{k}}^{\left(s\right)}$ is assumed to be centered
at the origin, hence $\tilde{F}_{\mathbf{k-k}_{0}}^{\left(s\right)}$
is centered at $\mathbf{k}_{0}$. Indeed it is this condition that
determines the value of $\mathbf{k}_{0}$.
\bibitem{null measure}The only problematic case\textbf{ }would be
when $\tilde{F}_{\mathbf{k-k}_{0}}$ equals the Dirac delta $\delta^{N}\left(\mathbf{k}-\mathbf{k}_{\mathrm{D}}\right)$.
This is however an unphysical case as it represents an unnormalizable
state.
\end{thebibliography}
\end{document}